\documentclass{aa}
\usepackage{graphicx}
\usepackage[varg]{txfonts}
\usepackage{natbib}
\usepackage[caption=false]{subfig}
\usepackage[linktocpage]{hyperref}
\usepackage{xcolor}
\usepackage{threeparttable}
\usepackage{etoolbox}
\usepackage{comment}
\usepackage{placeins}

\newcommand{\nombre}{{\sc UmScanGalactiK~}}

\begin{document}
   \title{On the application of dimensionality reduction and clustering algorithms for the classification of kinematic morphologies of galaxies}

    \author{M. S. Rosito \inst{1}, L. A. Bignone \inst{1}, P. B. Tissera \inst{2, 3}, S. E. Pedrosa \inst{1} } 

   \institute{Instituto de Astronom\'{\i}a y F\'{\i}sica del Espacio, 
CONICET-UBA, Casilla de Correos 67, Suc. 28, 1428, Buenos Aires, Argentina.
     \and
        Institute of Astronomy, Pontificia Universidad Cat\'{o}lica de Chile, Avenida Vicuña Mackena 4690, Santiago, Chile.
    \and
       Centro de Astro-Ingenier\'{\i}a, Pontificia Universidad Cat\'{o}lica de Chile, Avenida Vicuña Mackena 4690, Santiago, Chile. 
	}
	\abstract
   {The morphological classification of galaxies is considered a relevant issue and can be  approached from different points of view. The increasing growth in the size and accuracy of astronomical data sets brings with it the need for the use of automatic methods to perform these classifications.}
   {The aim of this work is to propose and evaluate a method for  automatic unsupervised classification of kinematic morphologies of galaxies that yields a meaningful clustering and captures the variations of the fundamental properties of galaxies. }
   {We obtain kinematic maps for a sample of 2064 galaxies from the largest simulation of the {\sc eagle} project that mimics integral field spectroscopy (IFS) images. These maps are the input of a dimensionality reduction algorithm followed by a clustering algorithm. We analyse the variation of physical and observational parameters among the clusters obtained from the application of this procedure to different inputs. The inputs studied in this paper are (a) line-of-sight velocity maps for the whole sample of galaxies observed at fixed inclinations, (b) line-of-sight velocity, dispersion, and flux maps together for the whole sample of galaxies observed at fixed inclinations, (c) line-of-sight velocity, dispersion, and flux maps together for two separate subsamples of edge-on galaxies with similar amount of rotation, and (d) line-of-sight velocity, dispersion, and flux maps together for galaxies from different observation angles mixed. }
   { The application of the method to solely line-of-sight velocity maps achieves a clear division between slow rotators (SRs) and fast rotators (FRs) and can differentiate rotation orientation. By adding the dispersion and flux information at the input, low rotation edge-on galaxies are separated according to their shapes and, at lower inclinations, the clustering using the three types of maps maintains the overall information obtained using only the line-of-sight velocity maps. This method still produces meaningful groups when applied to SRs and FRs separately, but, in the first case, the division into clusters is less clear than when the input includes a variety of morphologies. When applying the method to a mixture of galaxies observed from different inclinations, we obtain results that are similar to those in our previous experiments with the advantage that in this case the input is more realistic. In addition, our method has proven to be robust to consistently classify the same galaxies viewed from different inclinations. }
   {}


\authorrunning{M. S. Rosito et al.}
\titlerunning{Unsupervised algorithms for the classification of galaxies}
\maketitle
%------------

\section{Introduction}

Since galaxy morphology is the result of complex processes involved in the assembly history of galaxies, a morphological description is crucial to trace these processes and to achieve a better understanding of galaxy evolution \citep[e.g. see][for a recent review]{Conselice2014}. 
Historically, galaxies have been classified visually through their apparent morphology \citet{Hubble1926}.
The need of the human eye is an obvious difficulty \citep{Raddick2007} and the inaccuracies of visual classification are partially solved by the use of different quantitative measurements. However, while visual classification is still required \citep{galaxyzoo}, the vast photometric data sets that are being gathered requires the development of reliable, efficient, and fast classification methods.

The structure of galaxies has been quantified through parameters related to their light profiles \citep{deVaucouleurs1948, Sersic1968}. In particular, S\'ersic profile, which describes the shape of the surface density profiles, varies according to galaxy morphology and can be used to establish a separation between classical bulges and pseudo-bulges by using the S\'ersic index\citep[e.g.][]{Combes2009, Fisher2008, Tonini2016}.
Galaxy classification can also be based on a combination of parameters such as colours and S\'ersic index \citep{Vika2015}.

The disc (bulge)-to-total light ratio defines the position of a galaxy in the Hubble sequence \citep{Hubble1926} and is widely used in observational  works \citep[e.g.][]{Kormendy2012, Yoon2020}.
In numerical simulations, the availability of physical information allows the definition of the disc-to-total mass ratio ($D/T$), which is used to distinguish disc and spheroid dominated galaxies \citep[e.g.][]{Pedrosa2015, Tissera2016, Tissera2016b, Rosito2018, Rosito2019a, Rosito2019b}.
A negative correlation can be found between $D/T$ and the aforementioned S\'ersic index, as shown in \cite{Rosito2018}.

Another important parameter used for galaxy classification is the mean rotation-to-dispersion velocity ratio, \citep[$V/\sigma$, e.g.][]{Chisari2015, Dubois2016} which is tightly correlated to the disc fraction \citep{Rosito2021}.
Furthermore, the three dimensional axis of the ellipsoid enclosing a particular mass can be easily obtained from the simulations \citep{Tissera1998} providing, thus, a direct measurement of the shape of a galaxy.
The axis ratios are used to describe the prolateness/oblateness \citep{Artale2019, Cataldi2021, vandeven2021} and can also be correlated with rotation \citep{Rosito2019b}.

In contrast, non-parametric methods measure the distribution of light in galaxies without assumptions related to the stellar mass or light distributions.  
Galaxy structure can be quantitatively described by these methods through the CAS system, which combines concentration (C), asymmetry (A), and clumpiness (S) of the stellar light distribution  \citep{Conselice2003}, and through a number of similar parameters, such as the Gini index, the $M20$, and the  internal colour dispersion statistic \citep{Abraham2003, Lotz2004, Papovich2003}. These parameters can be used to define classification criteria \citep[e.g.][]{Deng2013}.
Besides, non-parametric statistics correlate with other morphology indicators: disc-to-total ratio \citep{scan2008}, the $\kappa_0$ parameter \citep{Correa2017}, and the ratio of rotation and dispersion velocities  \citep{Bignone2020}. 

Quantitative and accurate classification of galaxies according to kinematics has been possible with the advent of integral-field spectroscopy (IFS) techniques, being the SAURON project \citep{Bacon2001} disruptive and followed by galaxy surveys like CALIFA \citep{CALIFA2012}, MaNGA \citep{Bundy+2015} and SAMI \citep{Bryant2015}.
The distinction between slow rotators (SRs) and fast rotators (FRs) was first introduced by \cite{Emsellem2007} for early-type galaxies and has been widely studied from an observational point of view \citep[e.g.][]{Veale2017, Brough2017, Greene2018}.
\cite{Emsellem2007} and subsequent works \citep{Emsellem2011, cappellarireview2016, vdSande2021} propose classifications in SRs and FRs based on observational projected parameters.
Other authors consider definitions based on physical quantities, for instance by fixing a lower bound for the bulge-to-total mass ratio \citep[e.g.][]{Rosito2019b}.
The study of kinematics sheds light on fundamental concerns about galaxies.
Using the  {\sc eagle} cosmological simulations \citep{Crain2015, Schaye2015} and the HYDRANGEA zoom-in runs \citep{Brae2017}, \cite{Lagos2018} connect kinematic properties of galaxies with their formation paths. In particular, they find a correlation between the amount of rotation of galaxies and their merger histories, being the SRs more likely to have experienced dry major mergers, whereas wet mergers are more frequent in FRs.  
Other studies of numerical simulations find that mergers are able to transform kinematic properties which can provide clues to understand galaxy evolution \citep{Jesseit2009, Bois2011, Naab2014, Penoyre2017, Schulze2018}.

Due to the increasing depth and resolution of new surveys \citep[e.g.][]{Euclid2011, LSSTScienceBook, LSST2019} and the continuous increasing in numerical resolution of cosmological simulations, the use of automatic methods is becoming mandatory \citep[e.g.][]{Ball2010, Kremer2017, Howard2017}. 
The availability of more computational resources leads to the generation of large data sets from simulations as well as making possible to analyse data accurately and obtain robust conclusions regarding the underlying physics.

Machine learning (ML) is rapidly gaining ground in different fields of astronomy \citep[see][for a recent review]{Baron2019}.
The idea of the use of artificial neural networks for the morphological classification of galaxy images has been in place since the end of last century \citep{Storrie1992, Lahav1995}.
In the past few years, the application of several supervised ML techniques to galaxy classification has been widely studied and tested, obtaining increasing accuracy and performance.
For instance, \cite{Marin2013} achieve accuracies of 79 per-cent and 91 per-cent for naive Bayes and random forest classifiers, respectively. 
\cite{Selim2016} present a method for supervised classification based on non-negative matrix factorization obtaining an accuracy of 93 per-cent.
Convolutional neural networks, on the other hand, are commonly used in computer vision and are considered, thus, ideal tools to analyse galaxy images. As an example, \cite{Khalifa2017} achieve an accuracy of 97 per-cent for galaxy image classification. 
This can be improved through the addition of data augmentation techniques to overcome overfitting, as shown by \cite{Mittal2019}.
\cite{Tohill2021} use convolutional neural networks to predict non-parametric quantities from galaxy images. 
Their method outperforms other non-parametric measurement algorithms in the sense that it is more than 1000 times faster, as well as providing lower bias estimates with lower scatter.  
\cite{deDiego2020} show that deep neural networks are more accurate than other classification methods related to photometry and shape parameters (S\'ersic index and concentration index) when applied to modern deep surveys \citep{Bongiovanni2019}.
All these methods are based on supervised algorithms which rely deeply on the availability of training sets.

Unsupervised ML is particularly relevant to extract new information from a data set since unsupervised algorithms are trained with unlabelled data.
Since galaxies can be classified according to a variety of criteria, unsupervised classification can be considered ideal to automatically obtain meaningful groups.
A completely unsupervised technique presented by \cite{Hocking2018} is able to successfully separate images of early and late type galaxies besides being very computationally efficient. 
Dimensionality reduction algorithms based on principal components analysis (PCA) are used to study images from galaxy surveys with a number of applications such as outlier detection and prediction of missing data \citep{Uzeirbegovic2020}.
\cite{Potrillo2020} use variational autoencoders trained with a sample of high dimensional spectra and obtain an interpretable latent space able to separate galaxies with notably different properties related star formation and active galactic nuclei (AGN). In this work, they outperform PCA with the same number of components. 
Unsupervised clustering aims to group objects with similar properties and may follow dimensionality reduction algorithms.
Classical algorithms like k-means have been used to classify galaxy spectra being able to separate galaxies according to colours \citep{Almeida2010} but still have difficulties in identifying stars with different metallicities \citep{Almeida2013}.
In a recent work, \cite{Cheng2021} employ a combined technique which consists on a variational autoencoder followed by hierarchical clustering. They obtain a meaningful classification for the input images formed by 27 groups with well defined properties related to structure and shape finding also a correlation with physical properties.

An intermediate form between supervised and unsupervised learning is self-supervised learning.
Commonly applied to natural language processing and robotics, self-supervised learning is also used in astronomy. 
A novel self-supervised contrastive learning method presented by \cite{Sarmiento2021} takes kinematic and stellar population maps obtained from MaNGA \citep{Bundy+2015} and is able to produce two groups: one formed by old, metal-rich massive galaxies compatible with early-type galaxies and the other populated by low-mass, star-forming galaxies which can be associated with late-type galaxies.

The goal of our work is the automatic unsupervised classification of kinematic maps.
We propose a method based on the application of the Uniform Manifold Approximation and Projection ({\sc UMAP}) algorithm \citep{McInnesUMAP2018}, which is a non-linear dimensionality reduction technique, followed by the Hierarchical Density-Based Spatial Clustering of Applications with Noise \citep[][ \texttt{\sc HDBSCAN}]{CampelloHDBSCAN2013} algorithm with the aim at clustering a set of mock kinematic maps that mimic images obtained through IFS. Hereafter, we will refer to our method as \nombre.

For this purpose, we implement and test our  method in a galaxy catalogue constructed by \citet{Tissera2019} from the larger volume simulation of the {\sc eagle} project \citep{Schaye2015, Crain2015}, for which  visual classification is available.
The kinematic maps are generated with the R-package {\sc SimSpin} \citep{Harborne2020}.
Our method  is sensitive to both, physical and observational projected parameters, computed directly from the simulation.
Hence, \nombre is a simple and fast method to cluster galaxies which yields a classification that takes into account the main properties of galaxies.

This paper is organized as follows. In Section \ref{sec:simu} we describe the  {\sc eagle}  project  \citep{Schaye2015, Crain2015} and the simulation used for this study and our galaxy sample. The methodology we follow is presented in Section \ref{sec: method}. In Sections \ref{sec:vlos} and \ref{sec:call} we show and discuss our main results obtained through our method applied to different sets of kinematic maps.
In Section \ref{sec:SRFRs}, we study the clustering of SRs and FRs separately.
We discuss the robustness of this method and its applicability to more realistic cases in Section \ref{sec:mix} and we finally conclude in Section \ref{sec: conclusions}.

\section{Simulated galaxies}
\label{sec:simu}

In this study, we use a sample of galaxies selected from the larger volume simulation of the {\sc eagle} project \citep{Crain2015, Schaye2015}, which is a suite of hydrodynamic cosmological simulations that aims to follow the formation of structures and are consistent with a $\Lambda$-CDM universe\footnote{We use the {\sc eagle} public database \citep{McAlpine2016}.}.

The cosmological parameters are consistent with  the Planck cosmology \citep{Planck2014a, Planck2014b}: $\Omega_m = 0.307$, $\Omega_{\Lambda} = 0.693$, $\Omega_b = 0.04825$, $H_0 = 100 h$ km s$^{-1}$ Mpc$^{-1}$, being $h = 0.6777$. For this work we use the  simulation, so-called L100N1504, which corresponds to a  100 Mpc side box and it uses $1504^3$ dark matter and initial baryonic particles.
The mass resolution is  $9.70 \times 10^6$ M$_{\odot}$  and $1.81 \times 10^5$ M$_{\odot}$ for the dark matter and initial gas particles. 
The  gravitational softening ($0.7$ pkpc, proper kilo-parsecs) is kept constant in proper units below $z=2.8$; at higher $z$ the softening is kept constant in co-moving units
at 2.66$\sim$ckpc.

The {\sc eagle} simulations were performed by  using a modified version of the {\sc gadget-3} code described in \cite{Springel+2005}. 
This version includes radiate cooling \citep{wiersma2009}, reinonization \citep{Haardt2001}, stochastic star formation \citep{Schaye2008}, stellar \citep{Dallas2012} and AGN feedback \citep{Rosas2015}. A \cite{Chabrier2003} initial mass function (IMF) is used.
The dark matter haloes are identified  with a Friends-of-Friends algorithm \citep[{\sc FoF}]{Davis1987} and the gravitionally bound subhaloes are then selected  by using a {\sc SUBFIND} algorithm \citep{springel2001, Dolag2009}.

In this work, we analyse a subsample of galaxies  comprising 7482 central galaxies at $z=0$ selected by \cite{Tissera2019}. Central galaxies are defined as the most massive systems within the virial halo. 

To achieve a more accurate unsupervised classification, it is relevant to include in our sample galaxies with diverse morphologies and kinematic properties to assess, thus, the ability of our method to capture these features. 
Since the fraction of SRs, regardless of the criteria adopted to classify them, 
significantly increases with increasing stellar mass \citep{Lagos2018}, we choose galaxies with stellar masses above $10^{10}$ M$_{\odot}$ to ensure the inclusion of a non-negligible number of these type of galaxies.
The stellar masses are measured within 1.5 optical radius, defined as the radius that enclosed $\sim 80$ per-cent of the stellar mass of the galaxy \citep[][]{tissera2000}.
Our final sample consists, thus, on 2064 massive central galaxies.
As a morphological indicator, we compute the disc-to-total stellar mass ratio, $D/T$. This is done by means of the dynamical decomposition based on the binding energy and angular momentum content of the stellar particles as described by \cite{tissera2012}.

\section{Methodology}
\label{sec: method}

As mentioned in the Introduction, we propose a method that automatically classifies mock galaxy kinematic maps by the combination of the {\sc UMAP} dimensionality reduction algorithm \citep{McInnesUMAP2018} and the clustering algorithm {\sc HDBSCAN} \citep{CampelloHDBSCAN2013}.
The groups obtained as a result of the application of these algorithms can be used to classify galaxies according to morphology and kinematics.
Python implementation for {\sc UMAP}\footnote{https://github.com/lmcinnes/umap} and {\sc HDBSCAN}\footnote{https://github.com/scikit-learn-contrib/hdbscan} are publicly available on-line.

The data feeding the aforementioned algorithms is a set of mock kinematic maps of the sample of galaxies described in Section \ref{sec:simu}. 
These maps are obtained from IFS data cubes built with the R-package {\sc SimSpin}\footnote{https://github.com/kateharborne/SimSpin} presented by \cite{Harborne2020}.

Below, we briefly describe each algorithm as well as the generation of the kinematic maps and discuss how we use them throughout this work. 

\subsection{Kinematic maps}
\label{sec:kinematic}

Having selected the set of galaxies to be analysed in this work, we use the {\sc SimSpin} R-package \citep{Harborne2020} to build their IFS kinematic data cubes. 
This package is provided with information of each galaxy.
All galaxies in our sample were previously rotated so that the total angular momentum is parallel to the $z$-axis.
In this work, we use only stellar particles from which we extract their position, velocity, mass, age, metallicity, and initial mass.
The age, metallicity, and initial mass are used to model the spectral energy distributions for each stellar particle using the stellar population synthesis models of \cite{BC2003}.

For this study, we compute the line-of-sight velocity, velocity dispersion, and flux maps at different inclinations generated by {\sc SimSpin} from the kinematic data cubes by collapsing the cubes along the velocity axis ($z$-axis).
Pixels of the flux maps are computed by summing the contributions of each flux plane along the velocity axis, whereas line-of-sight velocity and dispersion maps represent the flux-weighted average of the velocity and the dispersion along the velocity axis of the data cubes \citep[see][for details]{Harborne2020}.
The images mimic the observational parameters of SAMI \citep{Scott2018}. In particular, the spatial pixel size is 0.5 arcsec.
We need our galaxies to be reasonably well resolved.
Hence, we choose a projected distance to each galaxy of $z=0.05$ obtaining, thus, a physical size or aperture of 15 kpc side.

We also use {\sc SimSpin} to compute physical and projected observational parameters.
The former include the 3D axis ratios $b/a$ and $c/a$ within half-mass, being $a \geq b \geq c$ and the spin parameter defined by \cite{Bullock2001}, $\lambda_B$, obtained from its radial profile computing the mass weighted average within the half-mass radius.
Regarding the observational parameters, we calculate the projected spin parameter \citep[$\lambda_R,$][]{Emsellem2007} and the projected ellipticity, $\varepsilon = 1 - b/a$, 
where $a$ and $b$ are the projected major and minor semi-axis for a given inclination, respectively.
Both parameters are measured within the projected effective radius.
It is important to clarify that we are not able to compute an accurate value of $\lambda_R$ for galaxies with projected effective radii larger than the aperture, so, we exclude those galaxies in the analysis of the projected properties. The number of removed galaxies varies with the inclination, but in all cases it is between 40 and 50.

Throughout this paper, we study the variations of these parameters, as well as $D/T$ (Sec.~\ref{sec:simu}) and the triaxiality parameter, $T = (1-b^2/a^2) / (1 -c^2/a^2)$ among the galaxy groups, classified with \nombre, with the goal of understanding the meaning of the clustering.

\begin{figure*}
  \centering
    \includegraphics[width=0.3\textwidth]{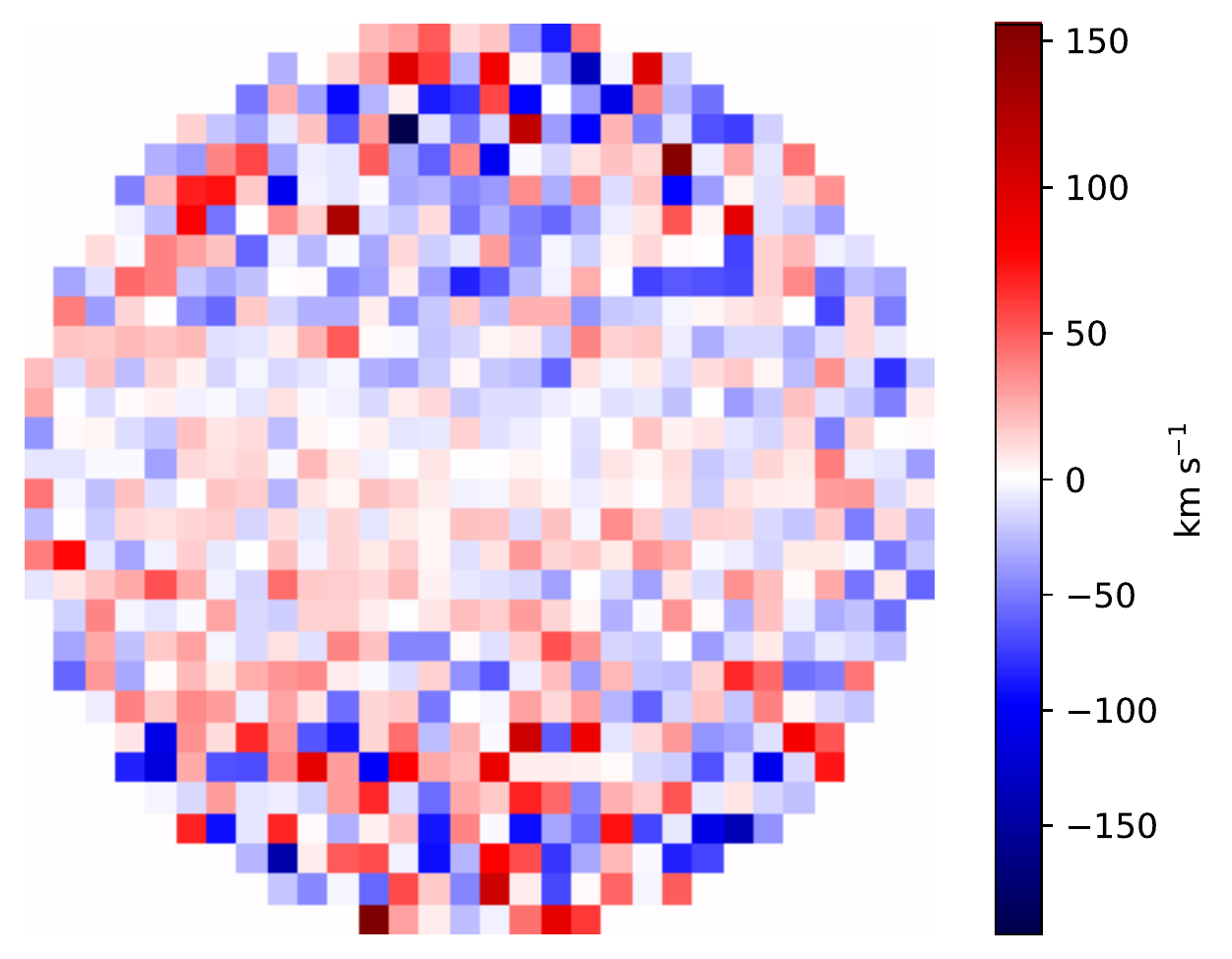}
    \includegraphics[width=0.29\textwidth]{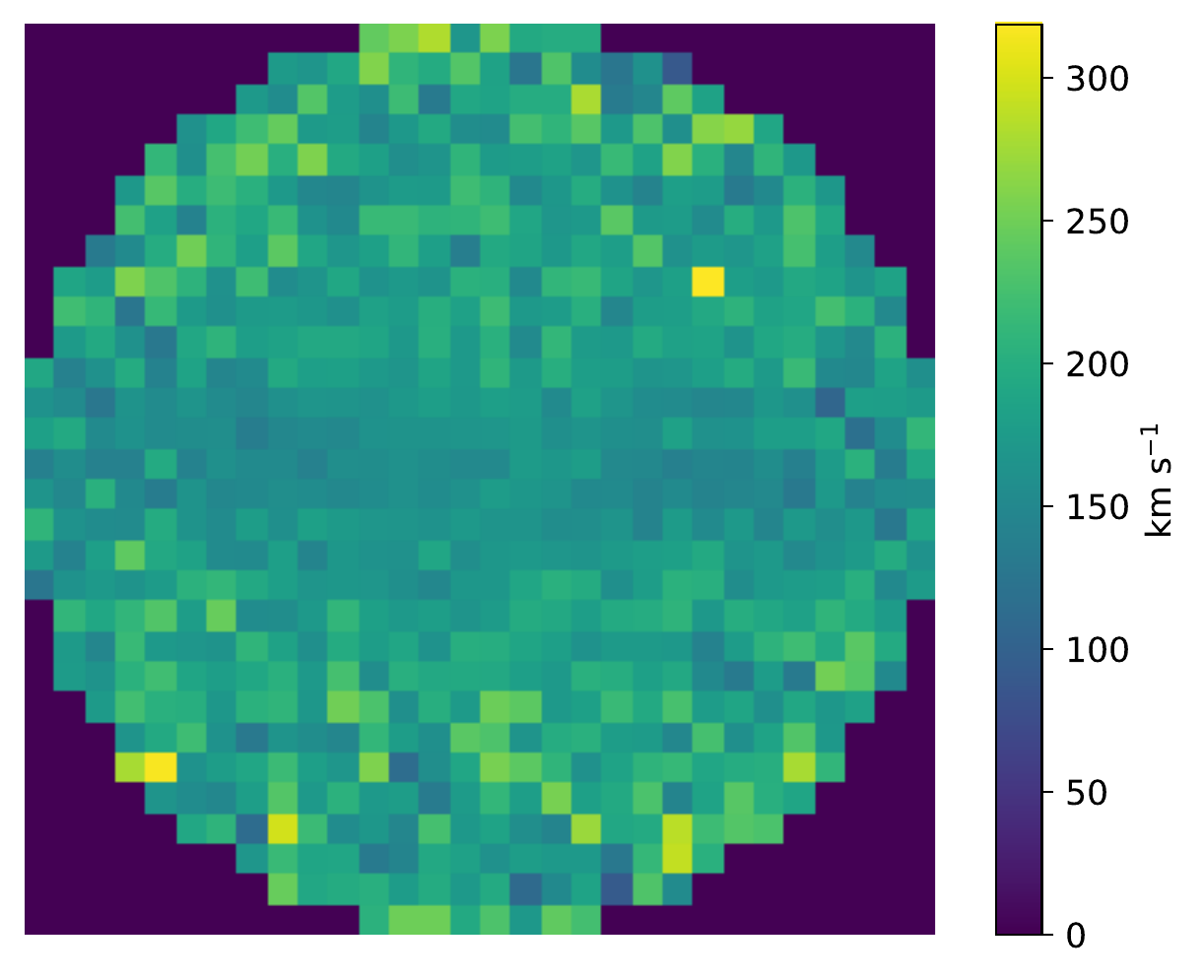}
    \includegraphics[width=0.245\textwidth]{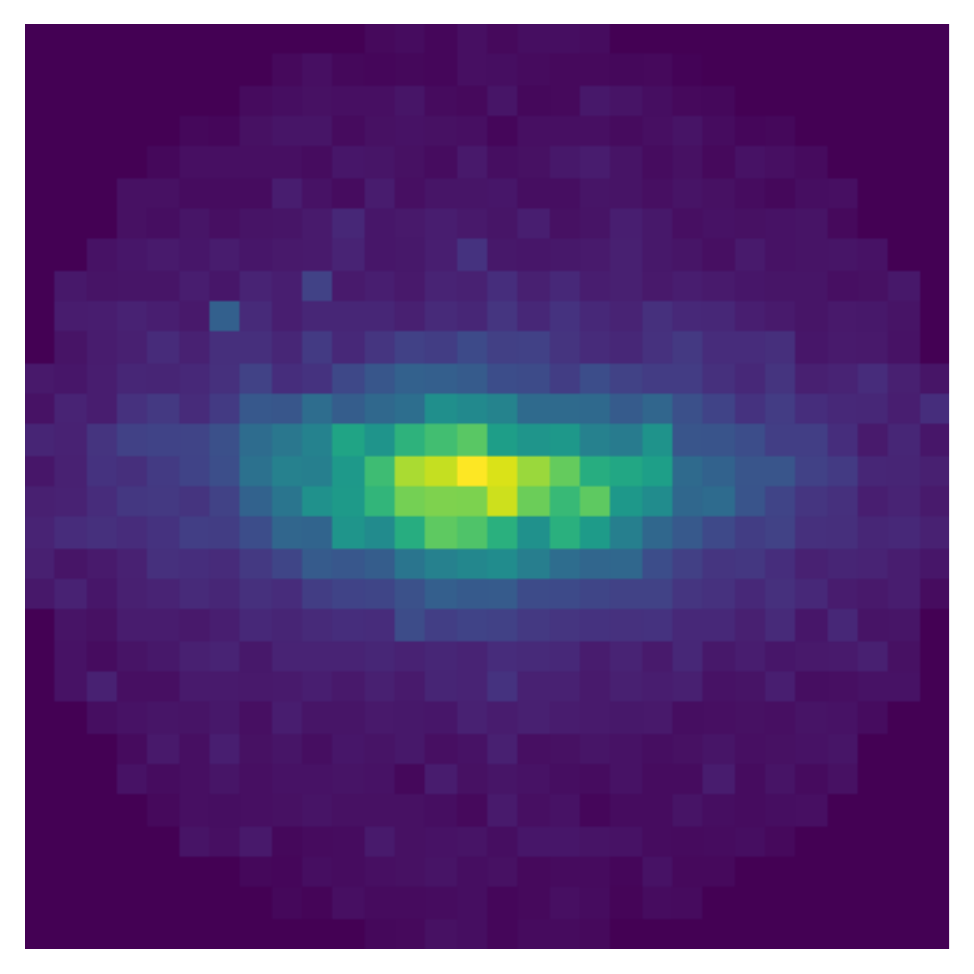} \\
    \includegraphics[width=0.3\textwidth]{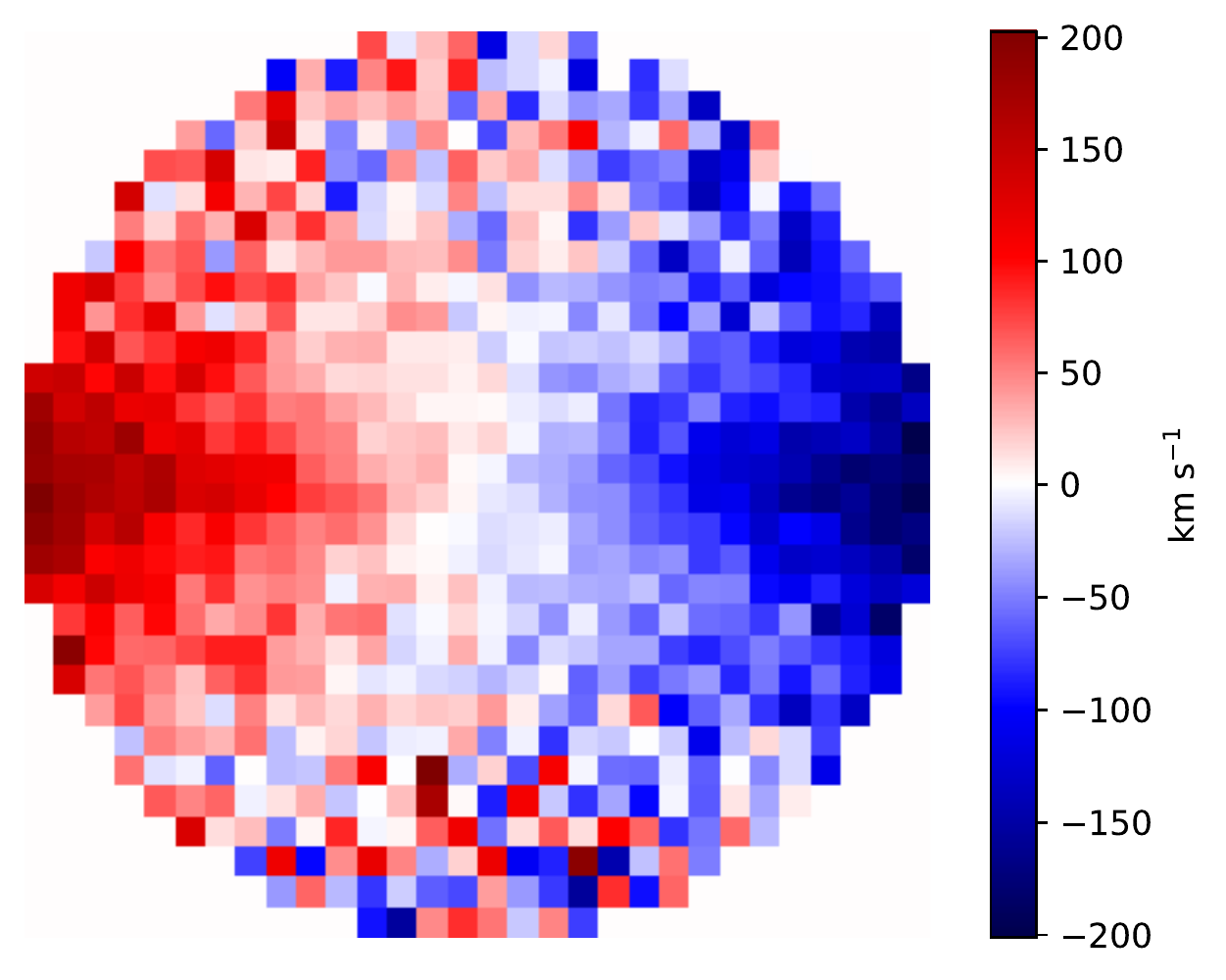}
    \includegraphics[width=0.29\textwidth]{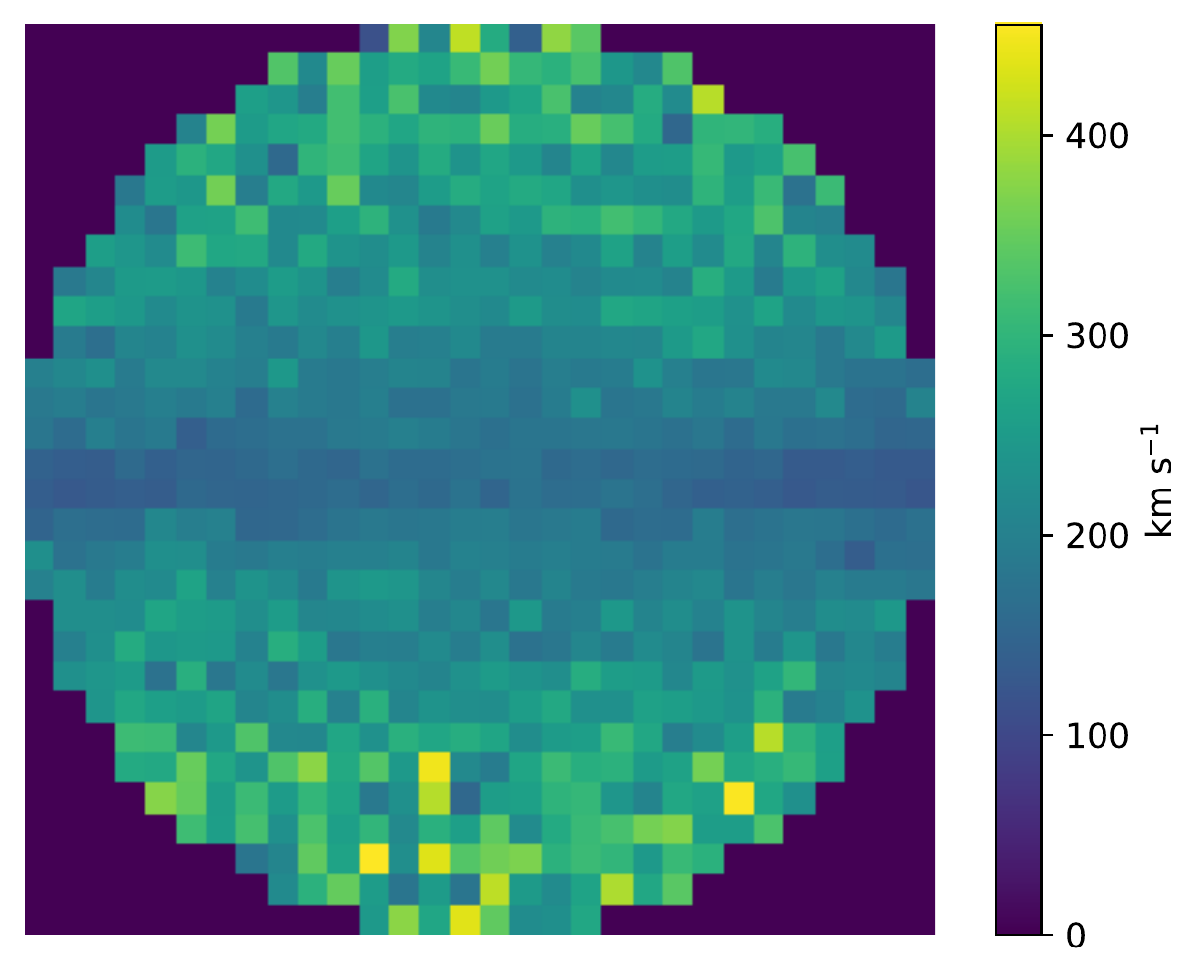}
    \includegraphics[width=0.245\textwidth]{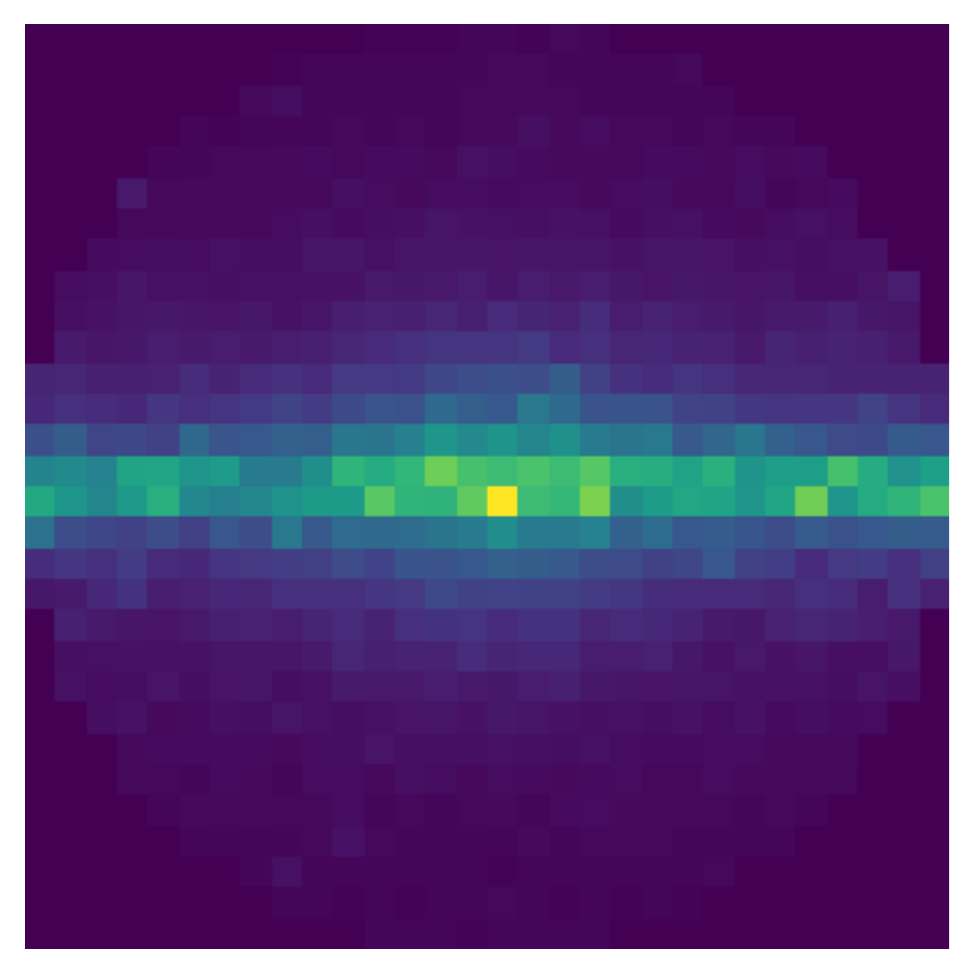} \\
    \includegraphics[width=0.3\textwidth]{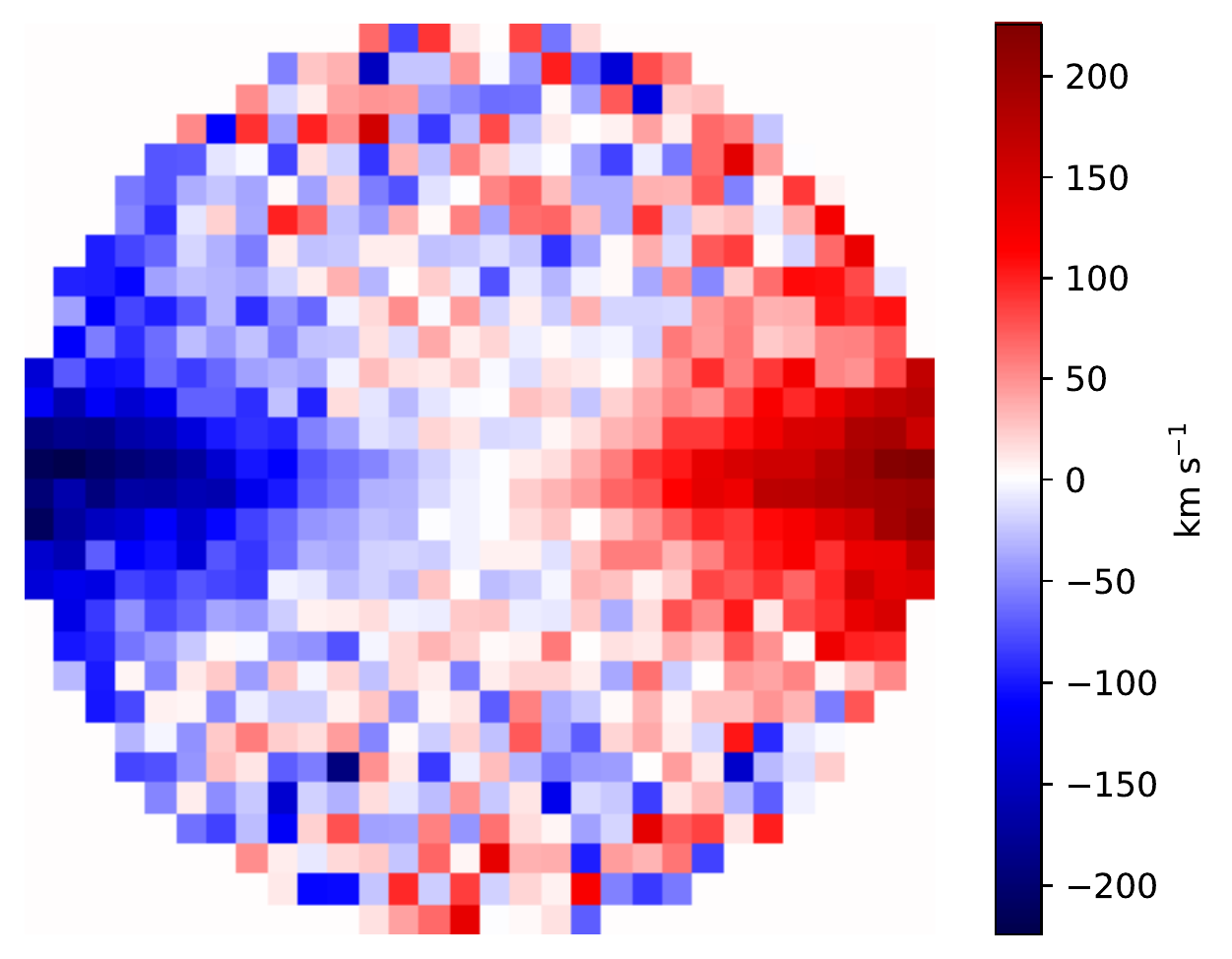}
    \includegraphics[width=0.29\textwidth]{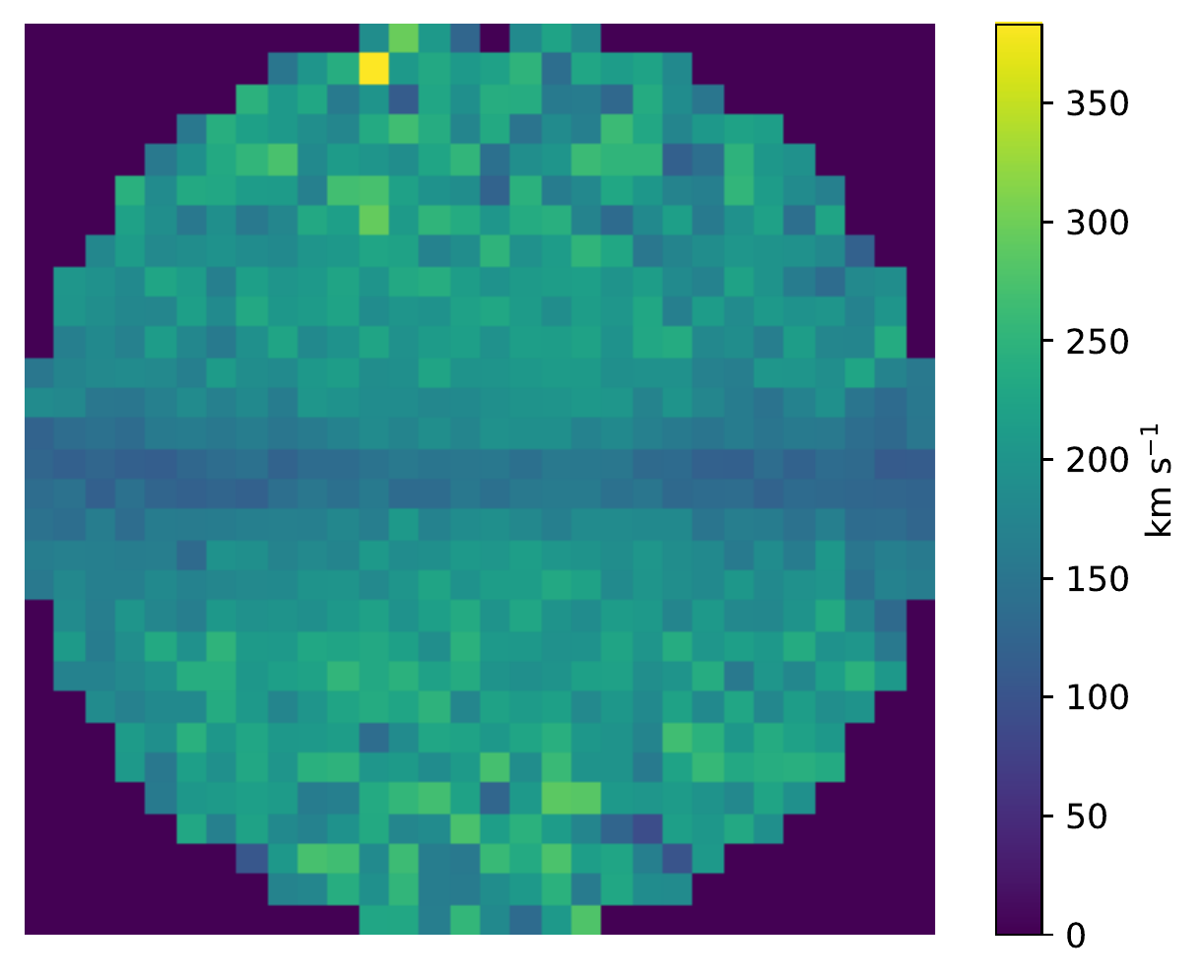}
    \includegraphics[width=0.245\textwidth]{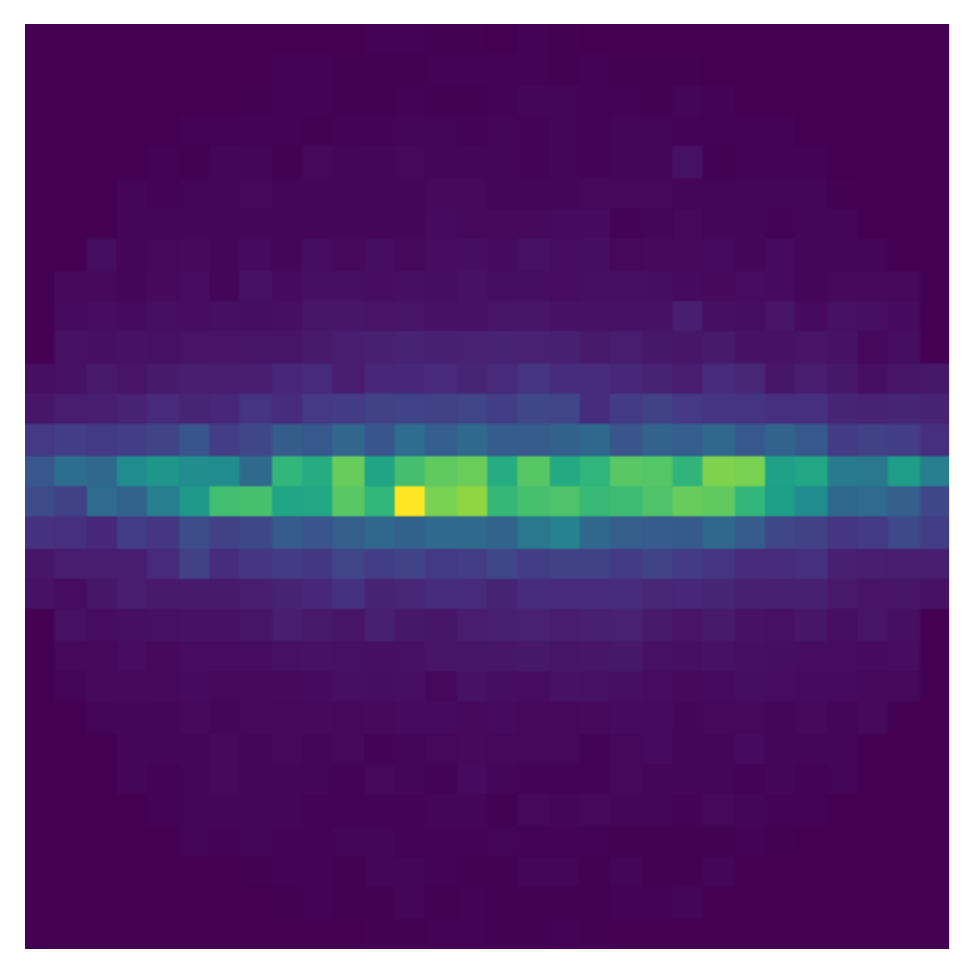} \\
  \caption{Examples of line-of-sight velocity (\textit{left panels}), dispersion (\textit{middle panels}), and flux (\textit{right panels}) maps for a galaxy with low rotation (\textit{upper panels}), a ``clockwise'' rotating galaxy (\textit{middle panels}), and a ``counterclockwise'' rotating galaxy (\textit{bottom panels}). The maps are obtained from the kinematic data cubes generated by {\sc SimSpin} by collapsing each data cube along the velocity axis, as described in \cite{Harborne2020}.}
  \label{fig:kinematicmaps}
\end{figure*}

In Fig.~\ref{fig:kinematicmaps} we show three examples of kinematic maps: a galaxy with little amount of rotation (upper panel) and two rotating galaxies with different directions.
The rotation orientation can be quantified by the sign of the average rotation velocity computed for each half of the line-of-sight velocity map, defining positive or negative whether the velocity is incoming or outgoing to the plane defined by the image.

\subsection{Dimensionality reduction algorithm}

High dimensional data may be difficult to handle and computationally expensive to process.
Dimensionality reduction techniques allow the transformation of high dimensional data into lower dimensional data minimizing the loss of 
information and facilitating, thus, the  visualisation and classification of the data sets.

{\sc UMAP} is a dimensionality reduction algorithm first presented by \cite{McInnesUMAP2018}. It serves as a manifold learning technique and, due to its good computational performance and scalability with dimensions and number of samples, it has a wide variety of applications in astronomy \citep[e.g. ][]{Reis2021, Kim2021}, as well as in other fields like bioinformatics and materials science. It is based on strong mathematical foundations that allow it to construct a representation in a low dimensional space, which is topologically equivalent to the representation of the high dimensional data \citep[see ][for details]{McInnesUMAP2018}.

In this work, we construct 30 px x 30 px kinematic images.
Each map is reshaped to obtained 900-dimensional vectors.
In Sec.~\ref{sec:vlos}, we study directly the line-of-sight velocity maps and therefore the input of the algorithm are 900-dimensional data points.
In Sec.~\ref{sec:call}, we include in the analysis the velocity dispersion and  the flux information by concatenating to the line-of-sight velocity maps, velocity dispersion and flux maps, increasing the input dimension to 2700. 
Because of the differences in the magnitudes among the different data, we apply a uniform linear normalization to each type of maps by dividing each value by the maximum absolute value of the components across the whole data set. Therefore the components of the velocity map are between -1 and 1,  whereas, for the dispersion and flux maps, the values lie in the interval [0,1]. 
In both cases, the high dimensional data is projected into a bidimensional plane obtaining thus an easy visualisation of the data and their subsequent clustering (see Appendix~\ref{app:dimensions} for a discussion about the number of components of the output space).

In Table~\ref{table:params} we show the {\sc UMAP} main (hyper)parameters we establish for each experiment in this work. In particular, n\_neighbors defines the number of points in the local neighborhood which the algorithm considers when it attempts to learn the manifold structure. Low values of n\_neighbors implies that the algorithm focuses on very local structure, whereas setting large values yields a more global representation of the data. Another (hyper)parameter is the minimum distance at which the projected points are allowed to be (min\_dist). When this (hyper)parameter is low, the embeddings are clumpier with small connected components, while higher values should be used to preserve the broad structure. The reduced dimension space in which the data will be embedded is set with the (hyper)parameter n\_components and the metrics used to compute distances in the input space defined by the metric (hyper)parameter.

\subsection{Clustering algorithm}

Clustering is the most commonly employed unsupervised ML algorithm and aims to group objects that are in some sense similar (high intra-cluster similarity) while objects belonging to different groups are not similar (low inter-cluster similarity).

We choose the {\sc HDBSCAN} \citep{CampelloHDBSCAN2013} algorithm as our second step. This method outperforms other density-based clustering algorithms and is often used in astronomy \citep[e.g.][]{Katz2016, Kimm2018}, as well as in a variety of disciplines like malware analysis, bioinformatics, and molecular dynamics. 
It is an ideal tool for unsupervised learning to define groups in data sets based on finding dense regions with the advantage that the number of clusters does not need to be specified. The set of significant clusters is obtained from the optimization of the stability of the clusters given a minimum cluster size set before so that components with a number of elements below this threshold are considered spurious.
The elements that cannot be assigned to any group are called outliers.

By using this algorithm, we find clear groups in the bidimensional projection depicting galaxies that share similar characteristics.
We study the properties of each group to evaluate the usefulness of our method.

\begin{table*}
\caption{{\sc UMAP} and {\sc HDBSCAN} (hyper)parameters set for each experiment. n\_neighbors: size of local neighborhood used for manifold approximation, min\_dist: minimum distance between embedded points, n\_components: dimension of the space to embed into, metric:  metric to use to compute distances in the high dimensional space,  min\_cluster\_size: minimum size of clusters, min\_samples: number of samples in a neighbourhood for a point to be considered a core point, cluster\_selection\_epsilon: distance threshold below   which clusters will be merged, alpha: distance scaling parameter.   } 
\label{table:params} 
\centering
\resizebox{0.75\textwidth}{!}{
\footnotesize
\hspace{-1. in}
\begin{tabular}{lcccccccc} 
\hline\hline 
 Experiment & \multicolumn{4}{c}{{\sc UMAP}}  & \multicolumn{4}{c}{{\sc HDBSCAN}} \\
  & n\_neighbors & min\_dist & n\_components & metric & min\_cluster\_size & min\_samples & cluster\_selection\_epsilon & alpha \\
  \hline
1 (Sec.~\ref{sec:vlos}) & 70 & 0. & 2 & 'euclidean' & 70 & 1 & 0. & 1.  \\
 2 (Sec.~\ref{sec:call}) & 70 & 0. & 2 & 'euclidean' & 70 & 1 & 0. & 1.  \\
 3 (Sec.~\ref{sec:SRFRs}, SRs) & 73 & 0.1 & 2 & 'euclidean' & 60 & 1 & 0. & 1.  \\
 4 (Sec.~\ref{sec:SRFRs}, FRs) & 72 & 0.9 & 2 & 'euclidean' & 70 & 1 & 0. & 1.  \\
 5 (Sec.~\ref{sec:mix}) & 45 & 0. & 2 & 'euclidean' & 150 & 1 & 0. & 1.  \\
  \hline 
\end{tabular} } \\
\end{table*}

The {\sc HDBSCAN} (hyper)parameters used in each experiment in this work are summarised in Table~\ref{table:params}. The (hyper)parameter min\_cluster\_size defines the smallest number of elements in a group. To define how conservative the clustering might be, we can modify min\_samples and alpha. Technically, min\_samples is the number of samples in a neighbourhood for a point to be considered a core point. Large values of this (hyper)parameter result in a more conservative clustering in which more points will be considered noise in comparison with the clustering using low min\_samples. Regarding alpha, it is a scale parameter related to the hierarchical algorithm and it is also related to how conservative the clustering is. The default value of this parameter is 1 and higher values yield more conservative results. The minimum distance between clusters is given by cluster\_selection\_epsilon in the sense that clusters at distances below this threshold are merged. 
For more details about the algorithm see \cite{CampelloHDBSCAN2013}.

We set all the (hyper)parameters prioritising a clear visualization of the trends among the clusters besides not being too restrictive, especially regarding the cluster sizes.
Since \nombre is an unsupervised method, we need human intervention in the analysis of the results that arise for different inputs, rather than relying on metrics like accuracy or sensitivity. The algorithms are sensitive to their (hyper)parameters and the number of combination of (hyper)parameters is potentially huge.
Hence, we can set their values for different inputs based on our astrophysical knowledge keeping the ‘human-in-the-loop’.
An important part of the challenges of this work is to find a combination of (hyper)parameters that allows us to find sensible and interesting results.
Exploring heuristic methods to improve the (hyper)parameters selection is a research question itself and is beyond the scope of this paper.

\section{Clustering of line-of-sight velocity maps}
\label{sec:vlos}

\subsection{Clustering of edge-on galaxies}
\label{sec:vlos1}

As a first approach, we use only the line-of-sight velocity maps of our galaxy sample as input of the algorithms described in Sec.~\ref{sec: method} and study the variation of either physical or observational galaxy properties among the clusters obtained by \nombre.
In this Subsection, we observe galaxies edge-on, that is, at an inclination of 90 degrees.
Velocity maps are sensitive to the inclination at which a galaxy is observed because IFUs can only measure line-of-sight velocity.
Therefore, rotation is best noticed at high inclinations. In particular, at 90 degrees the line-of-sight is perpendicular to the total angular momentum.

\begin{figure*}
    \centering
    \includegraphics[width=0.3\textwidth]{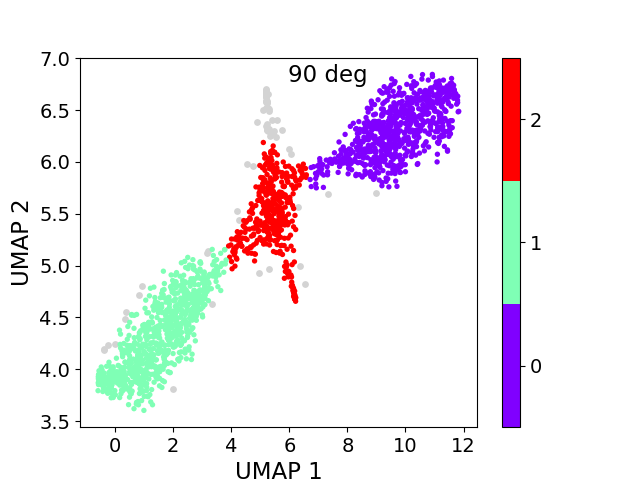}
    \includegraphics[width=0.3\textwidth]{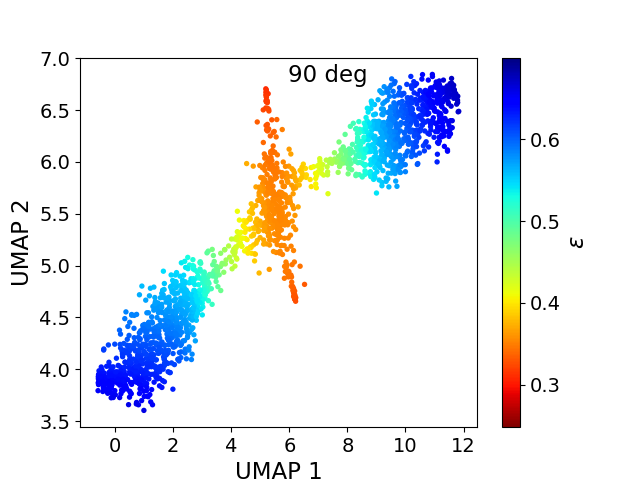}
    \includegraphics[width=0.3\textwidth]{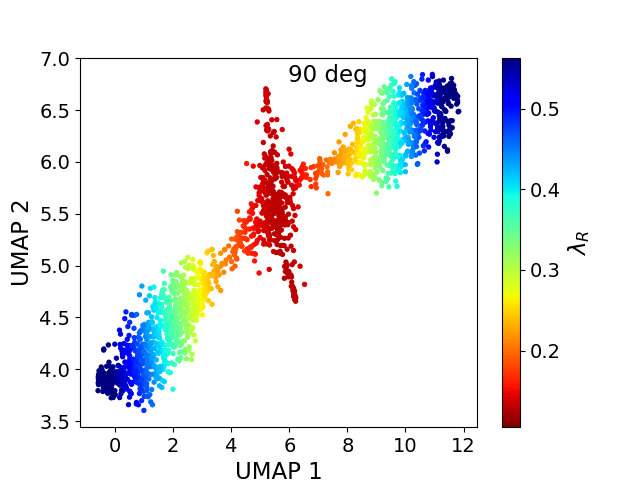} \\
    \includegraphics[width=0.3\textwidth]{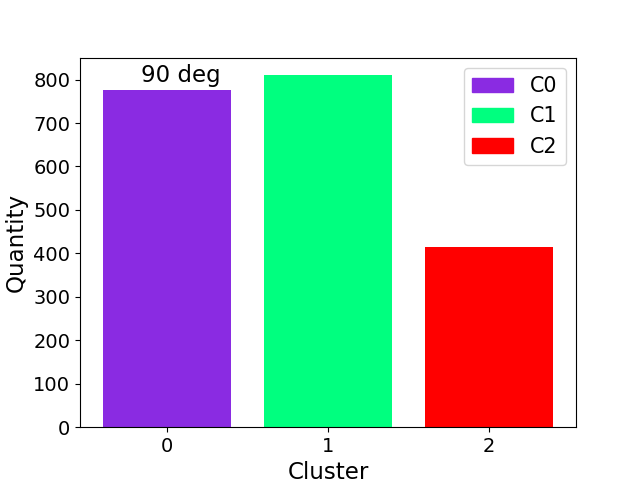}
    \includegraphics[width=0.3\textwidth]{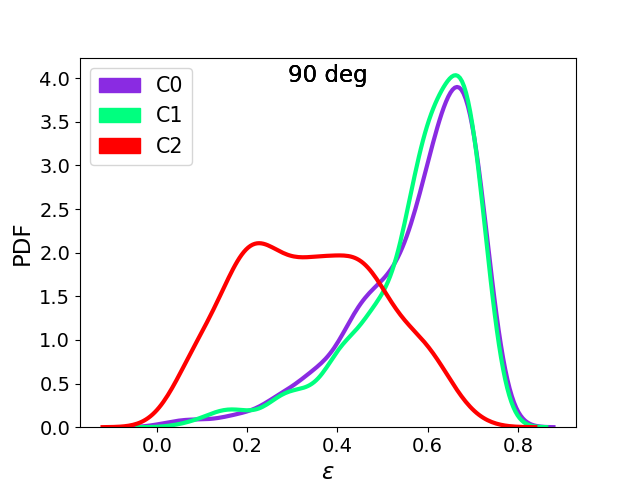}
    \includegraphics[width=0.3\textwidth]{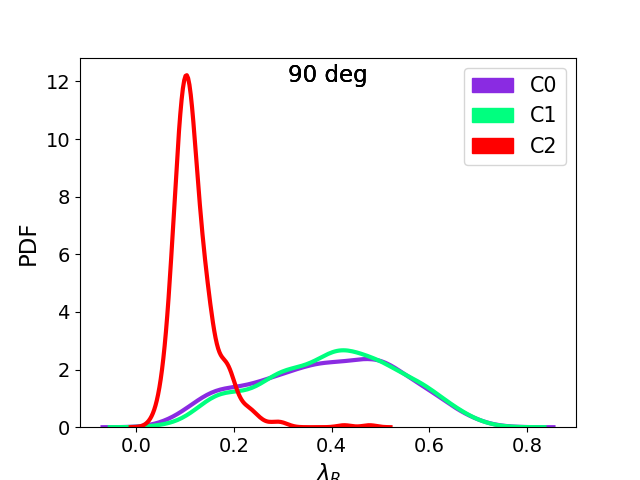} \\    \includegraphics[width=0.3\textwidth]{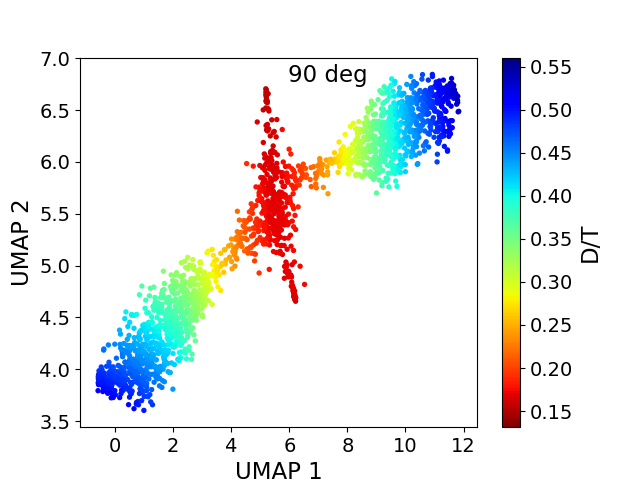}
    \includegraphics[width=0.3\textwidth]{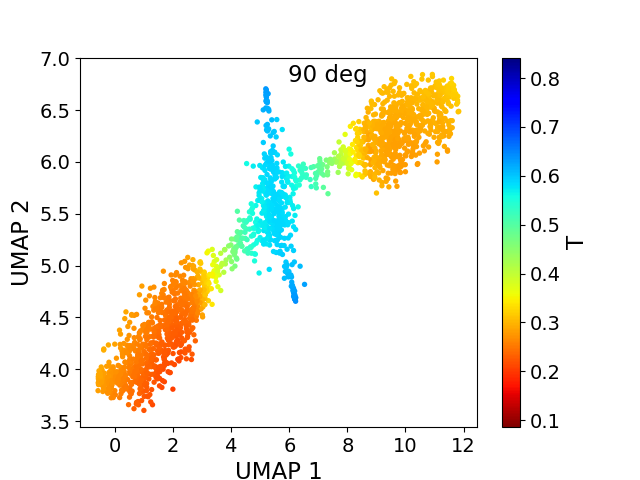}    \includegraphics[width=0.3\textwidth]{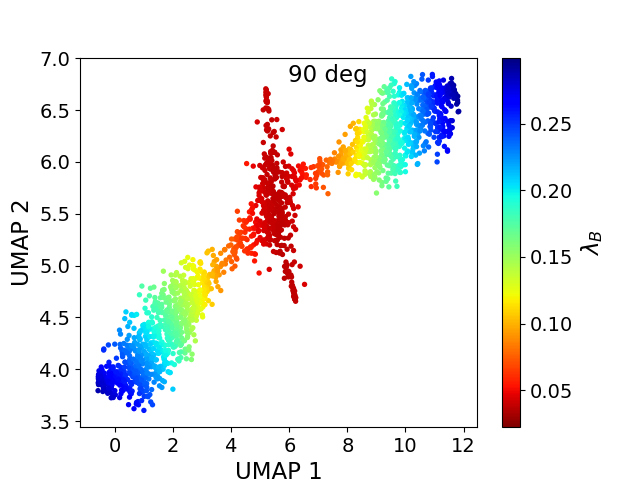} \\
    \includegraphics[width=0.3\textwidth]{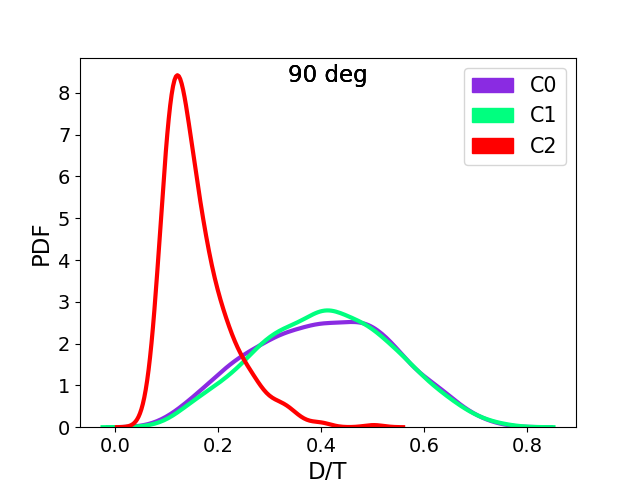}
    \includegraphics[width=0.3\textwidth]{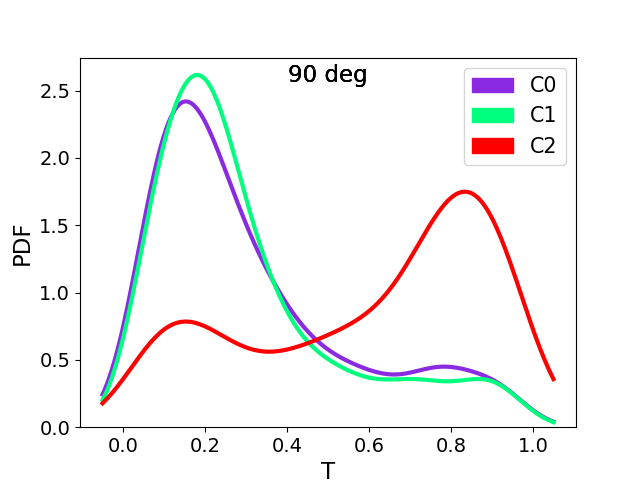}
    \includegraphics[width=0.3\textwidth]{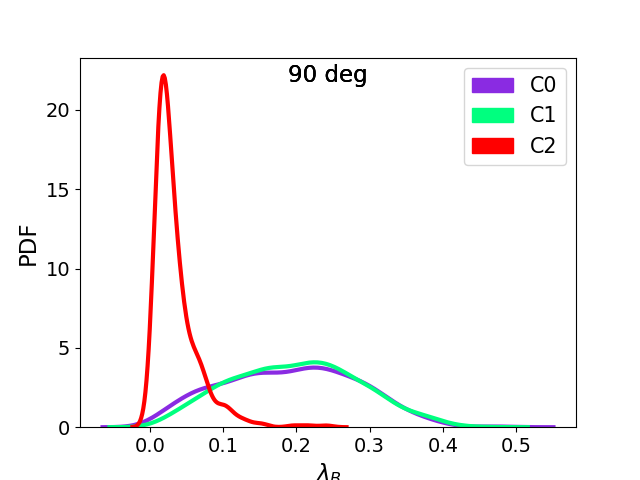}
    \caption{Results of \nombre applied to the line-of-sight velocity maps of our galaxy sample at inclination 90 degrees.
    \textit{Top row panels:} {\sc HDBSCAN} clusters in the {\sc UMAP} bidimensional projection including the outliers of the method as grey dots (\textit{left panel}), distributions of the projected parameters $\varepsilon$ (\textit{middle panel}) and $\lambda_R$ (\textit{right panel}) on the projection. \textit{Second row panels:} size of the clusters (\textit{left panel}), PDFs of $\varepsilon$  (\textit{middle panel}) and $\lambda_R$  (\textit{right panel}) for each cluster. \textit{Third row panels:} distribution of three dimensional parameters $D/T$  (\textit{left panel}), $T$  (\textit{middle panel}), and $\lambda_B$  (\textit{right panel}) on the projection. \textit{Bottom row panels:} PDFs of $D/T$  (\textit{left panel}), $T$  (\textit{middle panel}), and $\lambda_B$  (\textit{right panel}) for each cluster. Colorbar limits are fixed to $10^{\rm th}$ and $90^{\rm th}$ percentile of the variables.} 
    \label{fig:clustVLOS}
\end{figure*}

In Fig.~\ref{fig:clustVLOS} we show a summary of the application of \nombre to the the line-of-sight velocity maps of edge-on galaxies. Hereafter, the colorbar limits are determined by the $10^{\rm th}$ and $90^{\rm th}$ percentiles of the property according to which the symbols are coloured. 
We quantify the distributions of the parameters involved in this analysis by approximating a probability density function (PDF) from the normalised histograms.
The clustering method yields $\sim$ 3 per-cent of outliers (grey dots). It is important to remark that these outliers are located near the clustered points in the bidimensional embedding and that the variations of the physical and projected properties across the projection follow a continuous trend that includes the outliers.

We find three clusters, which we dubbed C0, C1, and C2. The central one, C2 (red), is populated by less disc-dominated galaxies with lower $\lambda_B$ and more prolate (higher values of $T$) than the galaxies in C0 (violet) and C1 (green). These properties are consistent with galaxies dominated by velocity dispersion and, in fact, the $D/T$ ratios of this population are low with most of the members having $D/T <0.2$. 
Regarding the triaxiality parameter, a bimodal distribution can be seen with peaks at $T = 0.15$ and $T = 0.8$ approximately. Galaxies in C2 with $T \leq 0.4$ (around the first peak) have also higher disc fractions compared to the remaining galaxies with a median $D/T$ of 0.21, being 0.16 and 0.26 the first and third quartiles, respectively. In contrast, the first, second (median), and third quartiles of $D/T$ of C2 galaxies with $T  > 0.4$ are 0.11, 0.13, and 0.16, respectively. 
It can also be seen from the figure that this ``low rotation cluster'' has the lowest values of $\varepsilon$ compared to the other two clusters, although a non-negligible number of galaxies in C2 have intermediate ellipticities overlapping with the distributions of $\varepsilon$ of C0 and C1. The situation is similar to that of $T$: the highest $\varepsilon$ galaxies in C2 have the highest $D/T$.
For instance, the median $D/T$ for galaxies in C2 with $\varepsilon < 0.5$ is 0.13 while the median $D/T$ for the remaining galaxies is 0.22. 
Low projected ellipticities mean that the projected axis are more similar and, thus, the galaxies are rounder. 
Galaxies in C2 have the lowest $\lambda_B$, as well as the lowest $\lambda_R$. This is what we expect since there is a significant correlation between both spin parameters, regardless of the inclination.  This is discussed in Appendix~\ref{app:spin} where we show that the correlations between the two spins are present down to 20 degrees of inclination. For lower angles the Spearman coefficients are below 0.5. 

\begin{figure}
    \centering
    \includegraphics[width = 0.45\textwidth]{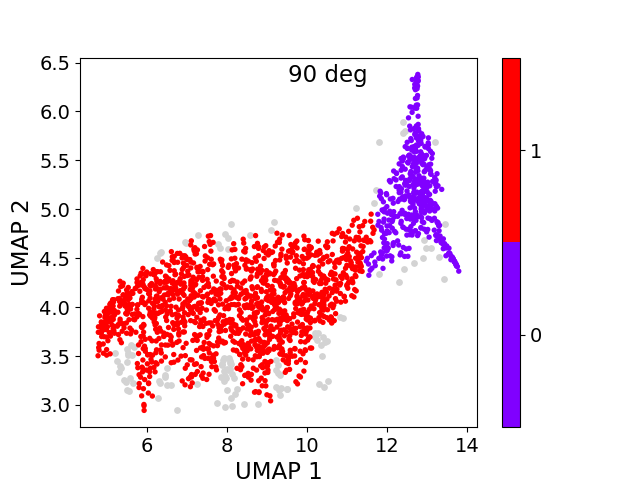} \\
    \includegraphics[width = 0.45\textwidth]{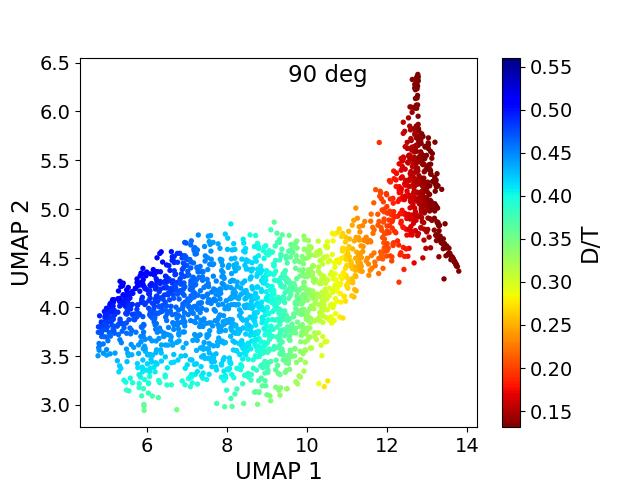} 
    \caption{\textit{Top panel:} Clusters obtained by \nombre when applied to the line-of-sight velocity maps in which every galaxy has the same rotation direction. We include the outliers of the method as grey dots. \textit{Bottom panel:} distribution of $D/T$ on the projection.}
    \label{fig:flipped}    
\end{figure}

It is clearly seen from Fig.\ref{fig:clustVLOS} that C0 and C1 are populated by oblate discy galaxies with the highest values of $\varepsilon$ and both spin parameters.
The disc fractions measured from the simulation of galaxies in C0 and C1 are systematically higher than those from C2.
On the other hand, no significant differences can be found between the $D/T$ of C0 and C1, as could be verified by a Brunner-Munzel test \citep{Brunner2000}\footnote{The Brunner-Munzel test is a statistical test used to assess stochastic equality of two samples without assuming that the shapes of the underlying distributions are the same.}
By comparing the PDFs in Fig.~\ref{fig:clustVLOS} for C0 and C1, we find very similar distributions of the parameters. This can also be quantified by the Brunner-Munzel test.
The difference we observe is that galaxies in C0 and in C1 rotate in opposite directions regarding the line-of-sight (see Sec.~\ref{sec: method}).
In order to explore the effect of the rotation direction in our classification, we repeat the procedure but this time flipping the kinematic images of ``counterclockwise'' rotating galaxies, following the sign convention mentioned in Sec.~\ref{sec:kinematic}. Thus, all input galaxies have the same direction. We show this in Fig~\ref{fig:flipped}. 
As can be seen, \nombre yields two clusters, being C1 (violet) associated with the highest disc-fractions. C0 and C1 from Fig.~\ref{fig:clustVLOS} are unified by \nombre in a single cluster.
We conclude that the method is able to distinguish orientation without loss of information relevant to galaxy morphology. 

Taking into account that we use kinematic information as input for our method, it is interesting to assess the applicability of this method to the classification of SRs and FRs. This separation is widely studied in the literature \citep[e.g.][]{Emsellem2007, Emsellem2011, Veale2017, Brough2017, Greene2018, Lagos2018, Rosito2019a}. Parametric classifications considering $\lambda_R$ are proposed first by \cite{Emsellem2007, Emsellem2011} and in subsequent works \citep[e.g.][]{cappellarireview2016, Graham2018,vdSande2021}. 
Hereafter, we consider the parametric classification for SRs of \cite{vdSande2021} shown in the Eq.~\ref{eq:SR}
\begin{equation}
    \label{eq:SR}
    \lambda_R < 0.12 + 0.25 \varepsilon, \text{ for } \varepsilon \le 0.5 
\end{equation}
where both parameters are measured within the half light radius, following also \cite{Lagos2022}.

\begin{figure*}
  \centering
\includegraphics[width=0.3\textwidth]{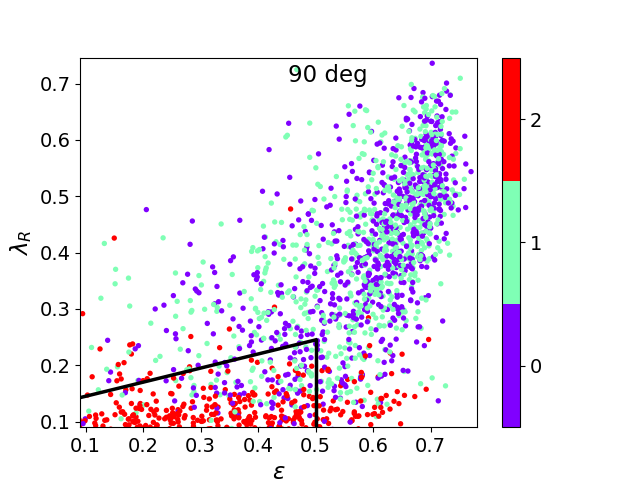}
\includegraphics[width=0.3\textwidth]{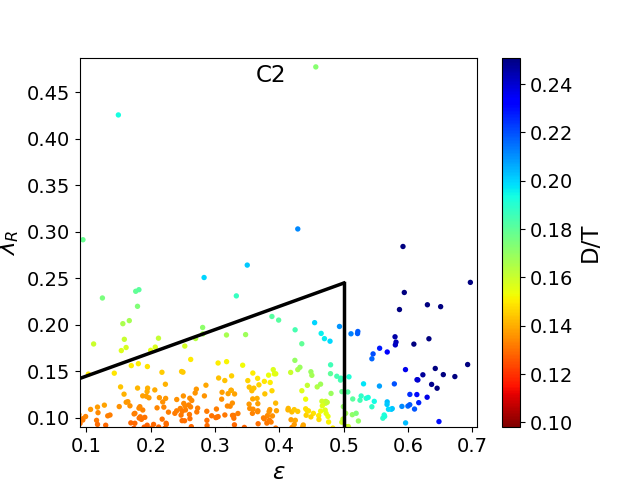}
\includegraphics[width=0.3\textwidth]{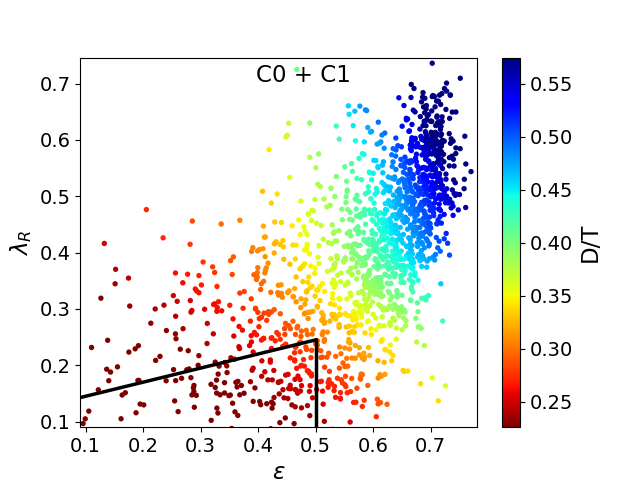}
  \caption{\textit{Left panel}: $\lambda_R-\varepsilon$ plane. Symbols are colored depicting clusters following the colour code of Fig.~\ref{fig:clustVLOS}. \textit{Middle panel}: $\lambda_R-\varepsilon$ plane for galaxies of C2 coloured by $D/T$.
  \textit{Right panel}: $\lambda_R-\varepsilon$ plane for galaxies of C0 and C1 coloured by $D/T$.
  Colorbar limits are fixed to 10$^{\rm th}$ and 90$^{\rm th}$ percentile of the variables. We include in all cases figures the criterion in Eq. \ref{eq:SR} depicted by the black solid lines.}
  \label{fig:lambdaeps_vLOS}
\end{figure*}

In Fig.~\ref{fig:lambdaeps_vLOS} (left panel) we show the $\lambda_R$-$\varepsilon$ plane for all clustered galaxies following the colour code of Fig.~\ref{fig:clustVLOS}. It is clear that the SRs region according to \cite{vdSande2021} is dominated by galaxies from C2, as expected. 
In fact, 71 per-cent of the SRs belong to C2, whereas the 17 per-cent and 12 per-cent belong to C0 and C1 respectively.
Our approach may outperform the parametric classification since a non-negligible number of galaxies with low $D/T$ belonging to C2 (23 per-cent) would be classified as FRs with the parametric definition of \cite{vdSande2021}. 
In the middle panel of Fig~\ref{fig:lambdaeps_vLOS} it can be seen that these galaxies are those with the highest $D/T$ within C2.
Similarly, galaxies in the SR region belonging to C0 or C1 have the lowest $D/T$ among these clusters (right panel).
We remark that physical parameters such as $D/T$ do not depend on the inclination and describe, thus, the fundamental properties of the galaxies.

Automatic ML methods for galaxy classification yielding groups that distinguish galaxy rotation have been previously employed. 
From an input consisting on stellar population and kinematic maps for galaxies from MaNGA survey \citep{Bundy+2015}, \cite{Sarmiento2021} obtain three galaxy clusters, one of which lies in the SR region of the $\lambda_R$-$\varepsilon$ plane (their figure 3, first row), albeit clusters significantly overlap.
These galaxies are also notably older than those in the other two clusters in agreement with galaxies from the {\sc eagle} simulation \citep{Rosito2019a}. 

We can conclude that, when using only the information about line-of-sight velocity, we can obtain a clear separation between SRs and FRs.
However, this requires that the rotation is clearly depicted in the kinematic maps as we explain in the next Subsection.

\subsection{Effects of the inclination}
\label{sec:vlos2}

Besides studying the clustering of edge-on galaxies, we
analyse the applicability of \nombre to galaxies observed at different inclinations. 
The rotation is poorly seen at low inclinations, hence, we assess how the method is affected by decreasing the angle of observation.
In Table~\ref{table:tclulstVLOS}, we summarise the properties of the clustering at inclinations down to 20 degrees.

\begin{figure*}
  \centering
\includegraphics[width=0.24\textwidth]{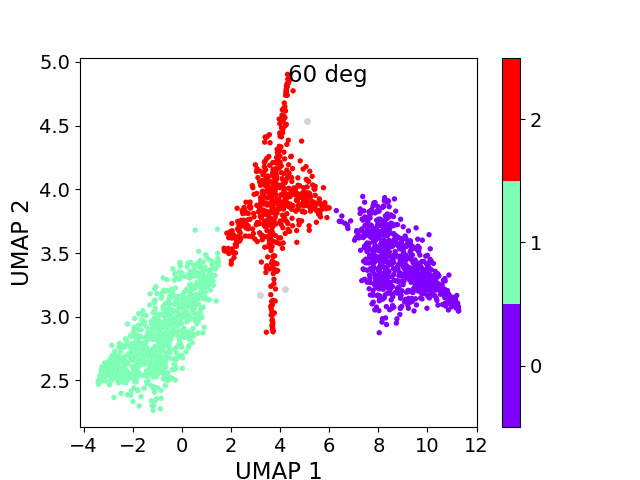}
    \includegraphics[width=0.24\textwidth]{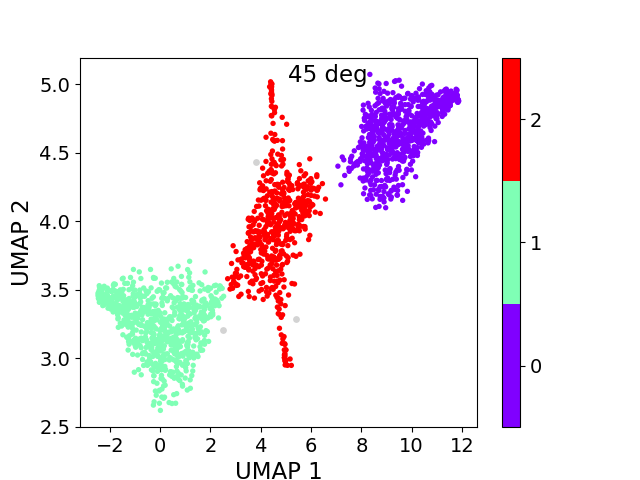}
    \includegraphics[width=0.24\textwidth]{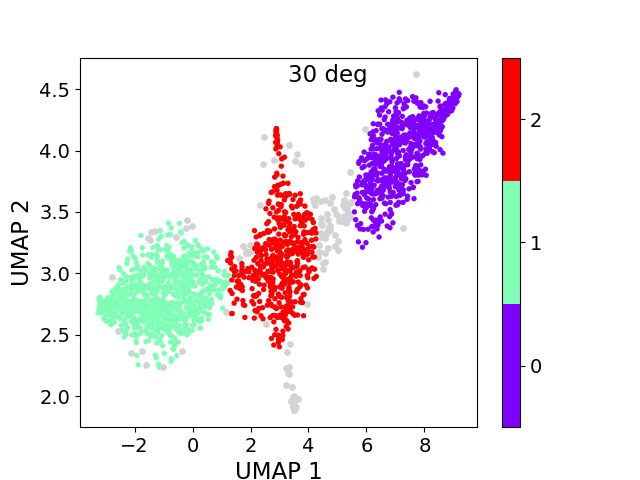}
    \includegraphics[width=0.24\textwidth]{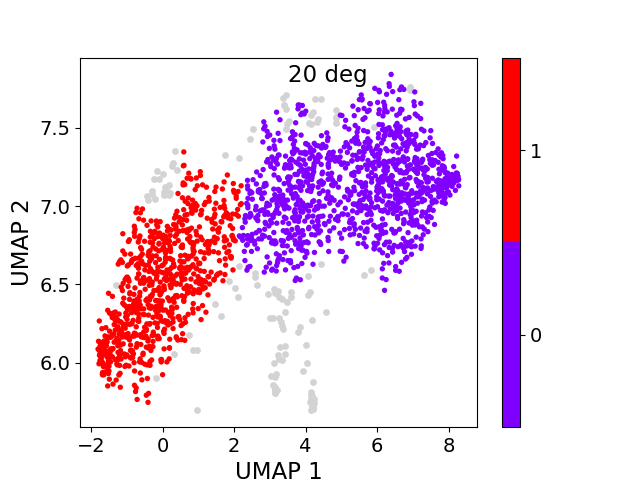} \\
    \includegraphics[width=0.24\textwidth]{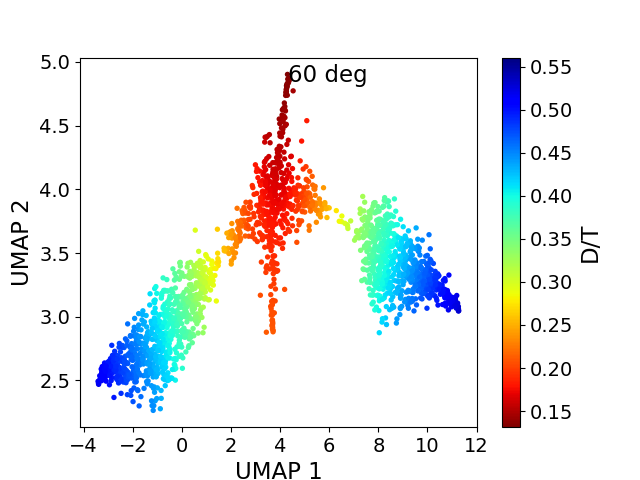}
    \includegraphics[width=0.24\textwidth]{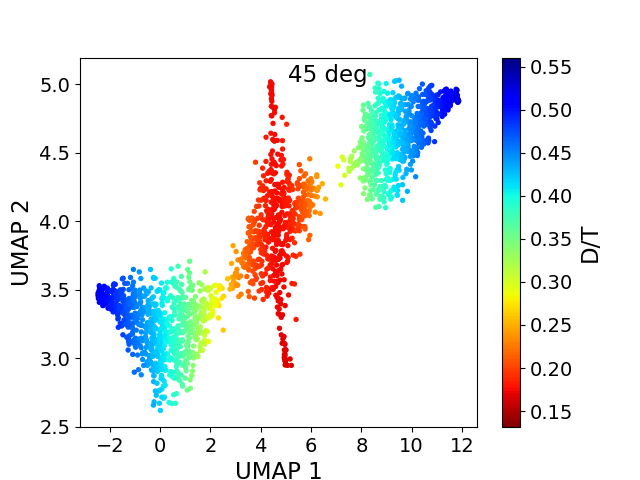}
    \includegraphics[width=0.24\textwidth]{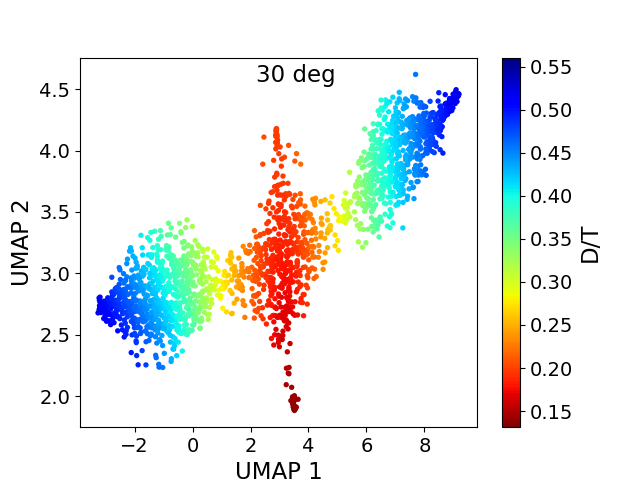}
    \includegraphics[width=0.24\textwidth]{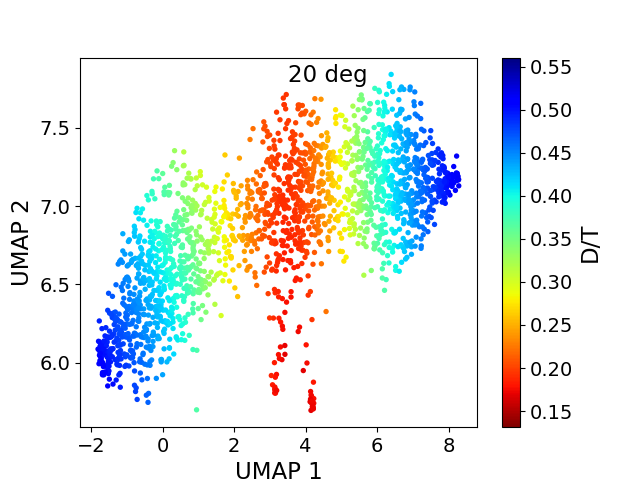} \\
    \includegraphics[width=0.24\textwidth]{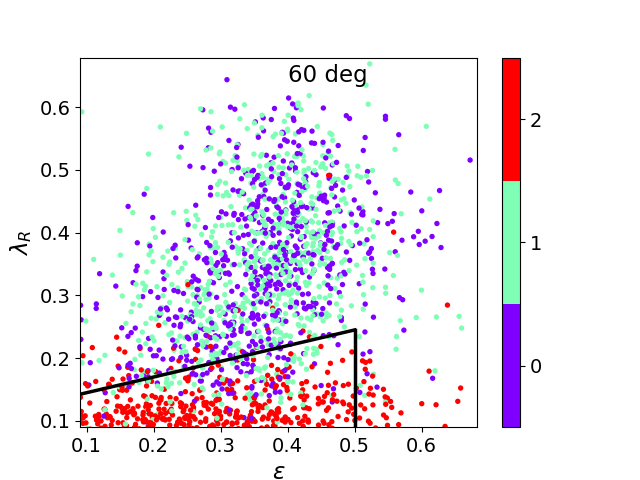}
    \includegraphics[width=0.24\textwidth]{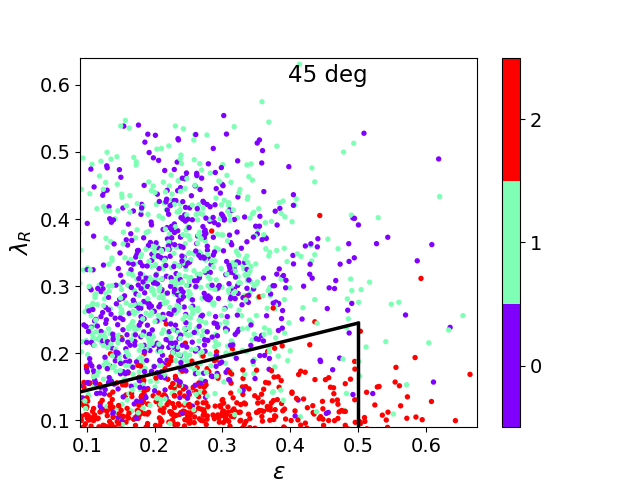}
    \includegraphics[width=0.24\textwidth]{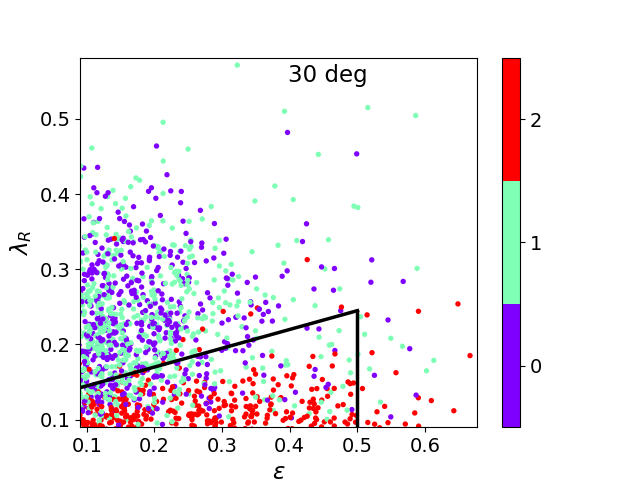}
    \includegraphics[width=0.24\textwidth]{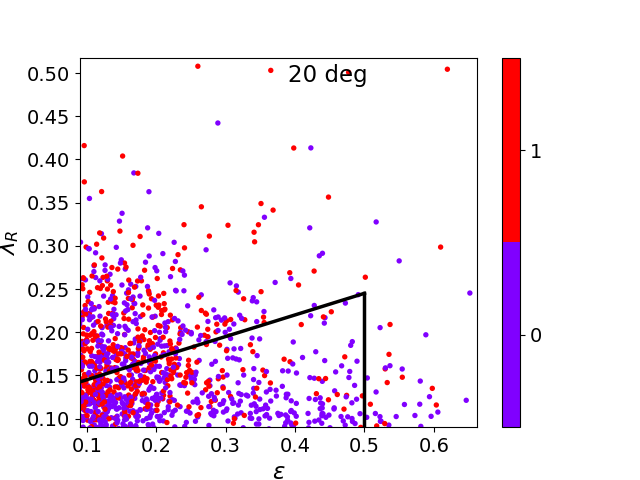}
  \caption{\textit{Top panels}: {\sc HDBSCAN} clusters in the {\sc UMAP} bidimensional projection of the line-of-sight velocity maps of galaxies observed at 60, 45, 30, and 20 degrees including the outliers of the method as grey dots. \textit{Middle panels}: distributions of $D/T$ on each projection. Colorbar limits are fixed to 10$^{\rm th}$ and 90$^{\rm th}$ percentile of the variables.
  \textit{Bottom panels}: $\lambda_R-\varepsilon$ plane. Symbols are colored depicting clusters following the same colour code of the top panels. We include the criterion in Eq. \ref{eq:SR} depicted by the black solid lines.}
  \label{fig:clus_inc_vLOS}
\end{figure*}

In Fig.~\ref{fig:clus_inc_vLOS} we show the application of \nombre to galaxies at inclinations of 60, 45, 30, and 20 degrees and the distributions of $D/T$ on the bidimensional projections (top and middle panels respectively). 
In the first three cases, we find three clusters among which the central one, C2 (red), seems to have notably less disc fraction. 
On the projection at 30 degrees, the division of C1 (green) and C2 (red) is less clear, albeit the clustering can identify the region with the lowest rotation.
\nombre cannot distinguish galaxies with the lowest $D/T$ in a particular group when galaxies are observed at angles below 20 degrees.

In Fig.~\ref{fig:clus_inc_vLOS} (bottom panels), we also show the $\lambda_R$-$\varepsilon$ plane. It is clear that the SRs region is dominated by galaxies from C2 in the first three cases, as expected.
For inclinations of 90, 60, and 45 degrees, C2 includes above 70 per-cent of the total SRs (according to Eq.~\ref{eq:SR}), whereas at 30 degrees this fraction decreases to 58 per-cent. At 20 degrees, most galaxies are considered SRs according to the same criterion.
The fraction of galaxies classified as SRs increases with decreasing inclination, as seen in Fig.~\ref{fig:clus_inc_vLOS} and Table~\ref{table:tclulstVLOS}.
The fact that the rotation becomes less noticeable at lower inclinations can be also seen from their lower values of $\varepsilon$.

When applied to the sample considered in this work, \nombre is robust enough to obtain a good classification of galaxies according to their level of rotation for inclinations greater or equal than 45 degrees.
To be conservative regarding the application of this method, we consider inclination of 30 degrees a borderline case, albeit the quantities on the projection at 30 and 20 degrees still present clear trends for the sample studied in this paper.
This is encouraging taking into account that our method relies on kinematic maps, which are sensitive to the inclination of the sources.

Regarding the outliers, we make the same observation as in Sec~\ref{sec:vlos1}: there is a continuous variation of the relevant properties across the bidimensional embeddings that involves both, clustered and not clustered galaxies. We can conclude this from the middle panels of Fig.~\ref{fig:clus_inc_vLOS} and the strong correlation between $D/T$ and the other physical and projected properties.
This pattern is also seen in the following sections. We emphasise that galaxies close to each other in the projection have similar properties. \nombre leverages this advantage of UMAP algorithm to obtain a more meaningful clustering.

\begin{table*}
\caption{Summary of the {\sc HDBSCAN} clustering at different inclinations using the line-of-sight velocity maps. SRs are defined by Eq.~\ref{eq:SR} among those for which $\lambda_R$ can be calculated.}             
\label{table:tclulstVLOS}      
\centering
\footnotesize
\begin{tabular}{cccccc} 
\hline\hline  
 Numbers & 90 degrees  & 60 degrees & 45 degrees & 30 degrees & 20 degrees \\
  \hline                      
 Clusters & 3 & 3 & 3 & 3 & 2\\
Clustered galaxies & 2001 & 2061 & 2061 & 1941 & 1929 \\
 Outliers & 63 & 3 & 3 & 123 & 135 \\
Clustered galaxies with $\lambda_R$ & 1955 & 2016 & 2018 & 1891 & 1891 \\
Clustered SRs with $\lambda_R$ & 441 & 613 & 673 & 747 & 1013 \\
  \hline 
\end{tabular} \\
\end{table*}

\section{Joint clustering of all kinematic maps types}
\label{sec:call}

\subsection{Clustering of edge-on galaxies}
\label{sec:call1}

Our results indicate that clustering of only line-of-sight velocity maps is sufficient to describe the differences between SRs and FRs. In this Section, we analyze the possibility of obtaining a meaningful clustering that captures more details about other features of galaxies, such as shape, by adding information also from the velocity dispersion and flux maps.
Hence, we apply \nombre as in the previous section, but first we concatenate the three types of kinematic maps (line-of-sight, dispersion, and flux maps) for galaxies observed at 90 degrees as input to perform the {\sc UMAP} bidimensional projections.

\begin{figure*}
    \centering
    \includegraphics[width=0.3\textwidth]{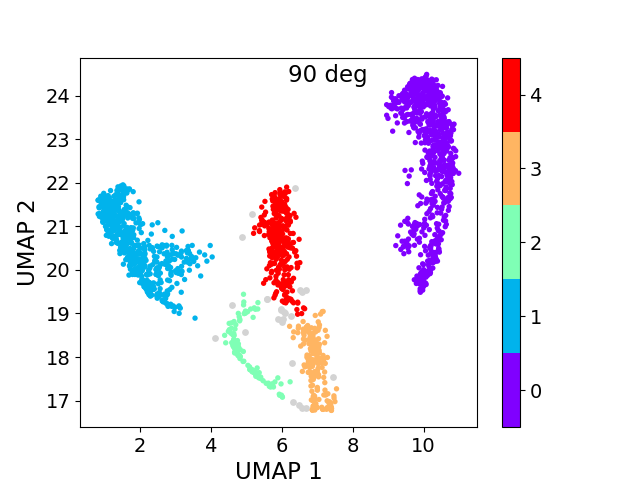}
    \includegraphics[width=0.3\textwidth]{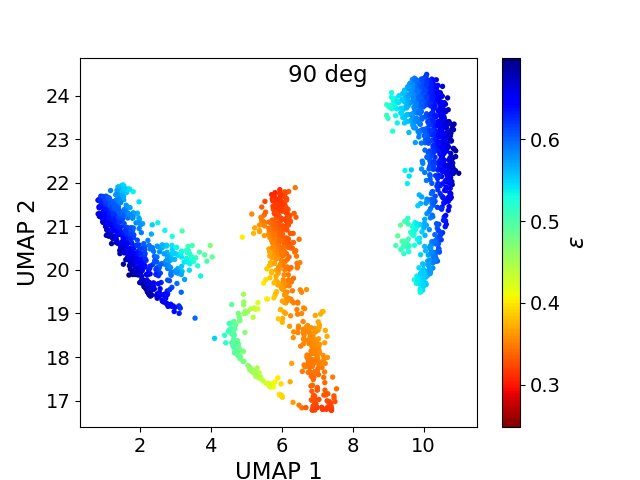}
    \includegraphics[width=0.3\textwidth]{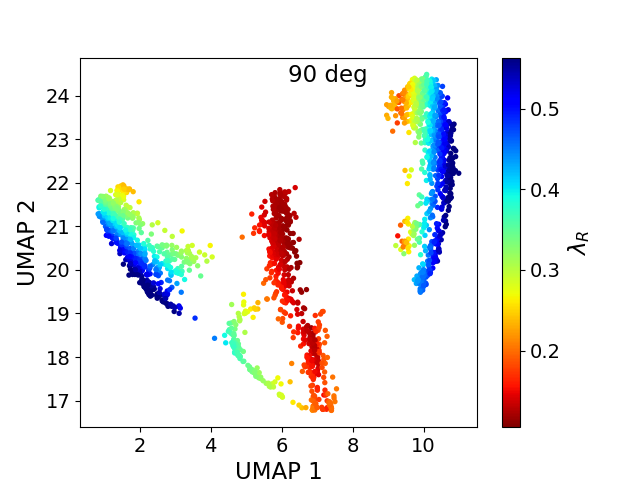} \\
    \includegraphics[width=0.3\textwidth]{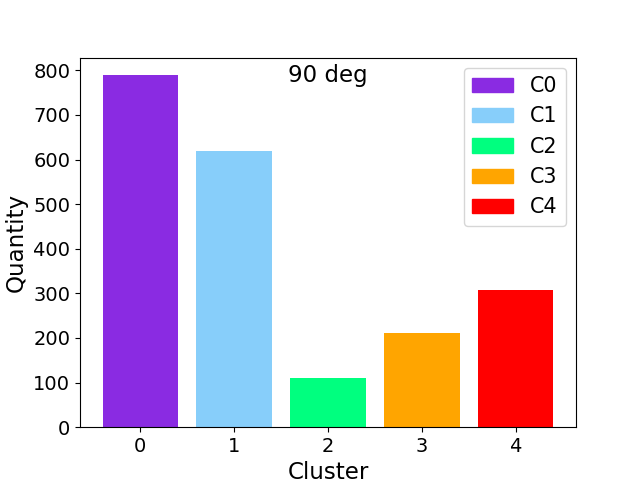}
    \includegraphics[width=0.3\textwidth]{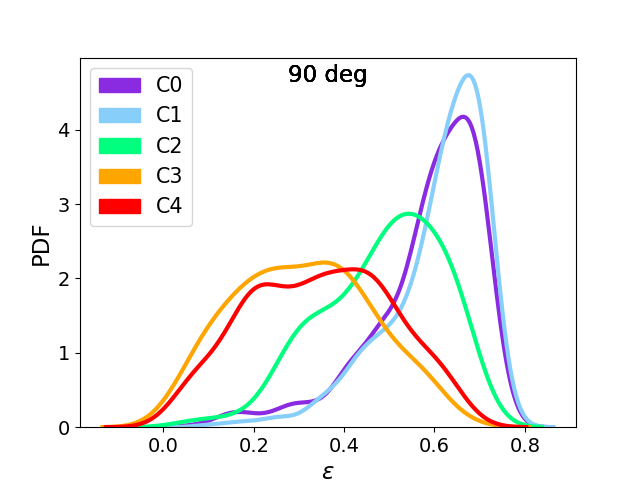}
    \includegraphics[width=0.3\textwidth]{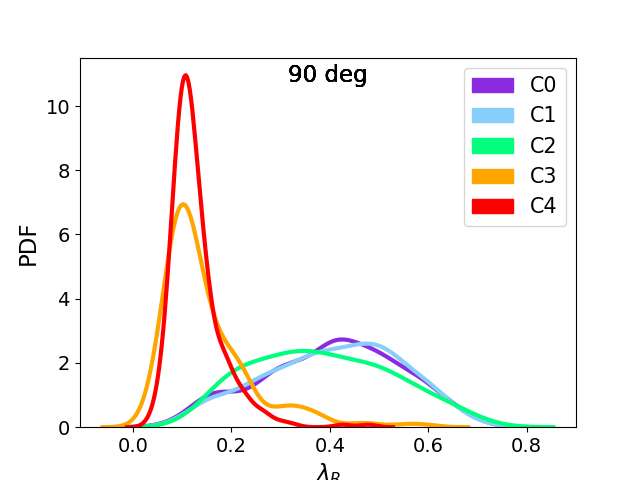} \\    \includegraphics[width=0.3\textwidth]{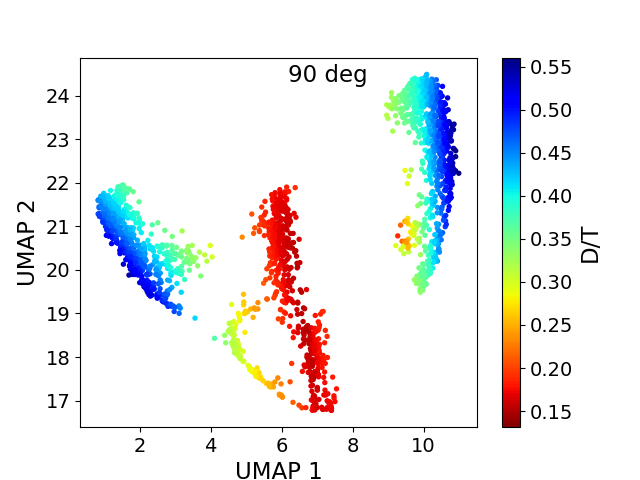}
    \includegraphics[width=0.3\textwidth]{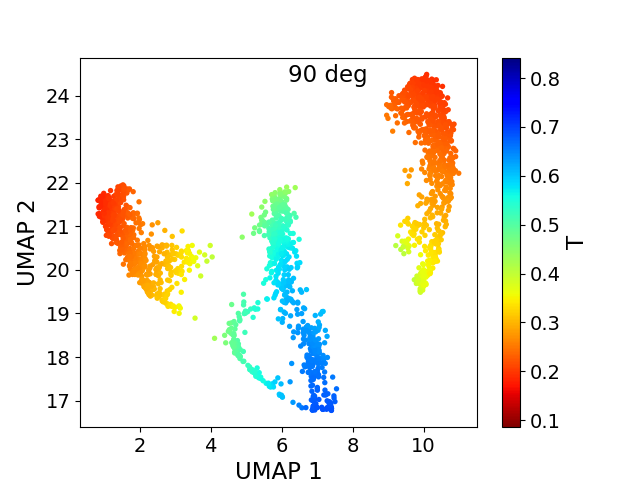}    \includegraphics[width=0.3\textwidth]{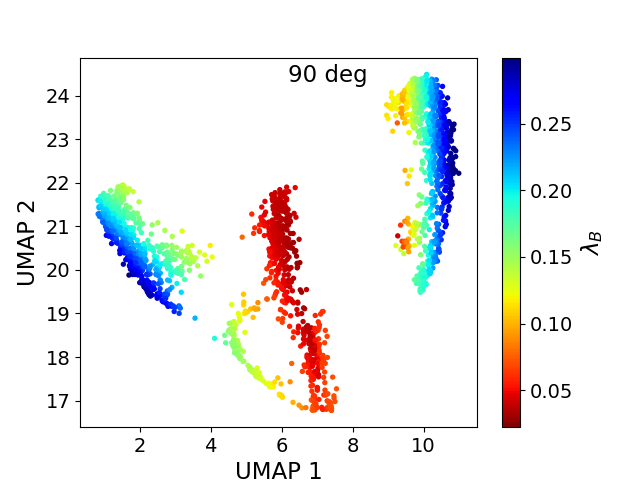} \\
    \includegraphics[width=0.3\textwidth]{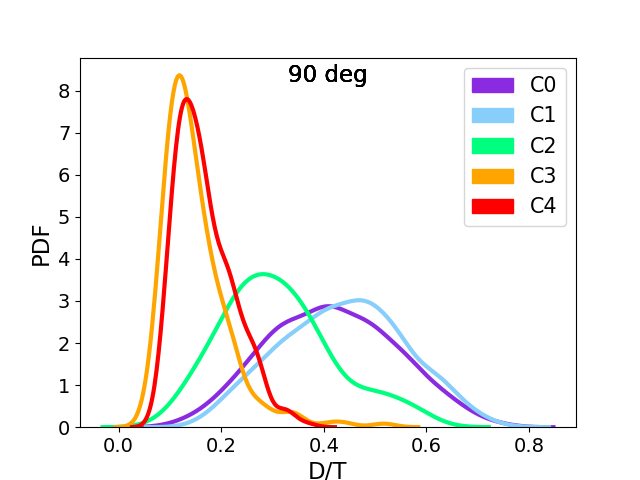}
    \includegraphics[width=0.3\textwidth]{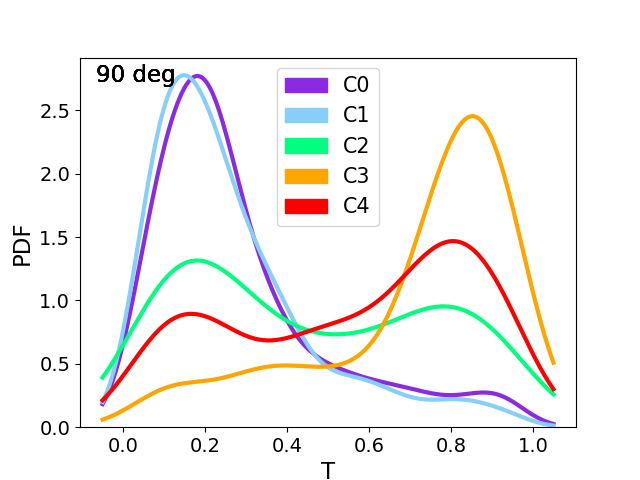}
    \includegraphics[width=0.3\textwidth]{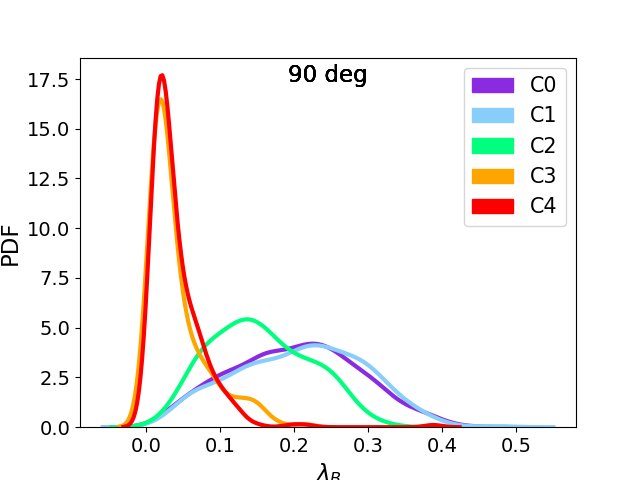}
    \caption{Results of \nombre applied to all kinematic map types of our galaxy sample at inclination 90 degrees.
    \textit{Top row panels:} {\sc HDBSCAN} clusters in the {\sc UMAP} bidimensional projection including the outliers of the method as grey dots (\textit{left panel}), distributions of the projected parameters $\varepsilon$ (\textit{middle panel}) and $\lambda_R$ (\textit{right panel}) on the projection. \textit{Second row panels:} size of the clusters (\textit{left panel}), PDFs of $\varepsilon$ (\textit{middle panel}) and $\lambda_R$ (\textit{right panel}) for each cluster. \textit{Third row panels:} distribution of three dimensional parameters $D/T$ (\textit{left panel}), $T$ (\textit{middle panel}), and $\lambda_B$ (\textit{right panel}) on the projection. \textit{Bottom row panels:} PDFs of $D/T$ (\textit{left panel}), $T$ (\textit{middle panel}), and $\lambda_B$ (\textit{right panel}) for each cluster. Colorbar limits are fixed to $10^{\rm th}$ and $90^{\rm th}$ percentile of the variables.} 
    \label{fig:clustall}
\end{figure*}

In Fig.~\ref{fig:clustall} we summarise the results of the application of \nombre to the set of kinematic maps mentioned above.
As can be seen from  Fig.~\ref{fig:clustall} the additional information brought in by the velocity dispersion and flux allow \nombre to cluster galaxies in 5 different groups. 
C3 (orange) and C4 (red) preferentially contain low rotation galaxies, including  80 per-cent of SRs defined by Eq.~\ref{eq:SR}.
The distributions of the parameters depicted in Fig.~\ref{fig:clustall} for these clusters with low-rotation galaxies show clear differences in comparison with the other clusters, having notably lower $D/T$, lower spin parameters, lower $\varepsilon$, and higher $T$.

By focusing on the comparison of the distributions of parameters from C3 and C4, in Fig.~\ref{fig:clustall} it can be seen clearly that the triaxiality parameter distributions in C3 and C4 differ. By applying the Brunner-Munzel test, we conclude that the values of $T$ in C3 are significantly higher than in C4, which means that galaxies are more prolate in the former. Through the same statistical test, it is possible to determine that $D/T$ in C4 are systematically higher, but the differences between the medians of each distribution is low (0.02).
In contrast, no significant differences among these two groups are seen regarding $\lambda_B$.
This situation is similar when analysing the projected parameters. There are not significant differences in the values of $\lambda_R$, which is what we expect due to the strong correlation of this parameter with $\lambda_B$. On the other hand, the values of $\varepsilon$ in C3 are lower and this is consistent with the fact that galaxies in C3 are less dominated by rotation and are seen, thus, rounder.
Therefore, for edge-on galaxies, including the information of dispersion and flux leads to a refinement of low rotation groups based primarily on galaxy shape.
The addition of extra kinematic information results in an unsupervised classification that takes into account shape as well as rotation leading to a more accurate and meaningful classification of SRs. This may be useful to assist qualitative predictions of galaxy features when applied to samples with missing information on some parameters.
Many efforts about the recovery of 3D shapes have been made recently, in particular those based on IFU, due to the importance of intrinsic shapes in the study of galaxies \citep{Weijmans2014, Foster2017, Li2018, Ene2018}.
\cite{Bassett2019} mention some of the difficulties these techniques face, such as the use of statistical methods to obtain distributions of three dimensional shapes, the generalisation of theoretical models to different data sets, or the need of more tests of IFS shape measurement methods applied to large samples. They suggest the use of ML methods.
Our approach, albeit qualitative, may be a direct way to predict shape parameters using the estimation of the distributions of those parameters for galaxies in the same group and further exploration of the variation of galaxy shapes among clusters may be useful to tackle this task.

The clusters C0 (violet), C1 (light blue), and C2 (green) also present differences.
From inspection of Fig.~\ref{fig:clustall}, it is clear that these clusters are populated by galaxies dominated by rotation, especially C0 and C1, and can be associated, thus, to FRs.
The differences between the distributions of parameters of C0 and C1 are not clear from the figure. By applying the Brunner-Munzel test, we find that the $D/T$ in C0 are lower than those in C1 and the opposite behaviour is seen for $T$ parameter. The second trend is significant\footnote{We consider a level of significance of 0.05, which is commonly used in hypothesis testing.} but weak, having a p-value of 0.03.
No significant differences between the values of $\lambda_B$  and $\lambda_R$ are found for galaxies in C0 and C1 and galaxies in C0 tend to have lower $\varepsilon$ than those in C1.
The main difference between C0 and C1 is the rotation orientation, in analogy to the situation presented in Sec.~\ref{sec:vlos} for high rotation galaxies.
On the other hand, the distributions of the parameters for C2 shown in Fig.~\ref{fig:clustall} are consistent with galaxies with intermediate rotation between SRs and the FRs.
However, the number of galaxies populating C2 is much smaller than those in C0 and C1.

\begin{figure}
    \centering
    \includegraphics[width=0.45\textwidth]{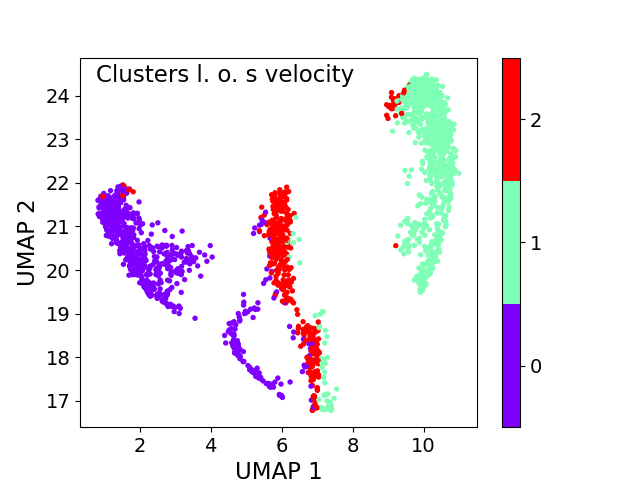} \\
    \includegraphics[width=0.45\textwidth]{Experiment2/umap_i90_clusters_vLOSsigmaflux_normalized.png} 
    \caption{\textit{Top panel}: {\sc UMAP} projection obtained from all kinematic maps coloured according to the clustering obtained for the line-of-sight velocity maps discussed in Sec.~\ref{sec:vlos1} following the colour code of Fig.~\ref{fig:clustVLOS}. \textit{Bottom panel}: clustering in Fig.~\ref{fig:clustall} is repeated for comparison.}
    \label{fig:clust_all_vlos}
\end{figure}

In Fig.~\ref{fig:clust_all_vlos} we show the {\sc UMAP} projections considered in this Section (Fig.~\ref{fig:clustall}) coloured by the clusters obtained using only the line-of-sight velocity maps studied in Sec.~\ref{sec:vlos}.
It can be appreciated that galaxies in C1 and C2 in this Section belong to C0 of the clustering obtained in Sec~\ref{sec:vlos1}, and have, thus, the same rotation orientation.

From Fig.~\ref{fig:clust_all_vlos}, it can be noticed that the ``extra'' clusters obtained with the information from all the kinematic maps types are approximately subsets of the groups obtained using only the line-of-sight velocity maps.
Therefore, these findings show that the refined clustering maintains the information provided by the line-of-sight velocity maps.

\subsection{Effects of the inclination}
\label{sec:call2}

In this Subsection, we study the robustness of \nombre applying it to the kinematic maps with different inclinations.
Based on the analysis performed in Sec.~\ref{sec:vlos2}, we compare the result of the method when applied to the set of all kinematic maps types at inclinations 90, 60, 45, and 30 degrees.
Table~\ref{table:tclulstall} summarises the main properties of each clustering.

 \begin{figure*}
  \centering
\includegraphics[width=0.3\textwidth]{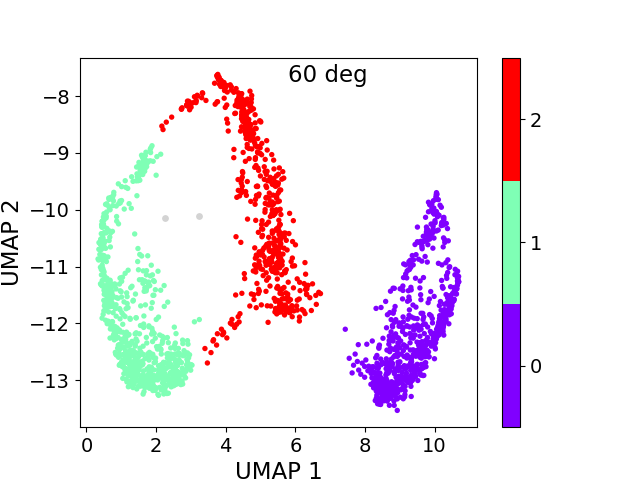}
\includegraphics[width=0.3\textwidth]{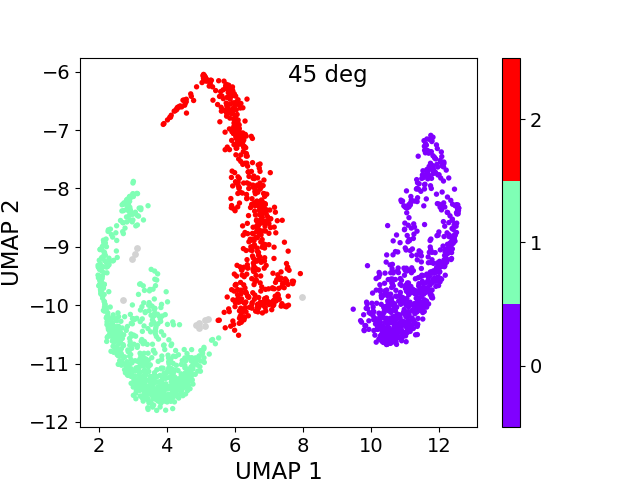}
\includegraphics[width=0.3\textwidth]{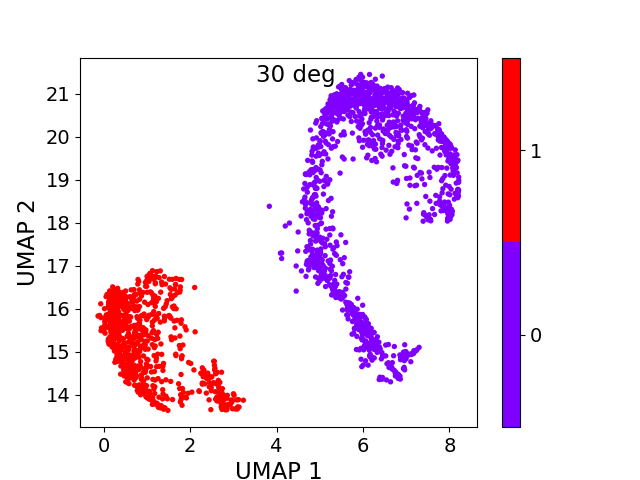} \\
\includegraphics[width=0.3\textwidth]{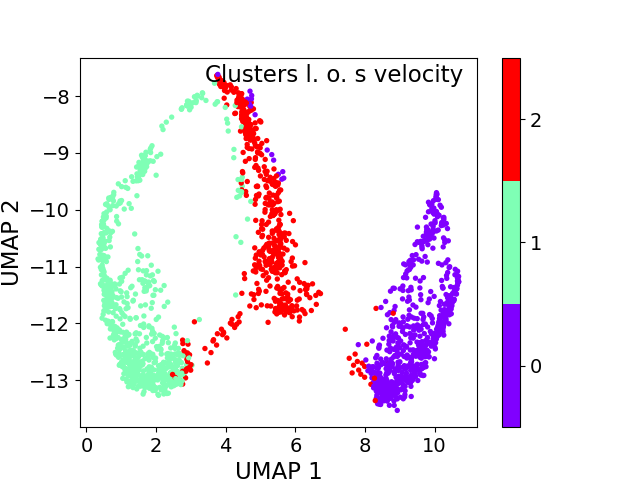}
\includegraphics[width=0.3\textwidth]{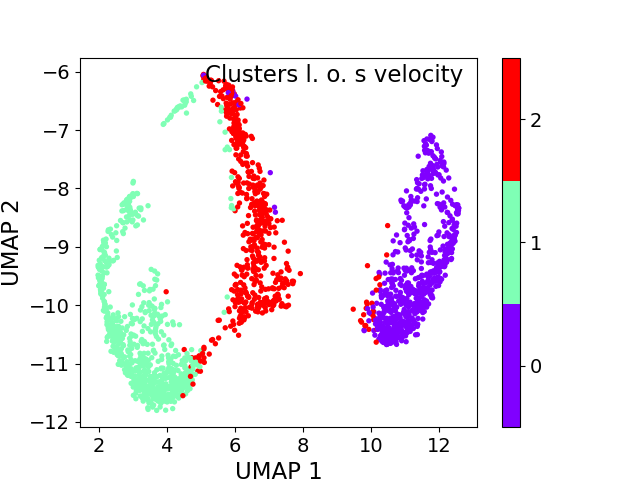}
\includegraphics[width=0.3\textwidth]{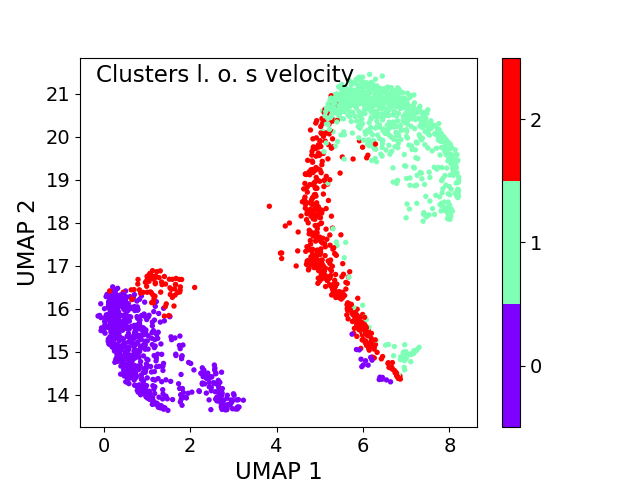} \\
\includegraphics[width=0.3\textwidth]{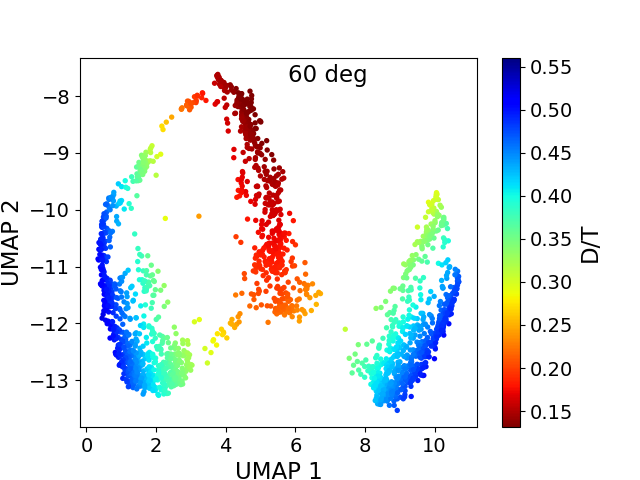}
\includegraphics[width=0.3\textwidth]{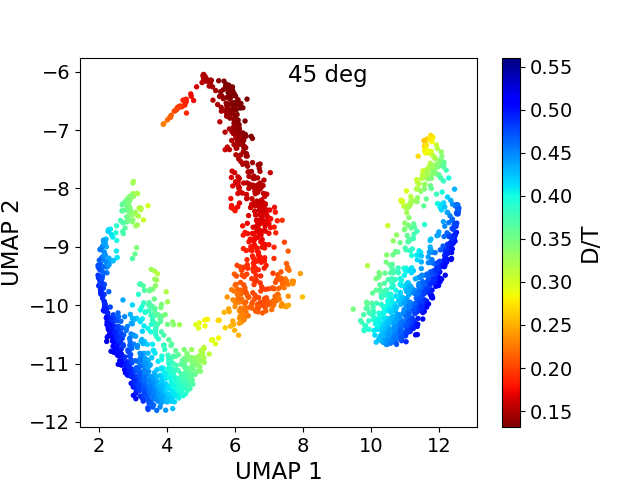}
\includegraphics[width=0.3\textwidth]{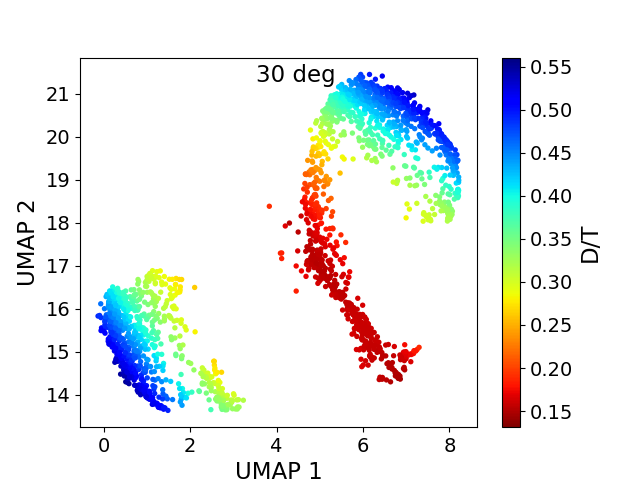} 
  \caption{\textit{Top panels}: {\sc HDBSCAN} clusters in {\sc UMAP} bidimensional projection of the set of all kinematic maps of galaxies observed at 60, 45, and 30 degrees including the outliers of the method as grey dots.
  \textit{Middle panels}: {\sc UMAP} projection obtained from all kinematic maps types coloured according to the clustering obtained for the line-of-sight velocity maps discussed in Section \ref{sec:vlos} following the colour code of Fig.~\ref{fig:clus_inc_vLOS}.
  \textit{Bottom panels}: distributions of $D/T$ on each projection. Colorbar limits are fixed to $10^{\rm th}$ and $90^{\rm th}$ percentile of the variables.
}
\label{fig:clus_inc_all}
\end{figure*}

\begin{table*}
\caption{Summary of the {\sc HDBSCAN} clustering at different inclinations using the three types of kinematic maps. SRs are defined by Eq.~\ref{eq:SR} among those for which $\lambda_R$ can be calculated.}      
\label{table:tclulstall}     
\centering
\footnotesize
\begin{tabular}{ccccc} 
\hline\hline  
  & 90 degrees  & 60 degrees & 45 degrees & 30 degrees \\
  \hline                       
Clusters & 5 & 3 & 3 & 3 \\
Clustered galaxies & 2039 & 2062 & 2053 & 2064 \\
Outliers & 25 & 2 & 11 & 0 \\
Clustered galaxies with $\lambda_R$ & 1993 & 2017 & 2010 & 2014\\
Clustered SRs with $\lambda_R$ & 461 & 615 & 670 & 838 \\
  \hline 
\end{tabular} \\
\end{table*}

In the top and middle panels of Fig.~\ref{fig:clus_inc_all} we show the clusters formed in the bidimensional projections of the set of all kinematic maps by the application of \nombre and the same projection coloured by the clustering analysed in Sec.~\ref{sec:vlos2}, respectively.
For inclinations of 60 and 45 degrees (bottom panels), there are well defined groups that present low rotation galaxies. Around 70 per-cent of the SRs classified according to Eq.~\ref{eq:SR} are included in these clusters at inclinations of 60 and 45 degrees, which is the same fraction observed in Sec~\ref{sec:vlos2}.
In both cases, the distributions of $D/T$ of the groups with low-rotation galaxies show differences with the other clusters, having notably lower values.
Differences can be appreciated at 30 degrees, at which \nombre using the only line-of-sight velocity can better identify low rotation galaxies, but, as we mentioned above, this is considered a borderline case and is not reliable to make robust conclusions.

Again, at inclinations greater or equal to 45 degrees, the addition of other kinematic types maps maintains the information obtained by the clustering using solely velocity maps with the advantage that SRs may be more accurately described at high inclinations.
Therefore, we suggest including all the information to achieve better classifications.

\section{Analysis of slow and fast rotators}
\label{sec:SRFRs}

In this Section, we assess the possibility of achieving more meaningful clustering for galaxies with lower (greater) amounts of rotation when considered them separately.
We select as input of our method the three types of kinematic map of galaxies with $\lambda_{\rm R, edge-on} \leq 0.2$ and $\lambda_{\rm R, edge-on} > 0.2$. 
\cite{Lagos2022} study a sample of the same simulation used in our work at $z=0$ and mention that most galaxies below a threshold of 0.2 in $\lambda_{\rm R, edge-on}$ have lower values of the spin parameter observed at any other inclination \citep{vdSande2021}.
This threshold is also considered as a suitable separation in SRs and FRs in previous works \citep{Lagos2018, Rosito2019a}.

\begin{figure*}
 \centering
\includegraphics[width=0.3\textwidth]{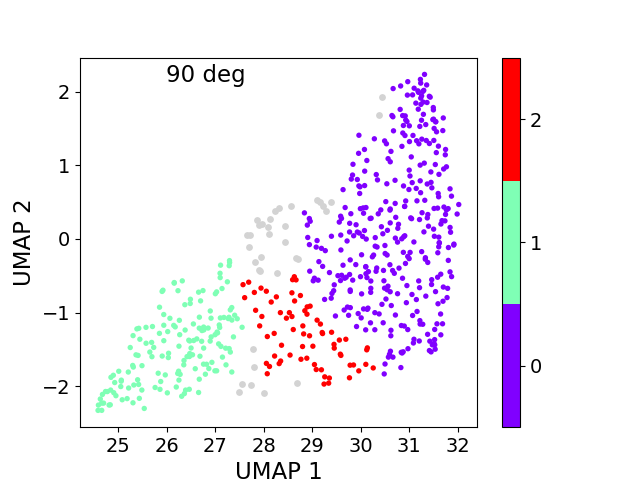}
\includegraphics[width=0.3\textwidth]{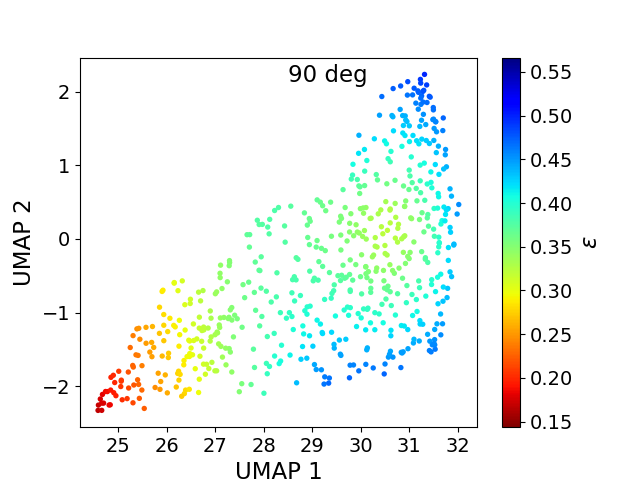}
\includegraphics[width=0.3\textwidth]{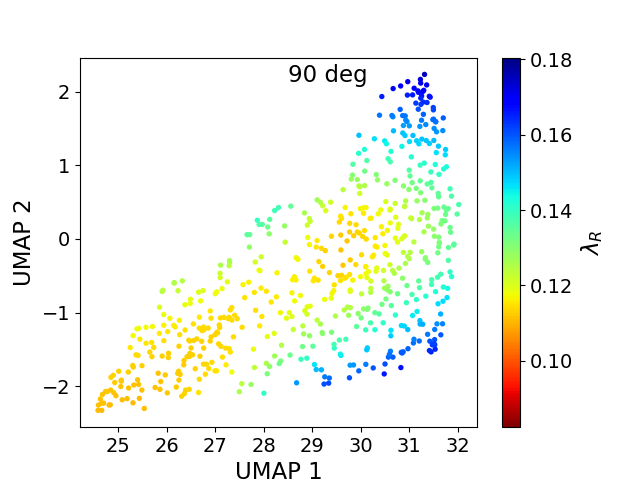} \\
\includegraphics[width=0.3\textwidth]{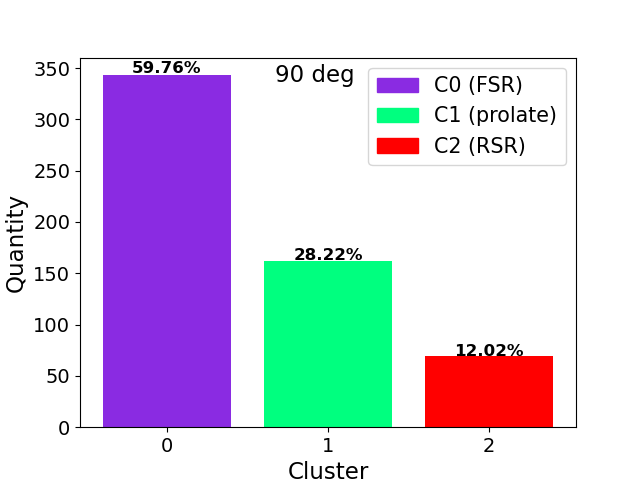}
\includegraphics[width=0.3\textwidth]{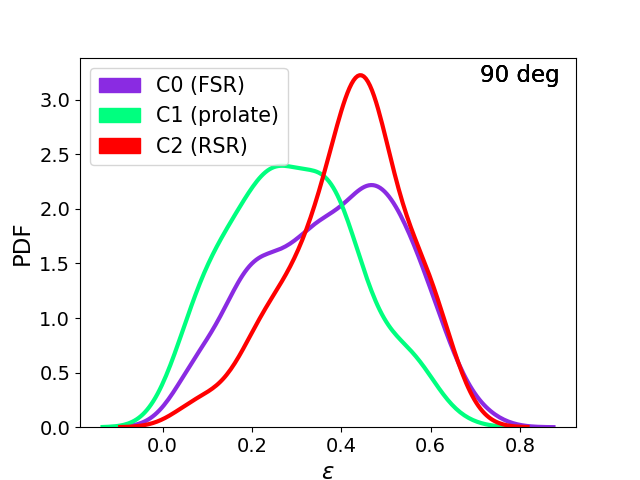}
\includegraphics[width=0.3\textwidth]{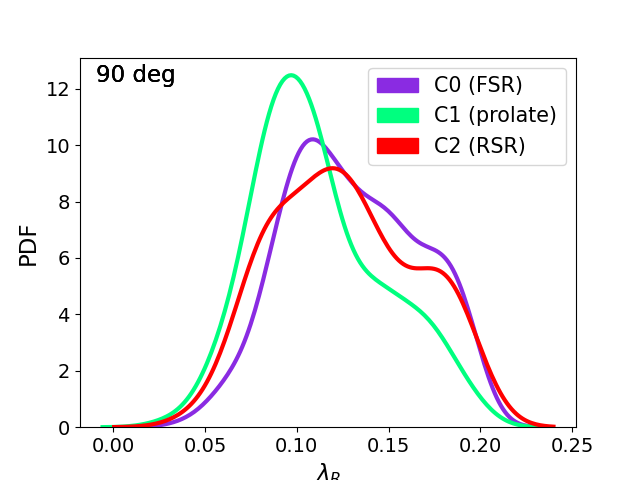} \\
\includegraphics[width=0.3\textwidth]{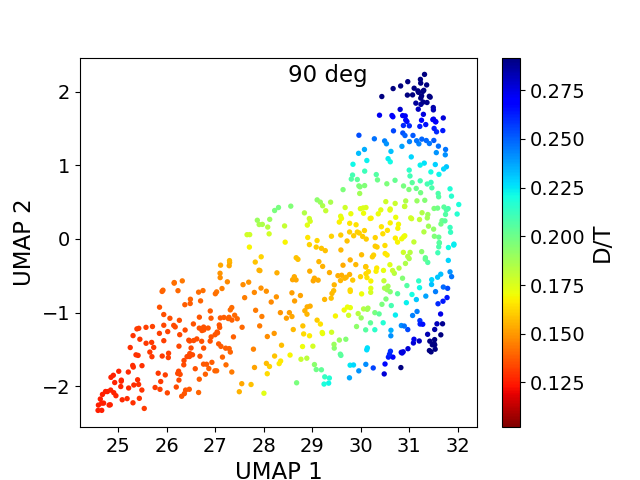}
\includegraphics[width=0.3\textwidth]{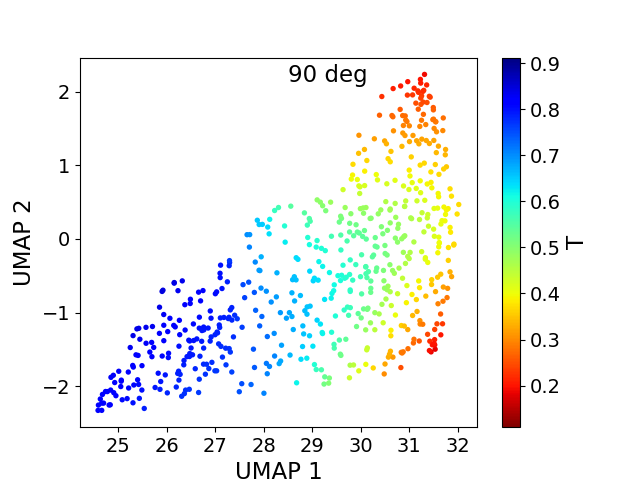}
\includegraphics[width=0.3\textwidth]{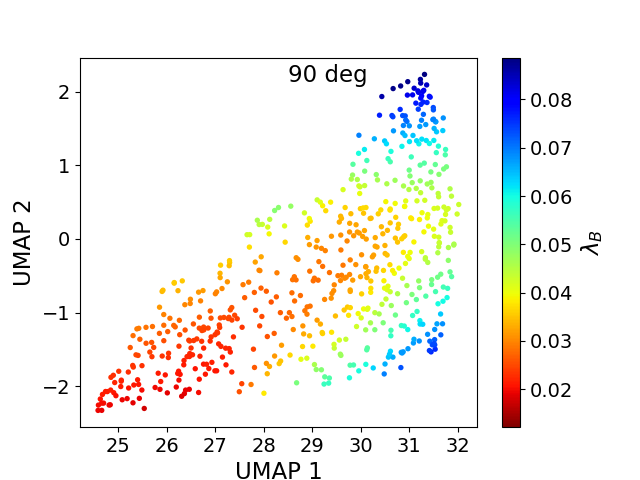} \\
\includegraphics[width=0.3\textwidth]{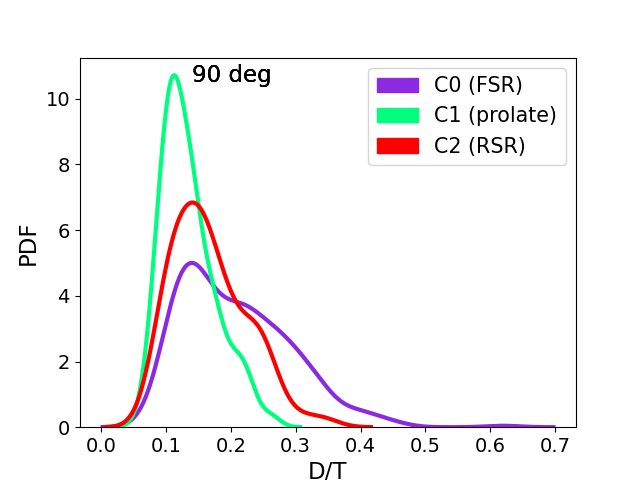}
\includegraphics[width=0.3\textwidth]{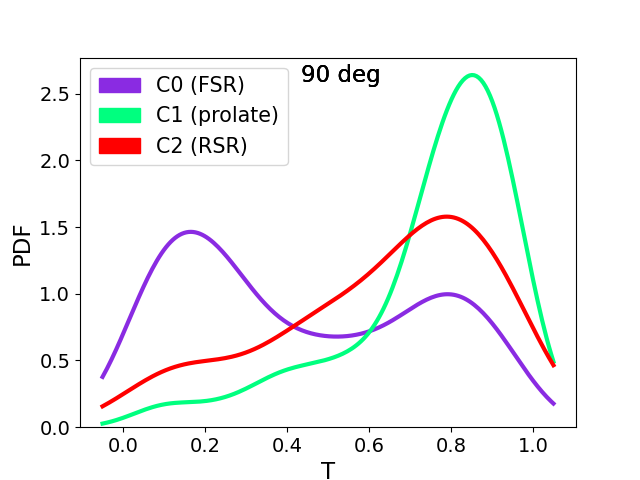}
\includegraphics[width=0.3\textwidth]{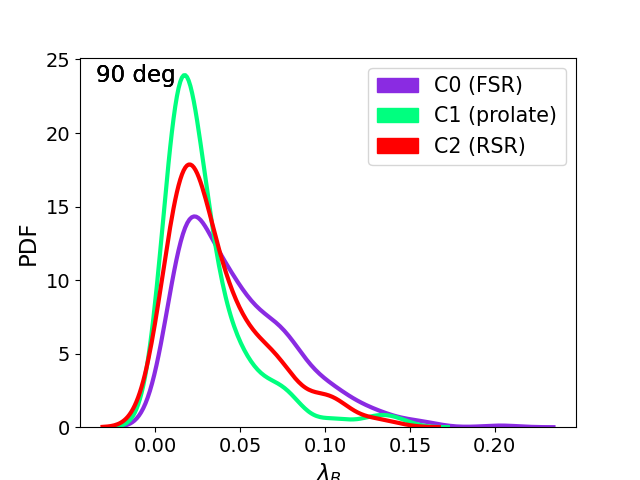} \\
 \caption{Results of \nombre applied to all the kinematic maps types of galaxies with $\lambda_{\rm R, edge-on} \leq 0.2$ observed at 90 degrees. 
    \textit{Top row panels:} {\sc HDBSCAN} clusters in the {\sc UMAP} bidimensional projection including the outliers of the method as grey dots (\textit{left panel}), distributions of the projected parameters $\varepsilon$ (\textit{middle panel}) and $\lambda_R$ (\textit{right panel}) on the projection. \textit{Second row panels:} size of the clusters (\textit{left panel}), PDFs of $\varepsilon$ (\textit{middle panel}) and  $\lambda_R$ (\textit{right panel}) for each cluster. \textit{Third row panels:} distributions of $D/T$ (\textit{left panel}), $T$ (\textit{middle panel}), and $\lambda_B$ (\textit{right panel}) on the projection, \textit{Bottom row panels:} PDFs of $D/T$ (\textit{left panel}), $T$ (\textit{middle panel}), and $\lambda_B$ (\textit{right panel}) for each cluster. Colorbar limits are fixed to 10$^{\rm th}$ and 90$^{\rm th}$ percentile of the variables.}
  \label{fig:clus_seplambdal02}
\end{figure*}

\begin{figure*}
    \centering
    \includegraphics[width=0.3\textwidth]{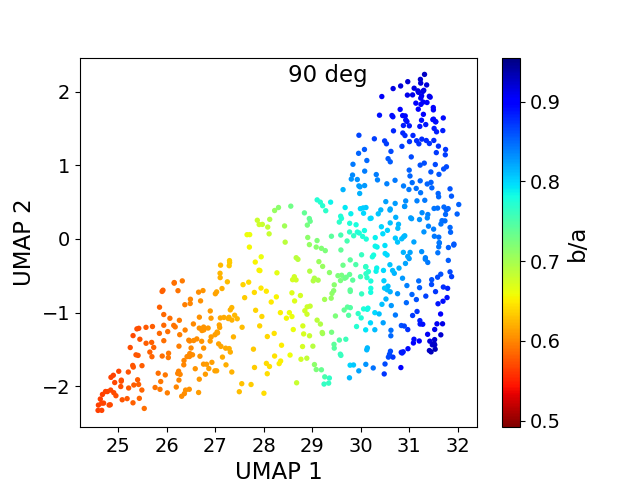}
    \includegraphics[width=0.3\textwidth]{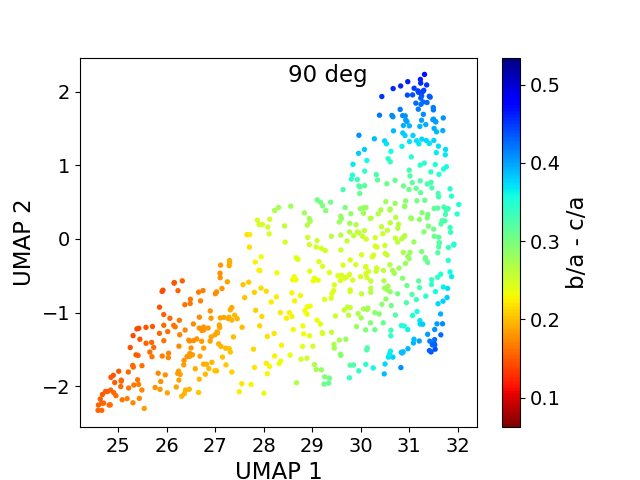}
    \includegraphics[width=0.3\textwidth]{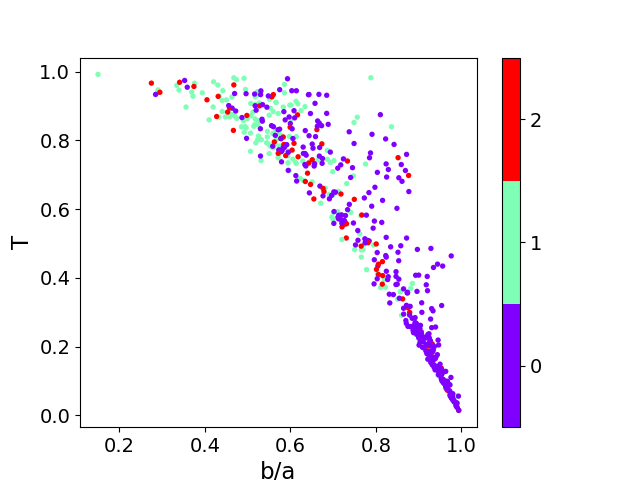} \\
    \includegraphics[width=0.3\textwidth]{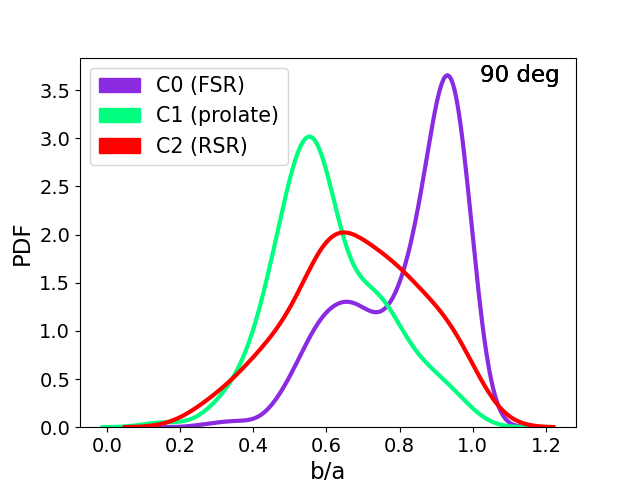}
    \includegraphics[width=0.3\textwidth]{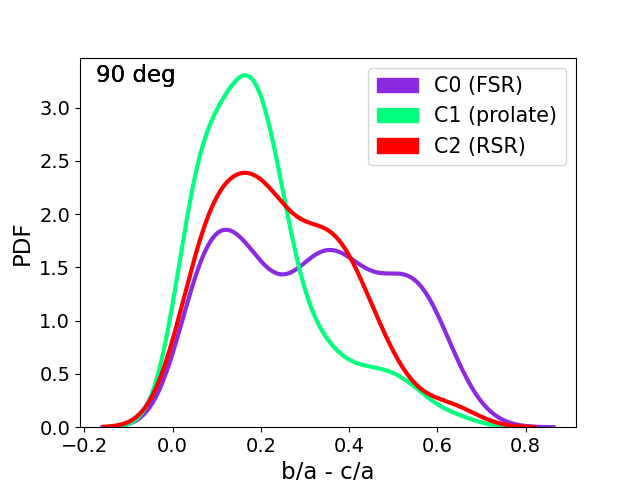}
    \includegraphics[width=0.3\textwidth]{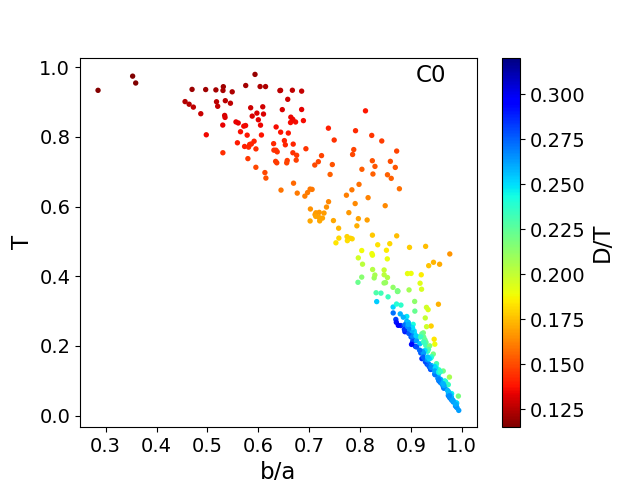}
    \caption{Analysis of the 3D axis of galaxies with $\lambda_{\rm R, edge-on} \leq 0.2$.
    \textit{Top panels:} distributions of $b/a$ (\textit{left panel}) and $b/a - c/a$ (\textit{middle panel}) in the {\sc UMAP} projection of the kinematic maps depicted in Fig.~\ref{fig:clus_seplambdal02}, $T$ vs. $b/a$ coloured according to the clusters following the colour code of Fig~\ref{fig:clus_seplambdal02} (\textit{right panel}). \textit{Bottom panels:} PDFs of $b/a$ (\textit{left panel}) and $b/a-c/a$ (\textit{middle panel}),  and $T$ vs. $b/a$ for C0 coloured according to $D/T$ (\textit{right panel}). Colorbar limits are fixed to 10$^{\rm th}$ and 90$^{\rm th}$ percentile of the variables.}
    \label{fig:clus_seplambdal02_axis}
\end{figure*}

In the first case, we find 609 galaxies in our sample with 
$\lambda_{\rm R, edge-on} \leq 0.2$.
This constraint in the edge-on projected spin parameter includes 96 per-cent SRs according to Eq.~\ref{eq:SR}.
We obtain three clusters and 35 outliers.
In Fig.~\ref{fig:clus_seplambdal02} we show an analysis of the clustering of the galaxies below this threshold.
It can be seen that the separation in clusters in the {\sc UMAP} projection is not as clear as those in the previous sections and that there is a continuous variation of the parameters among the three groups.
Since rotation in all these galaxies is low, in Fig.~\ref{fig:clus_seplambdal02_axis} we look in detail the 3D shapes in the different groups to analyse a meaningful clustering.

A more discy group, C0 (violet), can be seen from Fig.~\ref{fig:clus_seplambdal02} which also has notably lower values of $T$ and higher $\lambda_B$, $b/a$, and $b/a-c/a$ (Fig.~\ref{fig:clus_seplambdal02_axis}) with respect to the other two groups. 
This is consistent with the right panels of Fig.~\ref{fig:clus_seplambdal02_axis} in the sense that in the plane $T$-$b/a$ the region with low triaxiality parameters and high $b/a$ ratios is dominated by galaxies from C0. From that figure it can also be seen that galaxies from C0 with the highest $T$ and the lowest $b/a$ are the ones with the lowest $D/T$ within that cluster and hence we can associate this region with low amount of rotation.
Moreover, there are differences between the projected parameters in C0 and C1 (green) as can be appreciated from the PDFs in Fig.~\ref{fig:clus_seplambdal02}, being these parameters lower in C1.
However, the comparisons between the distributions of $\lambda_R$ and $\varepsilon$ from C0 and those of C2 (red) are unclear.
It can be seen by means of the Brunner-Munzel test that $\varepsilon$ from C0 have a weak trend (p-value $\sim$ 0.03) of being lower than those of C2 and that there is no systematic trends regarding $\lambda_R$.
On the other hand, although C1 and C2 are both associated with galaxies in which the rotational component is almost negligible, the distributions of the parameters present systematic differences as seen from the PDFs in Fig.~\ref{fig:clus_seplambdal02} and Fig.~\ref{fig:clus_seplambdal02_axis} and quantified through the Brunner-Munzel test.
C1 has the lowest $D/T$, $\varepsilon$, $\lambda_R, $ $b/a$, $b/a-c/a$, and the highest $T$. The spin parameter $\lambda_B$ can be considered lower in C1 than in C2, but this trend is weak (p-value 0.02).

\cite{Lagos2022} mention the importance of the visual classification of the kinematic maps to mitigate the inaccuracies of the parametric criteria based on the $\lambda_R-\varepsilon$ plane \citep{vdSande2021}.
\cite{Lagos2022} visually classify their galaxies in the following groups: ``flat slow rotators (FSRs)'', ``round slow rotators (RSRs)'', ``prolate'', ``$2 \sigma$'', ``rotators'', and ``unclear''. 
We show that we can perform automatically a classification similar to the visual classification in \cite{Lagos2022}.
%Our method \blue{requires } an input with a large variety of galaxy \blue{morphologies and kinematics}, which, on the other hand, is more similar to real observations.

The low triaxiality parameter for galaxies in C0 indicates that they are oblate and the fact that the major and middle axis have similar values but their difference with the minor axis is high evidences that these galaxies are flatter. 
The values of $T$ are high and $b/a-c/a$ are low for C1 and C2 but the $b/a$ ratios are significantly lower in C1, which means that C2 has rounder galaxies (axis of similar length) while galaxies in C1 are more prolate.
Hence, we can associate C0, C1, and C2 with the FSR, prolate, and RSR groups in \cite{Lagos2022} respectively.
The fraction of galaxies in each group differs (see percentages in Fig.~\ref{fig:clus_seplambdal02}), however, to those reported by \cite{Lagos2022}, who find a majority of FSR (48 per-cent) galaxies followed by RSRs (38 per-cent) and prolate (10 per-cent) galaxies.

\begin{figure*}
 \centering
\includegraphics[width=0.3\textwidth]{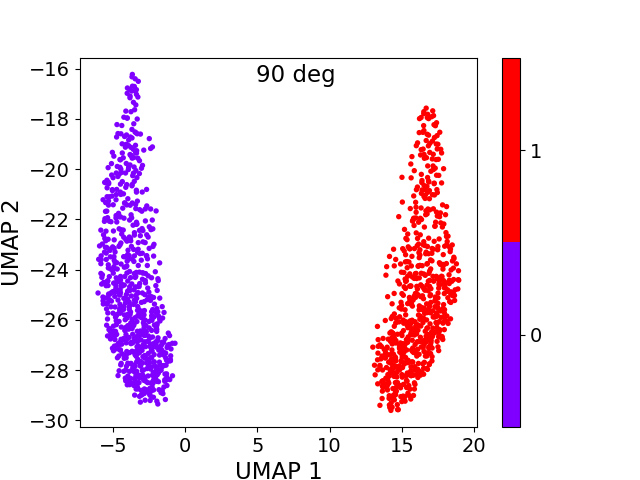}
\includegraphics[width=0.3\textwidth]{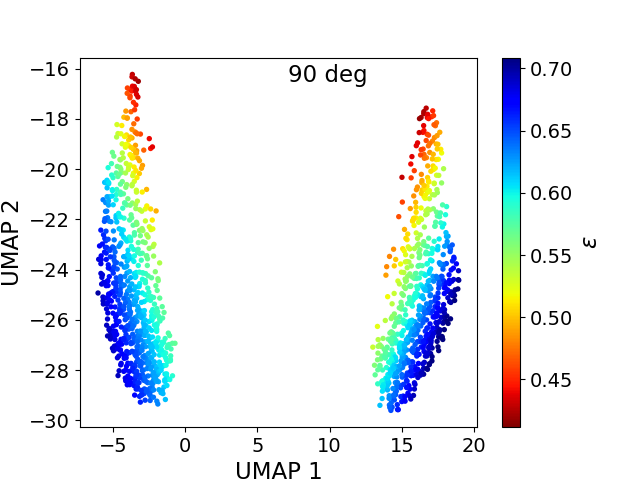}
\includegraphics[width=0.3\textwidth]{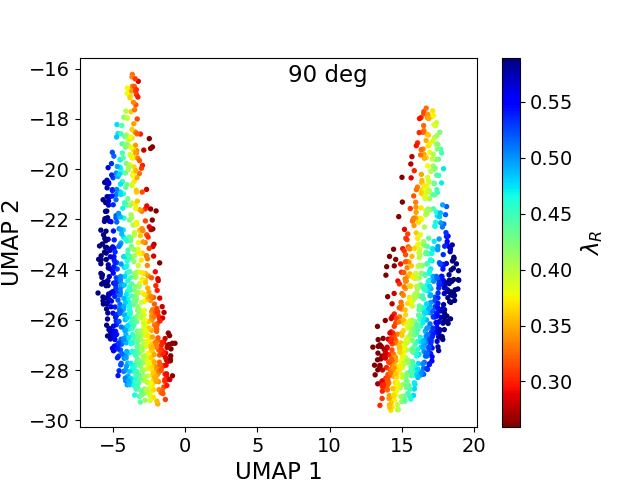} \\
\includegraphics[width=0.3\textwidth]{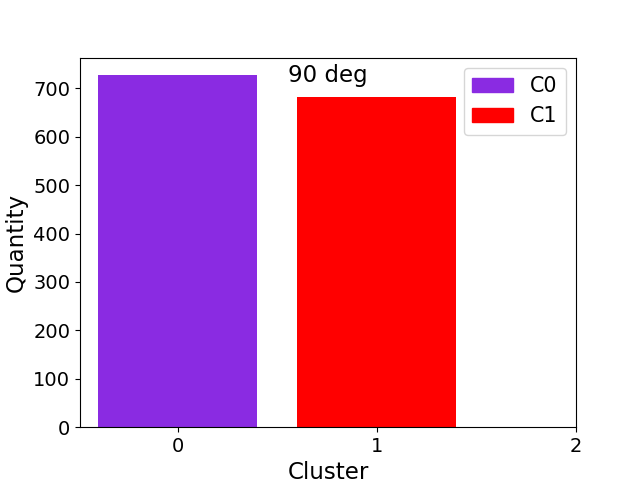}
\includegraphics[width=0.3\textwidth]{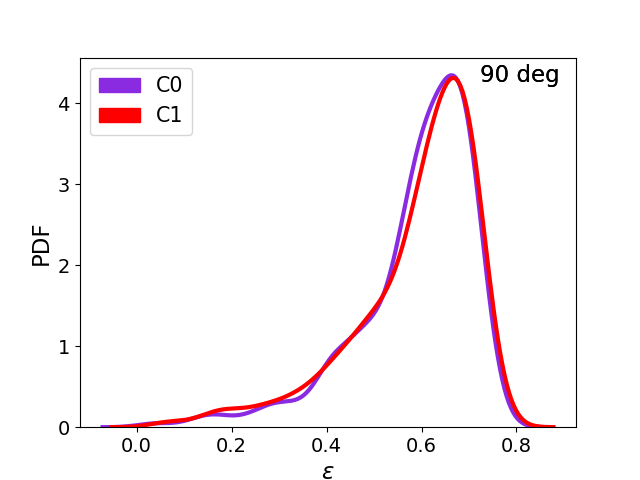}
\includegraphics[width=0.3\textwidth]{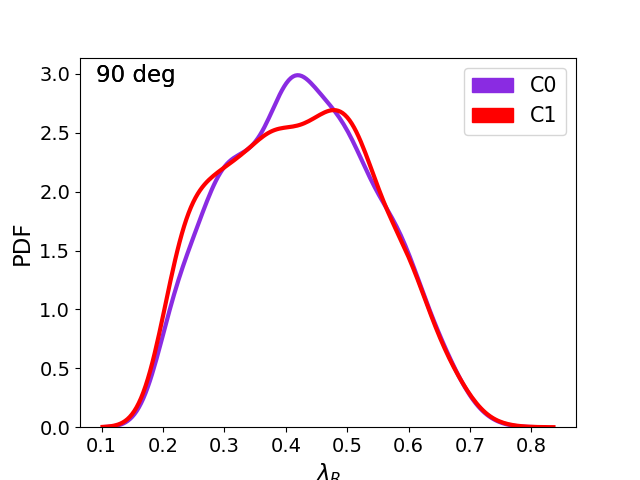} \\
\includegraphics[width=0.3\textwidth]{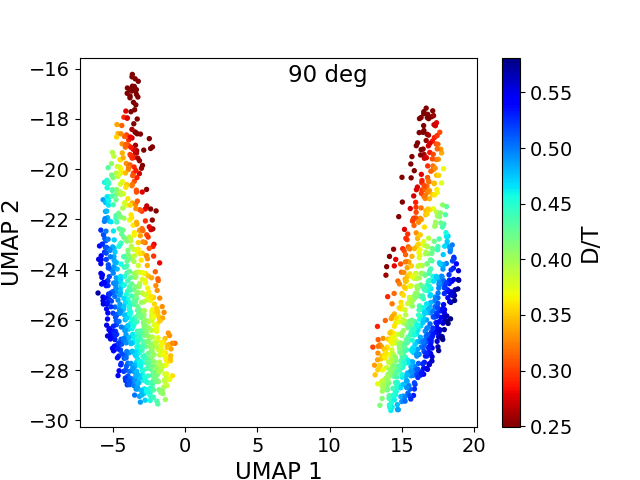}
\includegraphics[width=0.3\textwidth]{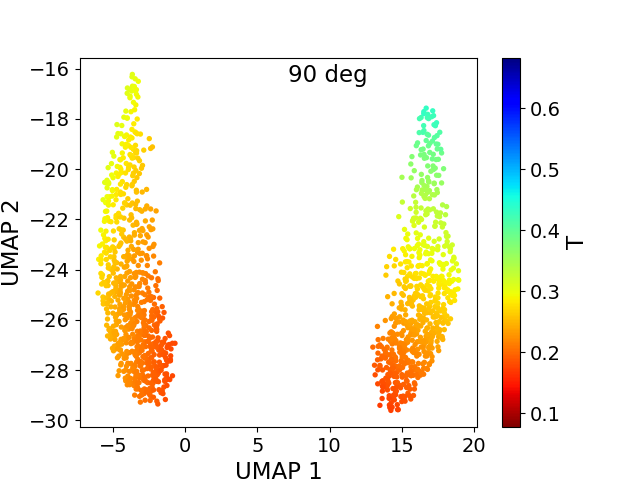}
\includegraphics[width=0.3\textwidth]{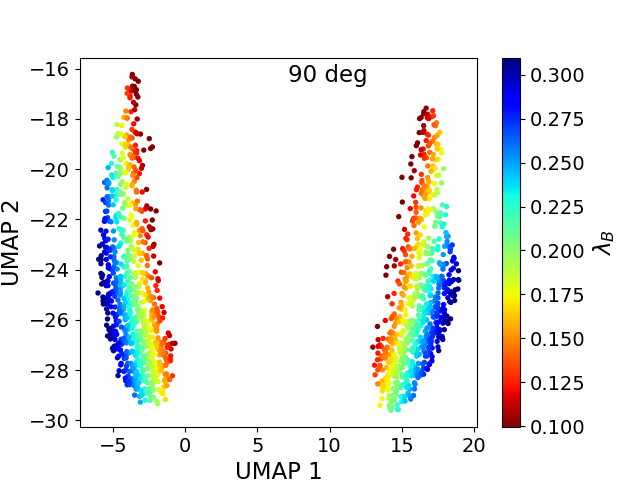} \\
\includegraphics[width=0.3\textwidth]{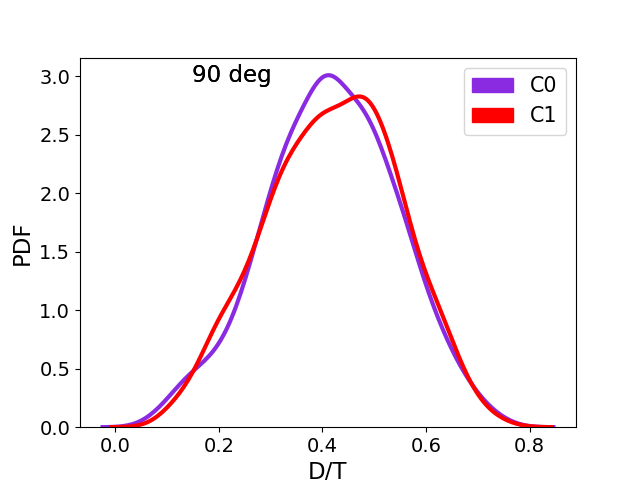}
\includegraphics[width=0.3\textwidth]{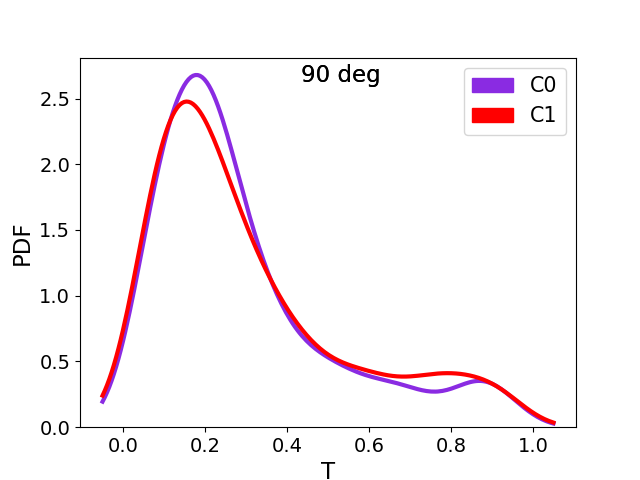}
\includegraphics[width=0.3\textwidth]{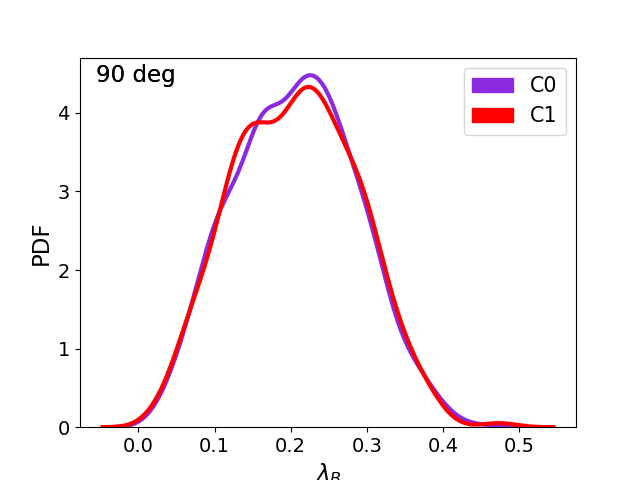} \\
 \caption{Results of \nombre applied to all the  kinematic maps types of galaxies with $\lambda_{\rm R, edge-on} > 0.2$ observed at 90 degrees. 
    \textit{Top row panels:} {\sc HDBSCAN} clusters in the {\sc UMAP} bidimensional projection (\textit{left panel}), distributions of the projected parameters $\varepsilon$ (\textit{middle panel}) and $\lambda_R$ (\textit{right panel}) on the projection. \textit{Second row panels:} size of the clusters (\textit{left panel}), PDFs of $\varepsilon$ (\textit{middle panel}) and  $\lambda_R$ (\textit{right panel}) for each cluster. \textit{Third row panels:} distributions of $D/T$ (\textit{left panel}), $T$ (\textit{middle panel}), and $\lambda_B$ (\textit{right panel}) on the projection, \textit{Bottom row panels:} PDFs of $D/T$ (\textit{left panel}), $T$ (\textit{middle panel}), and $\lambda_B$ (\textit{right panel}) for each cluster. Colorbar limits are fixed to 10$^{\rm th}$ and 90$^{\rm th}$ percentile of the variables.}
  \label{fig:clus_seplambdag02}
\end{figure*}

The second group of galaxies, $\lambda_{\rm R, edge-on}$ > 0.2,  consists on 1409 galaxies and \nombre yields a clustering depicting very clear groups with no outliers.
In Fig.~\ref{fig:clus_seplambdag02} we show the analysis of the two clusters obtained by our method. 
Very similar distributions of the parameters can be appreciated from the figure.
Due to the high p-values of the two-sided Brunner-Munzel tests (above 0.5) in all cases, we conclude that are no significant trends for the parameters in one cluster to be greater than those in the other.
Furthermore, the sizes of both groups are similar.
Galaxies in C0 and C1 rotate in opposite directions and this is what \nombre can capture.
%Galaxies in C0 rotate counterclockwise while those in C1 rotate clockwise and this direction is what \nombre can capture.

When using solely SRs as input, the method yields similar conclusions to those in Sec.~\ref{sec:call}, in which galaxies with little amount of rotation can be separated according to their shapes. Despite having identified three groups instead of the two in Sec.~\ref{sec:call1} (C3 and C4), our clustering is weaker when the input is not rich enough in variety of morphologies and kinematics.
If only FRs are clustered, our method still identifies two clear groups differentiated by the orientation of galaxy rotation, as happens when the whole sample is considered (Secs.~\ref{sec:vlos1} and \ref{sec:call1}).

\section{Mixing galaxies observed at different inclinations}
\label{sec:mix}

Throughout this work, we have clustered galaxies at a fixed inclination. 
Clustering kinematic maps from galaxies observed from different angles is more realistic than the procedures described in Sec.~\ref{sec:vlos} and Sec.~\ref{sec:call}.

In this Section, we assess the applicability and the robustness of \nombre when mixing galaxies at different inclinations.
We use all kinematic map types, as in Sec.~\ref{sec:call}.
The input of the method is the set of the kinematic maps constructed for each galaxy from different inclinations.
Due to the aforementioned reasons related to the reliability of the method as a function of inclination, we consider angles of 90, 60 and 45 degrees.
Hence, each galaxy is considered three times.

\subsection{Clustering of galaxies from different inclinations}
\label{sec:clust_mix}

We obtain a ``supersample'' that includes three different inclinations (90, 60, and 45 degrees) for each galaxy and consists, thus, on 2064 $\times$ 3 = 6192 elements.
Therefore, we need to project 6192 high dimensional points depicting the kinematic maps of each galaxy.
After the application of the method, we obtain six clusters with 229 outliers, among which there are 153 distinct galaxies. The number of galaxies appearing at least once in the set of clusters is 2042.

\begin{figure*}
    \centering
    \includegraphics[width=0.3\textwidth]{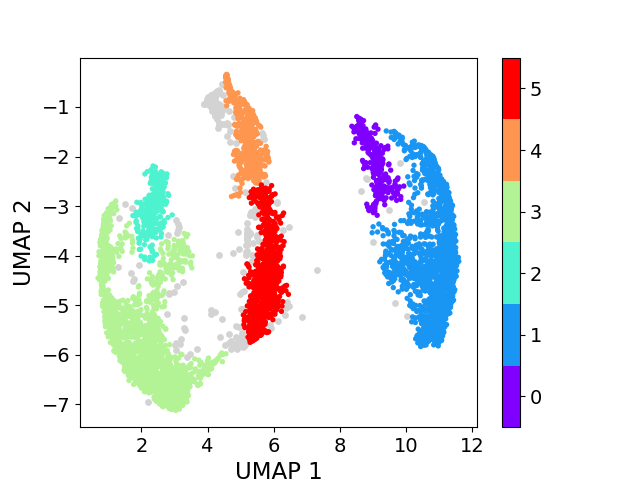}
    \includegraphics[width=0.3\textwidth]{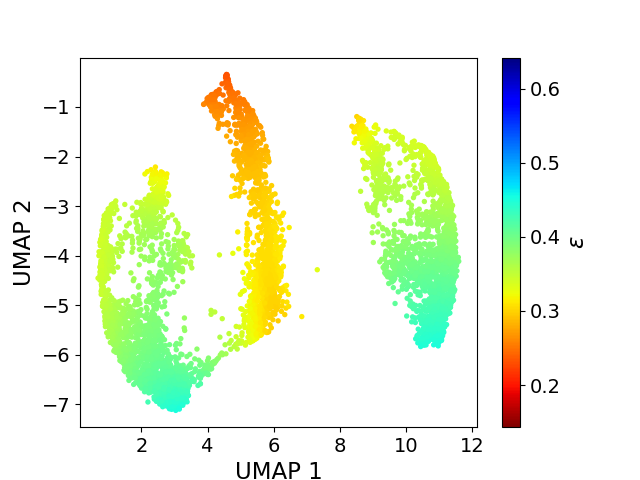}
    \includegraphics[width=0.3\textwidth]{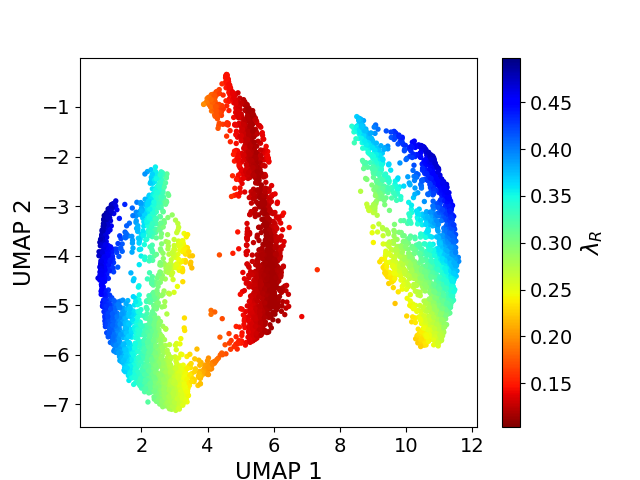} \\
    \includegraphics[width=0.3\textwidth]{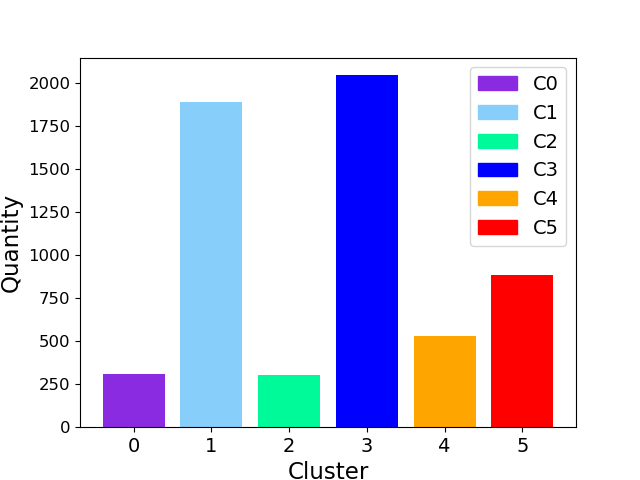}
    \includegraphics[width=0.3\textwidth]{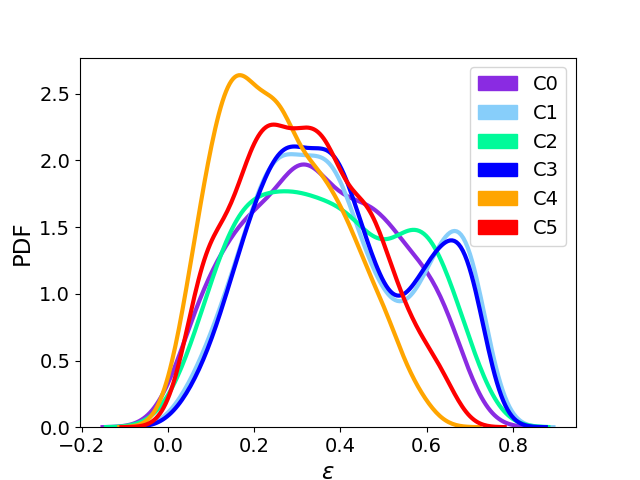}
    \includegraphics[width=0.3\textwidth]{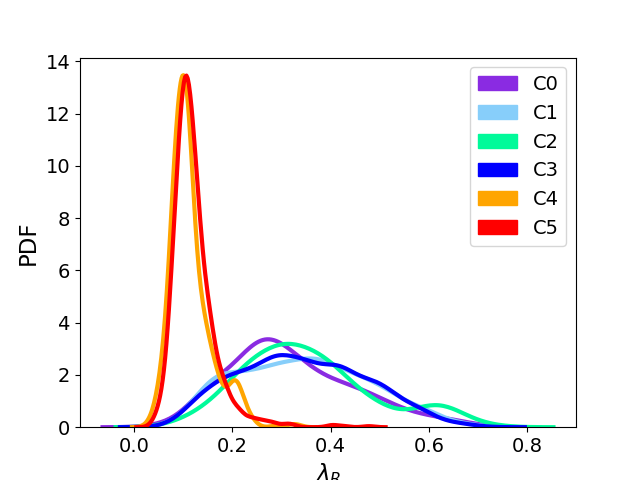} \\    \includegraphics[width=0.3\textwidth]{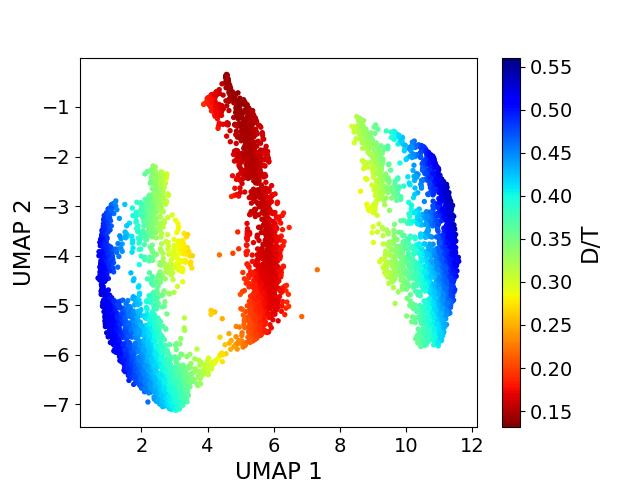}
    \includegraphics[width=0.3\textwidth]{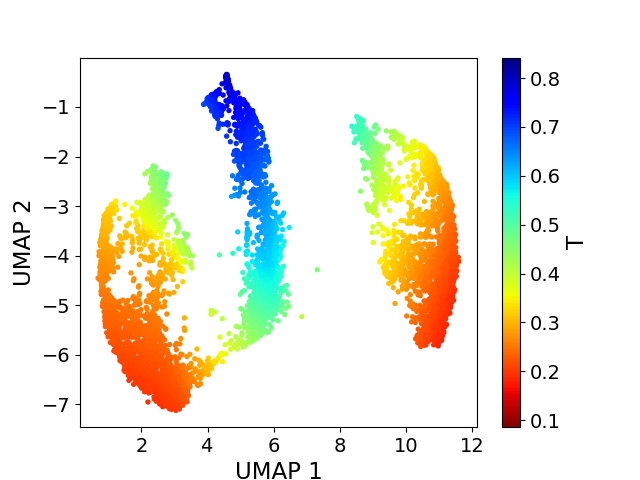}    
    \includegraphics[width=0.3\textwidth]{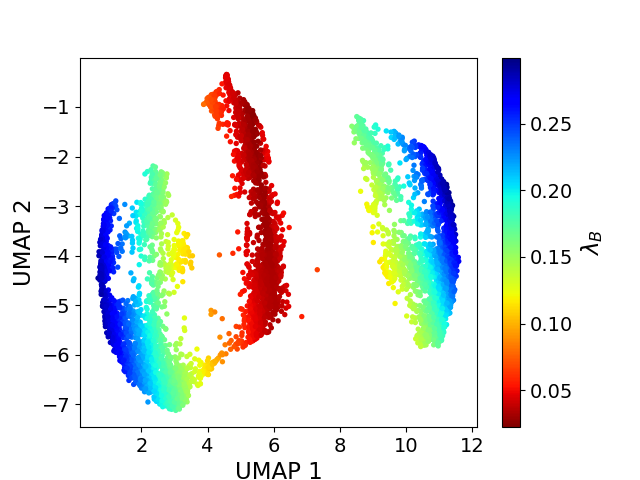} \\
    \includegraphics[width=0.3\textwidth]{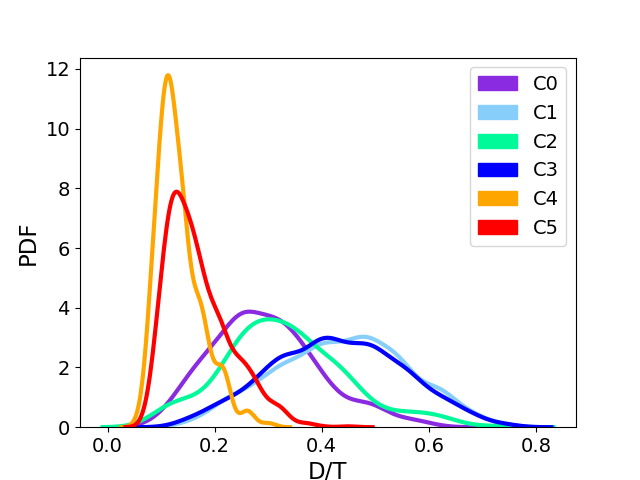}
    \includegraphics[width=0.3\textwidth]{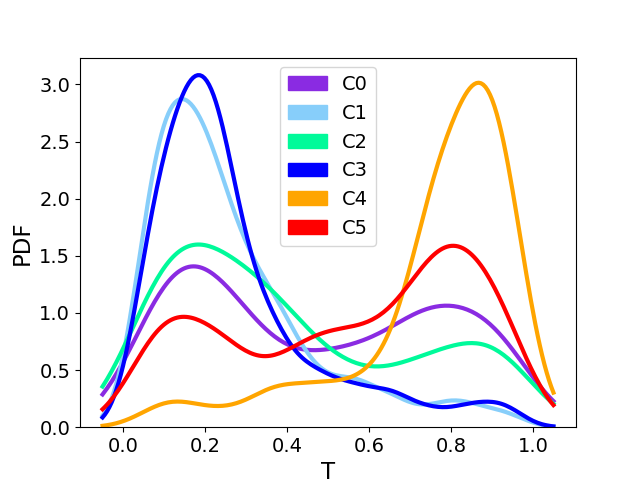}
    \includegraphics[width=0.3\textwidth]{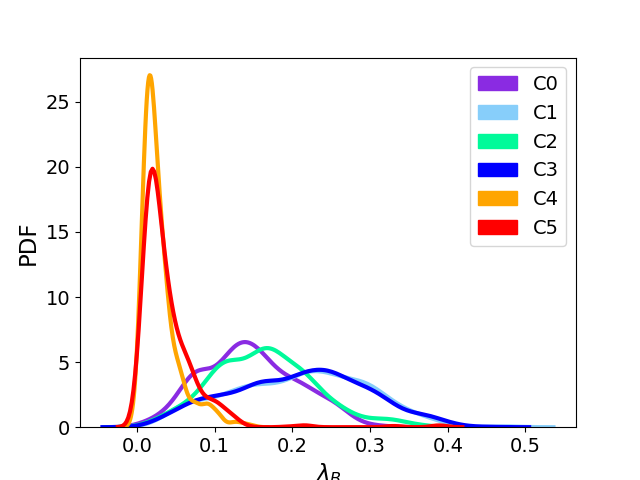}
    
    \caption{Results of \nombre applied to all the  kinematic maps types of our galaxy sample mixing three different inclinations including the outliers of the method as grey dots.
    \textit{Top row panels:} {\sc HDBSCAN} clusters in the {\sc UMAP} bidimensional projection (\textit{left panel}), distributions of the projected parameters $\varepsilon$ (\textit{middle panel}) and $\lambda_R$ (\textit{right panel}) on the projection. \textit{Second row panels:} size of the clusters (\textit{left panel}), PDFs of $\varepsilon$ (\textit{middle panel}) and  $\lambda_R$ (\textit{right panel}) for each cluster. \textit{Third row panels:} distributions of $D/T$ (\textit{left panel}), $T$ (\textit{middle panel}), and $\lambda_B$ (\textit{right panel}) on the projection, \textit{Bottom row panels:} PDFs of $D/T$ (\textit{left panel}), $T$ (\textit{middle panel}), and $\lambda_B$ (\textit{right panel}) for each cluster. Colorbar limits are fixed to 10$^{\rm th}$ and 90$^{\rm th}$ percentile of the variables.}
    \label{fig:clustmix}
\end{figure*}

In Fig.~\ref{fig:clustmix} we show the clustering obtained with the input studied in this Section and a summary of the properties of its galaxies.
We include physical and observational properties, being the latter sensitive to the angle of observation.
From the figure, it can be appreciated a division of galaxies with low level of rotation (C4 and C5) and faster rotators (C0, C1, C2, and C3), as occurs when the inclination is fixed.

Low rotation clusters present differences in the distributions of the parameters. By means of the Brunner-Munzel test, we find that galaxies in C4 have significantly lower $D/T$, lower $\lambda_B$, lower $\lambda_R$, lower $\varepsilon$, and higher $T$ than those from C5, which is consistent with having a smaller disc component.

The differences between the parameter distribution from C1 and C3 are less noticeable by inspection of Fig.~\ref{fig:clustmix}. Statistically, Brunner-Munzel test shows that disc fractions in C1 are greater than those in C3, but the trend is weak, since the p-value is 0.01. Similarly to Sec.~\ref{sec:call}, the smallest clusters C0 and C2 present parameters with intermediate values between SRs and FRs. The groups C0+C1 and C2+C3 are populated by galaxies rotating with opposite directions.

We show that when considering galaxies observed at different high inclinations, \nombre is still useful and the clustering meaningful, yielding similar conclusions to those of the edge-on galaxies study.

\subsection{On the robustness of our method}
\label{sec:robust}

In this Subsection, we check that, in most cases, the different instances of the same galaxy (that is, the kinematic maps for each galaxy observed at 90, 60, and 45 degrees) are clustered in the same group after the application of \nombre.
We detail in Table~\ref{table:incgroups} the number of elements within each group with the number of repetitions for each distinct galaxy. 

As stated in Table~\ref{table:incgroups}, in C1 and C3 the vast majority of the galaxies appears three times in the same cluster, that is, they receive the same classification regardless of the inclination.
The situation is similar in the low rotation clusters, C4 and C5: although the fraction of robustly classified galaxies is lower, it is still high.
In C0 and C2, these fractions of galaxies repeated three times are the lowest.
That means that galaxies on those groups are more susceptible to inclinations effects.

\begin{table*}
\caption{Number of galaxies and repetitions in each group obtained when applying \nombre to the mixture of the kinematic maps for the same galaxies observed at three different inclinations. We include the percentage of well classified galaxies.} 
\label{table:incgroups}      
\centering
\resizebox{\textwidth}{!}{%
\footnotesize
\begin{tabular}{ccccccc} 
\hline\hline  
 & C0  & C1 & C2 & C3 & C4 & C5\\
\hline 
Cluster size & 308 & 1890 & 303 & 2047 & 528 & 887 \\
N$^{\circ}$ of distinct galaxies & 154 & 672 & 174 & 749 & 224 & 369  \\
N$^{\circ}$ of galaxies not repeated & 65 & 29 & 102 & 39 & 53 & 73 \\
N$^{\circ}$ of galaxies repeated twice & 24 & 68 & 15 & 122 & 38 & 74 \\
N$^{\circ}$ of galaxies repeated three times & 65 (42 per-cent) & 575 (86 per-cent) & 57 (33 per-cent) & 588 (79 per-cent) & 133 (59 per-cent) & 222 (60 per-cent) \\
  \hline 
\end{tabular}} \\
\end{table*}

The result mentioned above yields a lower bound for robustly classified galaxies. We can affirm that 1640 out of 2064 galaxies (80 per-cent) are robustly classified by \nombre. 
This fraction may be higher if we take into account that not all 2064 galaxies have been clustered and that we are not considering galaxies repeated twice in a particular cluster.
This value is conservative but it is good enough to conclude that \nombre is robust even if the inclinations in the range between 45 deg and 90 deg are mixed.

The sizes of C0 and C2 are notably smaller than those of the other clusters as can be seen from Table~\ref{table:incgroups} and from the bar plot in Fig.~\ref{fig:clustmix}. In both clusters, the number of elements depicts the 5 per-cent of the total input size.
These clusters would be joined to C1 and C3, respectively, if we applied a more restrictive condition regarding the minimum cluster size.
Moreover, galaxies on those FRs clusters share similar properties in terms of shape and kinematic parameters, as mentioned above. 
If that was the case, we would obtain two larger clusters of sizes 2198 (C0 + C1) and 2350 (C2 + C3), from which 94 per-cent and 89 per-cent of their galaxies correctly classified regardless of their inclination.
In addition, taking into account that, as the inclination decreases, rotation is less noticeable and our method is then not able to separate SRs according to shape in different groups, as seen in Sec.~\ref{sec:call2}, joining C4 and C5 in a single cluster (C4 + C5) would also be reasonable when analysing the robustness of \nombre.
We would obtain 78 per-cent of properly classified SRs galaxies in C4 + C5.
Following this idea, by adding the number of robustly classified galaxies in C0 + C1, C2 + C3, and, C4 + C5, the aforementioned bound increases to 90 per-cent.

\section{Conclusions}
\label{sec: conclusions}

In this work, we define clear steps to perform an unsupervised kinematic morphology classification of a sample of galaxies from the {\sc eagle} simulation \citep{Schaye2015, Crain2015}.
Since the input of \nombre are kinematic maps mimics IFS observations, it may be applied either to observational data sets or to other simulations.
These maps are computed at different inclinations.
We project the maps (considered as high dimensional points) to a bidimensional space using the {\sc UMAP} algorithm \citep{McInnesUMAP2018} followed by the application of the {\sc HDBSCAN} clustering \citep{CampelloHDBSCAN2013}.
We identify the advantages and limitations of this methodology and conclude that it may be useful for defining unsupervised classifications of galaxies that capture their main morphological properties based on physical parameters.

The main results of this study are:

\begin{itemize}

    \item Line-of-sight velocity maps as input are sufficient to perform a good classification in SRs and FRs for edge-on galaxies. Most galaxies in the cluster with the lowest values of $D/T$ are classified as SRs according to the Eq.~\ref{eq:SR} \citep{vdSande2021}, as shown in Fig.~\ref{fig:lambdaeps_vLOS}. However, there is a fraction of these galaxies that would be considered FRs using this parametric criterion despite having low disc fractions.
    Therefore, \nombre may be considered more accurate than those based on the observational projected parameters ($\lambda_R$ and $\varepsilon$), since it provides a clearer separation regarding the rotational component. Furthermore, galaxies in clusters associated with high rotation are separated by the direction of rotation.
    We show that this is an advantage of \nombre in the sense that it allows the extraction of this information without introducing a false dimension that may leads to underperformance of the classification. Albeit not relevant to explain the kinematic morphologies of galaxies, this information could be useful to study the relative rotation between galaxies.
    
    \item By adding the velocity dispersion and flux maps to the analysis, edge-on galaxies with low rotation are divided in two groups (C3 and C4) which differ mainly in the triaxiality parameter yielding a refinement of the low rotation cluster found in Sec.~\ref{sec:vlos} (Fig.~\ref{fig:clustall}). Using hypothesis testing, we also find systematic differences in the values of $D/T$ and $\varepsilon$ from C3 and C4, and no significant difference regarding the spin parameters. 
    Among the clusters with more rotating galaxies (C0, C1, and C2), C0 and C1 differ mainly in the rotation orientation. 
    The distributions of the parameters for galaxies in C2 present graphically noticeable differences with respect to the others, but its size is too small to draw a robust conclusion. 
    
    \item Figs.~\ref{fig:clust_all_vlos} and \ref{fig:clus_inc_all} show that at inclinations greater or equal than 45 degrees, the clusters obtained in Sec.~\ref{sec:call} can be considered subsets of the groups obtained in the analysis performed in Sec.~\ref{sec:vlos}.
    Therefore, the addition of the dispersion and flux maps preserves the information provided by the velocity maps. Since the use of all kinematic map types yields a more detailed description of low rotation edge-on galaxies, we suggest including them when performing morphological classification using \nombre. 

    \item Due to the fact that the input data considered for our analysis consist on IFU kinematic maps, which are sensitive to the angle of observation,
    \nombre is useful when galaxies are observed at inclinations at which rotation can be clearly distinguished. When the inclination is 30 degrees, the fraction of SRs (according to the parameter definition) identified in the low rotation cluster decreases significantly with respect to higher inclinations and the division in clusters is less clear. We adopt a conservative criterion preferring angles greater to or equal than 45 degrees.
    
    \item When applying \nombre exclusively to slow rotating galaxies, we can identify  groups with different shapes
    However, the separation between groups in the {\sc UMAP} projection is not as clear as when galaxies with a variety of morphologies and kinematics are clustered together. On the other hand, the method divides a sample of FRs according to the rotation orientation and the parameter distributions present no significant differences among the two groups. 
    
    \item When the input consists on galaxies observed at different inclinations, \nombre still clearly differentiates SRs and FRs. The former are divided in two groups, as happens when only edge-on galaxies are considered (Sec.~\ref{sec:call1}). Galaxies with large amount of rotation are separated according to the direction of rotation. Furthermore, we find clusters with intermediate parameters between SRs and FRs, albeit these clusters are very small in comparison to the others. 
    An input that mixes galaxies observed at different inclinations is more realistic and the fact that our method still obtains a meaningful clustering is encouraging.
    Furthermore, \nombre is robust with respect to the inclinations of the input galaxies.
    It can be seen that in at least $\sim 80$ per-cent of the cases, the different instances of each galaxy are classified in the same group.
    If we apply a more restrictive clustering algorithm, this fraction can increase to $\sim 90$ per-cent. When allowing finer clustering, our method produces some mixing within groups, but is still able to clearly separate SRs and FRs.

\end{itemize}

The methodology presented in this paper has proven to be accurate for a sample of mock galaxies and has succeeded in identifying their intrinsic properties and their projected parameters.
The need for observing galaxies at high inclinations can be mitigated by predicting how the kinematic maps of galaxies observed in an arbitrary inclination would be seen edge-on.
Dynamical models \citep{Cappellari2008} have been used for similar tasks in the last decade. 

The next stage is to asses its applicability to observational data sets and to new generation simulations.
Testing \nombre on mock inputs that are more fairly comparable to observations would be an enriching and necessary step before applying it to real observations. 
We can obtain more diverse input samples by choosing different distances and inclination angles to build the {\sc SimSpin} data cubes besides changing the user-specified point spread function \citep{Harborne2020}.
There are other considerations to keep in mind when dealing IFS such as the need of a minimum signal-to-noise ratio to measure line-of-sight velocity.
Voronoi tessellations binning \citep{Cappellari2003} can be used for unbiased recovery of this kinematic information and it may be interesting to explore methods like the procedure proposed by \cite{WaloMartin2020}.

ML algorithms are becoming mandatory to study astronomy and the use of unsupervised techniques has the potential to contribute to scrutinize large data sets to extract information with physical meaning. 

\begin{acknowledgements}
PBT acknowledges partial funding by Fondecyt 1200703/2020 (ANID), ANID Basal Project FB210003, and Nucleo Milenio ANID ERIS. SP acknowledges support through PIP CONICET 11220170100638CO.
\end{acknowledgements}

\bibliographystyle{aa}
\bibliography{GalaxyClustersEAGLE}

\begin{thebibliography}{118}
\expandafter\ifx\csname natexlab\endcsname\relax\def\natexlab#1{#1}\fi

\bibitem[{{Abraham} {et~al.}(2003){Abraham}, {van den Bergh}, \&
  {Nair}}]{Abraham2003}
{Abraham}, R.~G., {van den Bergh}, S., \& {Nair}, P. 2003, \apj, 588, 218

\bibitem[{{Artale} {et~al.}(2019){Artale}, {Pedrosa}, {Tissera}, {Cataldi}, \&
  {Di Cintio}}]{Artale2019}
{Artale}, M.~C., {Pedrosa}, S.~E., {Tissera}, P.~B., {Cataldi}, P., \& {Di
  Cintio}, A. 2019, \aap, 622, A197

\bibitem[{{Bacon} {et~al.}(2001){Bacon}, {Copin}, {Monnet}, {Miller},
  {Allington-Smith}, {Bureau}, {Carollo}, {Davies}, {Emsellem}, {Kuntschner},
  {Peletier}, {Verolme}, \& {de Zeeuw}}]{Bacon2001}
{Bacon}, R., {Copin}, Y., {Monnet}, G., {et~al.} 2001, \mnras, 326, 23

\bibitem[{{Bah{\'e}} {et~al.}(2017){Bah{\'e}}, {Barnes}, {Dalla Vecchia},
  {Kay}, {White}, {McCarthy}, {Schaye}, {Bower}, {Crain}, {Theuns}, {Jenkins},
  {McGee}, {Schaller}, {Thomas}, \& {Trayford}}]{Brae2017}
{Bah{\'e}}, Y.~M., {Barnes}, D.~J., {Dalla Vecchia}, C., {et~al.} 2017, \mnras,
  470, 4186

\bibitem[{{Ball} \& {Brunner}(2010)}]{Ball2010}
{Ball}, N.~M. \& {Brunner}, R.~J. 2010, International Journal of Modern Physics
  D, 19, 1049

\bibitem[{{Baron}(2019)}]{Baron2019}
{Baron}, D. 2019, arXiv e-prints, arXiv:1904.07248

\bibitem[{{Bassett} \& {Foster}(2019)}]{Bassett2019}
{Bassett}, R. \& {Foster}, C. 2019, \mnras, 487, 2354

\bibitem[{{Bignone} {et~al.}(2020){Bignone}, {Pedrosa}, {Trayford}, {Tissera},
  \& {Pellizza}}]{Bignone2020}
{Bignone}, L.~A., {Pedrosa}, S.~E., {Trayford}, J.~W., {Tissera}, P.~B., \&
  {Pellizza}, L.~J. 2020, \mnras, 491, 3624

\bibitem[{{Binney}(1978)}]{Binney1978}
{Binney}, J. 1978, \mnras, 183, 501

\bibitem[{{Bois} {et~al.}(2011){Bois}, {Emsellem}, {Bournaud}, {Alatalo},
  {Blitz}, {Bureau}, {Cappellari}, {Davies}, {Davis}, {de Zeeuw}, {Duc},
  {Khochfar}, {Krajnovi{\'c}}, {Kuntschner}, {Lablanche}, {McDermid},
  {Morganti}, {Naab}, {Oosterloo}, {Sarzi}, {Scott}, {Serra}, {Weijmans}, \&
  {Young}}]{Bois2011}
{Bois}, M., {Emsellem}, E., {Bournaud}, F., {et~al.} 2011, \mnras, 416, 1654

\bibitem[{{Bongiovanni} {et~al.}(2019){Bongiovanni}, {Ram{\'o}n-P{\'e}rez},
  {P{\'e}rez Garc{\'\i}a}, {Cepa}, {Cervi{\~n}o}, {Nadolny}, {P{\'e}rez
  Mart{\'\i}nez}, {Alfaro}, {Casta{\~n}eda}, {de Diego}, {Ederoclite},
  {Fern{\'a}ndez-Lorenzo}, {Gallego}, {Gonz{\'a}lez}, {Gonz{\'a}lez-Serrano},
  {Lara-L{\'o}pez}, {Oteo G{\'o}mez}, {Padilla Torres}, {Pintos-Castro},
  {Povi{\'c}}, {S{\'a}nchez-Portal}, {Jones}, {Bland-Hawthorn}, \&
  {Cabrera-Lavers}}]{Bongiovanni2019}
{Bongiovanni}, {\'A}., {Ram{\'o}n-P{\'e}rez}, M., {P{\'e}rez Garc{\'\i}a},
  A.~M., {et~al.} 2019, \aap, 631, A9

\bibitem[{{Brough} {et~al.}(2017){Brough}, {van de Sande}, {Owers},
  {d'Eugenio}, {Sharp}, {Cortese}, {Scott}, {Croom}, {Bassett}, {Bekki},
  {Bland-Hawthorn}, {Bryant}, {Davies}, {Drinkwater}, {Driver}, {Foster},
  {Goldstein}, {L{\'o}pez-S{\'a}nchez}, {Medling}, {Sweet}, {Taranu}, {Tonini},
  {Yi}, {Goodwin}, {Lawrence}, \& {Richards}}]{Brough2017}
{Brough}, S., {van de Sande}, J., {Owers}, M.~S., {et~al.} 2017, \apj, 844, 59

\bibitem[{Brunner \& Munzel(2000)}]{Brunner2000}
Brunner, E. \& Munzel, U. 2000, Biometrical Journal, 42, 17

\bibitem[{{Bruzual} \& {Charlot}(2003)}]{BC2003}
{Bruzual}, G. \& {Charlot}, S. 2003, \mnras, 344, 1000

\bibitem[{{Bryant} {et~al.}(2015){Bryant}, {Owers}, {Robotham}, {Croom},
  {Driver}, {Drinkwater}, {Lorente}, {Cortese}, {Scott}, {Colless}, {Schaefer},
  {Taylor}, {Konstantopoulos}, {Allen}, {Baldry}, {Barnes}, {Bauer},
  {Bland-Hawthorn}, {Bloom}, {Brooks}, {Brough}, {Cecil}, {Couch}, {Croton},
  {Davies}, {Ellis}, {Fogarty}, {Foster}, {Glazebrook}, {Goodwin}, {Green},
  {Gunawardhana}, {Hampton}, {Ho}, {Hopkins}, {Kewley}, {Lawrence},
  {Leon-Saval}, {Leslie}, {McElroy}, {Lewis}, {Liske}, {L{\'o}pez-S{\'a}nchez},
  {Mahajan}, {Medling}, {Metcalfe}, {Meyer}, {Mould}, {Obreschkow}, {O'Toole},
  {Pracy}, {Richards}, {Shanks}, {Sharp}, {Sweet}, {Thomas}, {Tonini}, \&
  {Walcher}}]{Bryant2015}
{Bryant}, J.~J., {Owers}, M.~S., {Robotham}, A.~S.~G., {et~al.} 2015, \mnras,
  447, 2857

\bibitem[{{Bullock} {et~al.}(2001){Bullock}, {Dekel}, {Kolatt}, {Kravtsov},
  {Klypin}, {Porciani}, \& {Primack}}]{Bullock2001}
{Bullock}, J.~S., {Dekel}, A., {Kolatt}, T.~S., {et~al.} 2001, \apj, 555, 240

\bibitem[{{Bundy} {et~al.}(2015){Bundy}, {Bershady}, {Law}, {Yan}, {Drory},
  {MacDonald}, {Wake}, {Cherinka}, {S{\'a}nchez-Gallego}, {Weijmans}, {Thomas},
  {Tremonti}, {Masters}, {Coccato}, {Diamond-Stanic}, {Arag{\'o}n-Salamanca},
  {Avila-Reese}, {Badenes}, {Falc{\'o}n-Barroso}, {Belfiore}, {Bizyaev},
  {Blanc}, {Bland-Hawthorn}, {Blanton}, {Brownstein}, {Byler}, {Cappellari},
  {Conroy}, {Dutton}, {Emsellem}, {Etherington}, {Frinchaboy}, {Fu}, {Gunn},
  {Harding}, {Johnston}, {Kauffmann}, {Kinemuchi}, {Klaene}, {Knapen},
  {Leauthaud}, {Li}, {Lin}, {Maiolino}, {Malanushenko}, {Malanushenko}, {Mao},
  {Maraston}, {McDermid}, {Merrifield}, {Nichol}, {Oravetz}, {Pan}, {Parejko},
  {Sanchez}, {Schlegel}, {Simmons}, {Steele}, {Steinmetz}, {Thanjavur},
  {Thompson}, {Tinker}, {van den Bosch}, {Westfall}, {Wilkinson}, {Wright},
  {Xiao}, \& {Zhang}}]{Bundy+2015}
{Bundy}, K., {Bershady}, M.~A., {Law}, D.~R., {et~al.} 2015, \apj, 798, 7

\bibitem[{{Campello} {et~al.}(2013){Campello}, {Moulavi}, \&
  {Sander}}]{CampelloHDBSCAN2013}
{Campello}, R. J. G.~B., {Moulavi}, D., \& {Sander}, J. 2013, in Advances in
  Knowledge Discovery and Data Mining, ed. J.~{Pei}, V.~S. {Tseng}, L.~{Cao},
  H.~{Motoda}, \& G.~{Xu} (Berlin, Heidelberg: Springer Berlin Heidelberg),
  160--172

\bibitem[{{Cappellari}(2008)}]{Cappellari2008}
{Cappellari}, M. 2008, \mnras, 390, 71

\bibitem[{{Cappellari}(2016)}]{cappellarireview2016}
{Cappellari}, M. 2016, \araa, 54, 597

\bibitem[{{Cappellari} \& {Copin}(2003)}]{Cappellari2003}
{Cappellari}, M. \& {Copin}, Y. 2003, \mnras, 342, 345

\bibitem[{{Cataldi} {et~al.}(2021){Cataldi}, {Pedrosa}, {Tissera}, \&
  {Artale}}]{Cataldi2021}
{Cataldi}, P., {Pedrosa}, S.~E., {Tissera}, P.~B., \& {Artale}, M.~C. 2021,
  \mnras, 501, 5679

\bibitem[{{Chabrier}(2003)}]{Chabrier2003}
{Chabrier}, G. 2003, \pasp, 115, 763

\bibitem[{{Chadha}(2007)}]{galaxyzoo}
{Chadha}, K.~S. 2007, Astronomy Now, 21, 28

\bibitem[{Cheng {et~al.}(2021)Cheng, Huertas-Company, Conselice,
  Aragón-Salamanca, Robertson, \& Ramachandra}]{Cheng2021}
Cheng, T.-Y., Huertas-Company, M., Conselice, C.~J., {et~al.} 2021, \mnras,
  503, 4446

\bibitem[{{Chisari} {et~al.}(2015){Chisari}, {Codis}, {Laigle}, {Dubois},
  {Pichon}, {Devriendt}, {Slyz}, {Miller}, {Gavazzi}, \&
  {Benabed}}]{Chisari2015}
{Chisari}, N., {Codis}, S., {Laigle}, C., {et~al.} 2015, \mnras, 454, 2736

\bibitem[{{Combes}(2009)}]{Combes2009}
{Combes}, F. 2009, in Astronomical Society of the Pacific Conference Series,
  Vol. 419, Galaxy Evolution: Emerging Insights and Future Challenges, ed.
  S.~{Jogee}, I.~{Marinova}, L.~{Hao}, \& G.~A. {Blanc}, 31

\bibitem[{{Conselice}(2003)}]{Conselice2003}
{Conselice}, C.~J. 2003, \apjs, 147, 1

\bibitem[{{Conselice}(2014)}]{Conselice2014}
{Conselice}, C.~J. 2014, \araa, 52, 291

\bibitem[{{Correa} {et~al.}(2017){Correa}, {Schaye}, {Clauwens}, {Bower},
  {Crain}, {Schaller}, {Theuns}, \& {Thob}}]{Correa2017}
{Correa}, C.~A., {Schaye}, J., {Clauwens}, B., {et~al.} 2017, \mnras, 472, L45

\bibitem[{{Crain} {et~al.}(2015){Crain}, {Schaye}, {Bower}, {Furlong},
  {Schaller}, {Theuns}, {Dalla Vecchia}, {Frenk}, {McCarthy}, {Helly},
  {Jenkins}, {Rosas-Guevara}, {White}, \& {Trayford}}]{Crain2015}
{Crain}, R.~A., {Schaye}, J., {Bower}, R.~G., {et~al.} 2015, \mnras, 450, 1937

\bibitem[{{Dalla Vecchia} \& {Schaye}(2012)}]{Dallas2012}
{Dalla Vecchia}, C. \& {Schaye}, J. 2012, \mnras, 426, 140

\bibitem[{{Davies} {et~al.}(1983){Davies}, {Efstathiou}, {Fall}, {Illingworth},
  \& {Schechter}}]{Davies1983}
{Davies}, R.~L., {Efstathiou}, G., {Fall}, S.~M., {Illingworth}, G., \&
  {Schechter}, P.~L. 1983, \apj, 266, 41

\bibitem[{{de Diego} {et~al.}(2020){de Diego}, {Nadolny}, {Bongiovanni},
  {Cepa}, {Povi{\'c}}, {P{\'e}rez Garc{\'\i}a}, {Padilla Torres},
  {Lara-L{\'o}pez}, {Cervi{\~n}o}, {P{\'e}rez Mart{\'\i}nez}, {Alfaro},
  {Casta{\~n}eda}, {Fern{\'a}ndez-Lorenzo}, {Gallego}, {Gonz{\'a}lez},
  {Gonz{\'a}lez-Serrano}, {Pintos-Castro}, {S{\'a}nchez-Portal}, {Cedr{\'e}s},
  {Gonz{\'a}lez-Otero}, {Heath Jones}, \& {Bland-Hawthorn}}]{deDiego2020}
{de Diego}, J.~A., {Nadolny}, J., {Bongiovanni}, {\'A}., {et~al.} 2020, \aap,
  638, A134

\bibitem[{{de Vaucouleurs}(1948)}]{deVaucouleurs1948}
{de Vaucouleurs}, G. 1948, Annales d'Astrophysique, 11, 247

\bibitem[{{Deng}(2013)}]{Deng2013}
{Deng}, X.-F. 2013, Research in Astronomy and Astrophysics, 13, 651

\bibitem[{{Djorgovski} \& {Davis}(1987)}]{Davis1987}
{Djorgovski}, S. \& {Davis}, M. 1987, \apj, 313, 59

\bibitem[{{Dolag} {et~al.}(2009){Dolag}, {Borgani}, {Murante}, \&
  {Springel}}]{Dolag2009}
{Dolag}, K., {Borgani}, S., {Murante}, G., \& {Springel}, V. 2009, \mnras, 399,
  497

\bibitem[{{Dubois} {et~al.}(2016){Dubois}, {Peirani}, {Pichon}, {Devriendt},
  {Gavazzi}, {Welker}, \& {Volonteri}}]{Dubois2016}
{Dubois}, Y., {Peirani}, S., {Pichon}, C., {et~al.} 2016, \mnras, 463, 3948

\bibitem[{{Emsellem} {et~al.}(2011){Emsellem}, {Cappellari}, {Krajnovi{\'c}},
  {Alatalo}, {Blitz}, {Bois}, {Bournaud}, {Bureau}, {Davies}, {Davis}, {de
  Zeeuw}, {Khochfar}, {Kuntschner}, {Lablanche}, {McDermid}, {Morganti},
  {Naab}, {Oosterloo}, {Sarzi}, {Scott}, {Serra}, {van de Ven}, {Weijmans}, \&
  {Young}}]{Emsellem2011}
{Emsellem}, E., {Cappellari}, M., {Krajnovi{\'c}}, D., {et~al.} 2011, \mnras,
  414, 888

\bibitem[{{Emsellem} {et~al.}(2007){Emsellem}, {Cappellari}, {Krajnovi{\'c}},
  {van de Ven}, {Bacon}, {Bureau}, {Davies}, {de Zeeuw}, {Falc{\'o}n-Barroso},
  {Kuntschner}, {McDermid}, {Peletier}, \& {Sarzi}}]{Emsellem2007}
{Emsellem}, E., {Cappellari}, M., {Krajnovi{\'c}}, D., {et~al.} 2007, \mnras,
  379, 401

\bibitem[{{Ene} {et~al.}(2018){Ene}, {Ma}, {Veale}, {Greene}, {Thomas},
  {Blakeslee}, {Foster}, {Walsh}, {Ito}, \& {Goulding}}]{Ene2018}
{Ene}, I., {Ma}, C.-P., {Veale}, M., {et~al.} 2018, \mnras, 479, 2810

\bibitem[{{Fisher} \& {Drory}(2008)}]{Fisher2008}
{Fisher}, D.~B. \& {Drory}, N. 2008, \aj, 136, 773

\bibitem[{{Foster} {et~al.}(2017){Foster}, {van de Sande}, {D'Eugenio},
  {Cortese}, {McDermid}, {Bland-Hawthorn}, {Brough}, {Bryant}, {Croom},
  {Goodwin}, {Konstantopoulos}, {Lawrence}, {L{\'o}pez-S{\'a}nchez}, {Medling},
  {Owers}, {Richards}, {Scott}, {Taranu}, {Tonini}, \& {Zafar}}]{Foster2017}
{Foster}, C., {van de Sande}, J., {D'Eugenio}, F., {et~al.} 2017, \mnras, 472,
  966

\bibitem[{{Graham} {et~al.}(2018){Graham}, {Cappellari}, {Li}, {Mao},
  {Bershady}, {Bizyaev}, {Brinkmann}, {Brownstein}, {Bundy}, {Drory}, {Law},
  {Pan}, {Thomas}, {Wake}, {Weijmans}, {Westfall}, \& {Yan}}]{Graham2018}
{Graham}, M.~T., {Cappellari}, M., {Li}, H., {et~al.} 2018, \mnras, 477, 4711

\bibitem[{{Greene} {et~al.}(2018){Greene}, {Leauthaud}, {Emsellem}, {Ge},
  {Arag{\'o}n-Salamanca}, {Greco}, {Lin}, {Mao}, {Masters}, {Merrifield},
  {More}, {Okabe}, {Schneider}, {Thomas}, {Wake}, {Pan}, {Bizyaev}, {Oravetz},
  {Simmons}, {Yan}, \& {van den Bosch}}]{Greene2018}
{Greene}, J.~E., {Leauthaud}, A., {Emsellem}, E., {et~al.} 2018, \apj, 852, 36

\bibitem[{{Haardt} \& {Madau}(2001)}]{Haardt2001}
{Haardt}, F. \& {Madau}, P. 2001, in Clusters of Galaxies and the High Redshift
  Universe Observed in X-rays, ed. D.~M. {Neumann} \& J.~T.~V. {Tran}, 64

\bibitem[{{Harborne} {et~al.}(2020){Harborne}, {Power}, \&
  {Robotham}}]{Harborne2020}
{Harborne}, K.~E., {Power}, C., \& {Robotham}, A. S.~G. 2020, \pasa, 37, e016

\bibitem[{{Harborne} {et~al.}(2019){Harborne}, {Power}, {Robotham}, {Cortese},
  \& {Taranu}}]{Harborne2019}
{Harborne}, K.~E., {Power}, C., {Robotham}, A.~S.~G., {Cortese}, L., \&
  {Taranu}, D.~S. 2019, \mnras, 483, 249

\bibitem[{{Hocking} {et~al.}(2018){Hocking}, {Geach}, {Sun}, \&
  {Davey}}]{Hocking2018}
{Hocking}, A., {Geach}, J.~E., {Sun}, Y., \& {Davey}, N. 2018, \mnras, 473,
  1108

\bibitem[{{Howard}(2017)}]{Howard2017}
{Howard}, E.~M. 2017, in Astronomical Society of the Pacific Conference Series,
  Vol. 512, Astronomical Data Analysis Software and Systems XXV, ed. N.~P.~F.
  {Lorente}, K.~{Shortridge}, \& R.~{Wayth}, 245

\bibitem[{{Hubble}(1926)}]{Hubble1926}
{Hubble}, E.~P. 1926, \apj, 64, 321

\bibitem[{{Illingworth}(1977)}]{Illingworth1977}
{Illingworth}, G. 1977, \apjl, 218, L43

\bibitem[{{Ivezi{\'c}} {et~al.}(2019){Ivezi{\'c}}, {Kahn}, {Tyson}, {Abel},
  {Acosta}, {Allsman}, {Alonso}, {AlSayyad}, {Anderson}, {Andrew}, {Angel},
  {Angeli}, {Ansari}, {Antilogus}, {Araujo}, {Armstrong}, {Arndt}, {Astier},
  {Aubourg}, {Auza}, {Axelrod}, {Bard}, {Barr}, {Barrau}, {Bartlett}, {Bauer},
  {Bauman}, {Baumont}, {Bechtol}, {Bechtol}, {Becker}, {Becla}, {Beldica},
  {Bellavia}, {Bianco}, {Biswas}, {Blanc}, {Blazek}, {Blandford}, {Bloom},
  {Bogart}, {Bond}, {Booth}, {Borgland}, {Borne}, {Bosch}, {Boutigny},
  {Brackett}, {Bradshaw}, {Brandt}, {Brown}, {Bullock}, {Burchat}, {Burke},
  {Cagnoli}, {Calabrese}, {Callahan}, {Callen}, {Carlin}, {Carlson},
  {Chandrasekharan}, {Charles-Emerson}, {Chesley}, {Cheu}, {Chiang}, {Chiang},
  {Chirino}, {Chow}, {Ciardi}, {Claver}, {Cohen-Tanugi}, {Cockrum}, {Coles},
  {Connolly}, {Cook}, {Cooray}, {Covey}, {Cribbs}, {Cui}, {Cutri}, {Daly},
  {Daniel}, {Daruich}, {Daubard}, {Daues}, {Dawson}, {Delgado}, {Dellapenna},
  {de Peyster}, {de Val-Borro}, {Digel}, {Doherty}, {Dubois},
  {Dubois-Felsmann}, {Durech}, {Economou}, {Eifler}, {Eracleous}, {Emmons},
  {Fausti Neto}, {Ferguson}, {Figueroa}, {Fisher-Levine}, {Focke}, {Foss},
  {Frank}, {Freemon}, {Gangler}, {Gawiser}, {Geary}, {Gee}, {Geha}, {Gessner},
  {Gibson}, {Gilmore}, {Glanzman}, {Glick}, {Goldina}, {Goldstein}, {Goodenow},
  {Graham}, {Gressler}, {Gris}, {Guy}, {Guyonnet}, {Haller}, {Harris},
  {Hascall}, {Haupt}, {Hernandez}, {Herrmann}, {Hileman}, {Hoblitt}, {Hodgson},
  {Hogan}, {Howard}, {Huang}, {Huffer}, {Ingraham}, {Innes}, {Jacoby}, {Jain},
  {Jammes}, {Jee}, {Jenness}, {Jernigan}, {Jevremovi{\'c}}, {Johns}, {Johnson},
  {Johnson}, {Jones}, {Juramy-Gilles}, {Juri{\'c}}, {Kalirai}, {Kallivayalil},
  {Kalmbach}, {Kantor}, {Karst}, {Kasliwal}, {Kelly}, {Kessler}, {Kinnison},
  {Kirkby}, {Knox}, {Kotov}, {Krabbendam}, {Krughoff}, {Kub{\'a}nek},
  {Kuczewski}, {Kulkarni}, {Ku}, {Kurita}, {Lage}, {Lambert}, {Lange},
  {Langton}, {Le Guillou}, {Levine}, {Liang}, {Lim}, {Lintott}, {Long},
  {Lopez}, {Lotz}, {Lupton}, {Lust}, {MacArthur}, {Mahabal}, {Mandelbaum},
  {Markiewicz}, {Marsh}, {Marshall}, {Marshall}, {May}, {McKercher}, {McQueen},
  {Meyers}, {Migliore}, {Miller}, {Mills}, {Miraval}, {Moeyens}, {Moolekamp},
  {Monet}, {Moniez}, {Monkewitz}, {Montgomery}, {Morrison}, {Mueller},
  {Muller}, {Mu{\~n}oz Arancibia}, {Neill}, {Newbry}, {Nief}, {Nomerotski},
  {Nordby}, {O'Connor}, {Oliver}, {Olivier}, {Olsen}, {O'Mullane}, {Ortiz},
  {Osier}, {Owen}, {Pain}, {Palecek}, {Parejko}, {Parsons}, {Pease},
  {Peterson}, {Peterson}, {Petravick}, {Libby Petrick}, {Petry},
  {Pierfederici}, {Pietrowicz}, {Pike}, {Pinto}, {Plante}, {Plate}, {Plutchak},
  {Price}, {Prouza}, {Radeka}, {Rajagopal}, {Rasmussen}, {Regnault}, {Reil},
  {Reiss}, {Reuter}, {Ridgway}, {Riot}, {Ritz}, {Robinson}, {Roby}, {Roodman},
  {Rosing}, {Roucelle}, {Rumore}, {Russo}, {Saha}, {Sassolas}, {Schalk},
  {Schellart}, {Schindler}, {Schmidt}, {Schneider}, {Schneider}, {Schoening},
  {Schumacher}, {Schwamb}, {Sebag}, {Selvy}, {Sembroski}, {Seppala}, {Serio},
  {Serrano}, {Shaw}, {Shipsey}, {Sick}, {Silvestri}, {Slater}, {Smith},
  {Smith}, {Sobhani}, {Soldahl}, {Storrie-Lombardi}, {Stover}, {Strauss},
  {Street}, {Stubbs}, {Sullivan}, {Sweeney}, {Swinbank}, {Szalay}, {Takacs},
  {Tether}, {Thaler}, {Thayer}, {Thomas}, {Thornton}, {Thukral}, {Tice},
  {Trilling}, {Turri}, {Van Berg}, {Vanden Berk}, {Vetter}, {Virieux},
  {Vucina}, {Wahl}, {Walkowicz}, {Walsh}, {Walter}, {Wang}, {Wang}, {Warner},
  {Wiecha}, {Willman}, {Winters}, {Wittman}, {Wolff}, {Wood-Vasey}, {Wu},
  {Xin}, {Yoachim}, \& {Zhan}}]{LSST2019}
{Ivezi{\'c}}, {\v{Z}}., {Kahn}, S.~M., {Tyson}, J.~A., {et~al.} 2019, \apj,
  873, 111

\bibitem[{{Jesseit} {et~al.}(2009){Jesseit}, {Cappellari}, {Naab}, {Emsellem},
  \& {Burkert}}]{Jesseit2009}
{Jesseit}, R., {Cappellari}, M., {Naab}, T., {Emsellem}, E., \& {Burkert}, A.
  2009, \mnras, 397, 1202

\bibitem[{Katz {et~al.}(2016)Katz, Lelli, McGaugh, Di~Cintio, Brook, \&
  Schombert}]{Katz2016}
Katz, H., Lelli, F., McGaugh, S.~S., {et~al.} 2016, \mnras, 466, 1648

\bibitem[{{Khalifa} {et~al.}(2017){Khalifa}, {Taha}, {Hassanien}, \&
  {Selim}}]{Khalifa2017}
{Khalifa}, N. E.~M., {Taha}, M. H.~N., {Hassanien}, A.~E., \& {Selim}, I.~M.
  2017, arXiv e-prints, arXiv:1709.02245

\bibitem[{{Kim} {et~al.}(2021){Kim}, {Telea}, {Trager}, \&
  {Roerdink}}]{Kim2021}
{Kim}, Y., {Telea}, A.~C., {Trager}, S.~C., \& {Roerdink}, J. B.~T.~M. 2021,
  arXiv e-prints, arXiv:2110.00317

\bibitem[{Kimm {et~al.}(2018)Kimm, Haehnelt, Blaizot, Katz, Michel-Dansac,
  Garel, Rosdahl, \& Teyssier}]{Kimm2018}
Kimm, T., Haehnelt, M., Blaizot, J., {et~al.} 2018, \mnras, 475, 4617

\bibitem[{{Kormendy} \& {Bender}(2012)}]{Kormendy2012}
{Kormendy}, J. \& {Bender}, R. 2012, \apjs, 198, 2

\bibitem[{{Kremer} {et~al.}(2017){Kremer}, {Stensbo-Smidt}, {Gieseke},
  {Steenstrup Pedersen}, \& {Igel}}]{Kremer2017}
{Kremer}, J., {Stensbo-Smidt}, K., {Gieseke}, F., {Steenstrup Pedersen}, K., \&
  {Igel}, C. 2017, arXiv e-prints, arXiv:1704.04650

\bibitem[{{Lagos} {et~al.}(2022){Lagos}, {Emsellem}, {van de Sande},
  {Harborne}, {Cortese}, {Davison}, {Foster}, \& {Wright}}]{Lagos2022}
{Lagos}, C. d.~P., {Emsellem}, E., {van de Sande}, J., {et~al.} 2022, \mnras,
  509, 4372

\bibitem[{{Lagos} {et~al.}(2018){Lagos}, {Schaye}, {Bah{\'e}}, {Van de Sande},
  {Kay}, {Barnes}, {Davis}, \& {Dalla Vecchia}}]{Lagos2018}
{Lagos}, C.~d.~P., {Schaye}, J., {Bah{\'e}}, Y., {et~al.} 2018, \mnras, 476,
  4327

\bibitem[{{Lahav} {et~al.}(1995){Lahav}, {Naim}, {Buta}, {Corwin}, {de
  Vaucouleurs}, {Dressler}, {Huchra}, {van den Bergh}, {Raychaudhury}, {Sodre},
  \& {Storrie-Lombardi}}]{Lahav1995}
{Lahav}, O., {Naim}, A., {Buta}, R.~J., {et~al.} 1995, Science, 267, 859

\bibitem[{{Laureijs} {et~al.}(2011){Laureijs}, {Amiaux}, {Arduini},
  {Augu{\`e}res}, {Brinchmann}, {Cole}, {Cropper}, {Dabin}, {Duvet}, {Ealet},
  {Garilli}, {Gondoin}, {Guzzo}, {Hoar}, {Hoekstra}, {Holmes}, {Kitching},
  {Maciaszek}, {Mellier}, {Pasian}, {Percival}, {Rhodes}, {Saavedra Criado},
  {Sauvage}, {Scaramella}, {Valenziano}, {Warren}, {Bender}, {Castander},
  {Cimatti}, {Le F{\`e}vre}, {Kurki-Suonio}, {Levi}, {Lilje}, {Meylan},
  {Nichol}, {Pedersen}, {Popa}, {Rebolo Lopez}, {Rix}, {Rottgering},
  {Zeilinger}, {Grupp}, {Hudelot}, {Massey}, {Meneghetti}, {Miller}, {Paltani},
  {Paulin-Henriksson}, {Pires}, {Saxton}, {Schrabback}, {Seidel}, {Walsh},
  {Aghanim}, {Amendola}, {Bartlett}, {Baccigalupi}, {Beaulieu}, {Benabed},
  {Cuby}, {Elbaz}, {Fosalba}, {Gavazzi}, {Helmi}, {Hook}, {Irwin}, {Kneib},
  {Kunz}, {Mannucci}, {Moscardini}, {Tao}, {Teyssier}, {Weller}, {Zamorani},
  {Zapatero Osorio}, {Boulade}, {Foumond}, {Di Giorgio}, {Guttridge}, {James},
  {Kemp}, {Martignac}, {Spencer}, {Walton}, {Bl{\"u}mchen}, {Bonoli},
  {Bortoletto}, {Cerna}, {Corcione}, {Fabron}, {Jahnke}, {Ligori}, {Madrid},
  {Martin}, {Morgante}, {Pamplona}, {Prieto}, {Riva}, {Toledo}, {Trifoglio},
  {Zerbi}, {Abdalla}, {Douspis}, {Grenet}, {Borgani}, {Bouwens}, {Courbin},
  {Delouis}, {Dubath}, {Fontana}, {Frailis}, {Grazian}, {Koppenh{\"o}fer},
  {Mansutti}, {Melchior}, {Mignoli}, {Mohr}, {Neissner}, {Noddle}, {Poncet},
  {Scodeggio}, {Serrano}, {Shane}, {Starck}, {Surace}, {Taylor},
  {Verdoes-Kleijn}, {Vuerli}, {Williams}, {Zacchei}, {Altieri}, {Escudero
  Sanz}, {Kohley}, {Oosterbroek}, {Astier}, {Bacon}, {Bardelli}, {Baugh},
  {Bellagamba}, {Benoist}, {Bianchi}, {Biviano}, {Branchini}, {Carbone},
  {Cardone}, {Clements}, {Colombi}, {Conselice}, {Cresci}, {Deacon}, {Dunlop},
  {Fedeli}, {Fontanot}, {Franzetti}, {Giocoli}, {Garcia-Bellido}, {Gow},
  {Heavens}, {Hewett}, {Heymans}, {Holland}, {Huang}, {Ilbert}, {Joachimi},
  {Jennins}, {Kerins}, {Kiessling}, {Kirk}, {Kotak}, {Krause}, {Lahav}, {van
  Leeuwen}, {Lesgourgues}, {Lombardi}, {Magliocchetti}, {Maguire}, {Majerotto},
  {Maoli}, {Marulli}, {Maurogordato}, {McCracken}, {McLure}, {Melchiorri},
  {Merson}, {Moresco}, {Nonino}, {Norberg}, {Peacock}, {Pello}, {Penny},
  {Pettorino}, {Di Porto}, {Pozzetti}, {Quercellini}, {Radovich}, {Rassat},
  {Roche}, {Ronayette}, {Rossetti}, {Sartoris}, {Schneider}, {Semboloni},
  {Serjeant}, {Simpson}, {Skordis}, {Smadja}, {Smartt}, {Spano}, {Spiro},
  {Sullivan}, {Tilquin}, {Trotta}, {Verde}, {Wang}, {Williger}, {Zhao},
  {Zoubian}, \& {Zucca}}]{Euclid2011}
{Laureijs}, R., {Amiaux}, J., {Arduini}, S., {et~al.} 2011, arXiv e-prints,
  arXiv:1110.3193

\bibitem[{{Li} {et~al.}(2018){Li}, {Mao}, {Cappellari}, {Ge}, {Long}, {Li},
  {Mo}, {Li}, {Zheng}, {Bundy}, {Thomas}, {Brownstein}, {Roman Lopes}, {Law},
  \& {Drory}}]{Li2018}
{Li}, H., {Mao}, S., {Cappellari}, M., {et~al.} 2018, \mnras, 476, 1765

\bibitem[{{Lotz} {et~al.}(2004){Lotz}, {Primack}, \& {Madau}}]{Lotz2004}
{Lotz}, J.~M., {Primack}, J., \& {Madau}, P. 2004, \aj, 128, 163

\bibitem[{{LSST Science Collaboration} {et~al.}(2009){LSST Science
  Collaboration}, {Abell}, {Allison}, {Anderson}, {Andrew}, {Angel}, {Armus},
  {Arnett}, {Asztalos}, {Axelrod}, {Bailey}, {Ballantyne}, {Bankert},
  {Barkhouse}, {Barr}, {Barrientos}, {Barth}, {Bartlett}, {Becker}, {Becla},
  {Beers}, {Bernstein}, {Biswas}, {Blanton}, {Bloom}, {Bochanski}, {Boeshaar},
  {Borne}, {Bradac}, {Brandt}, {Bridge}, {Brown}, {Brunner}, {Bullock},
  {Burgasser}, {Burge}, {Burke}, {Cargile}, {Chandrasekharan}, {Chartas},
  {Chesley}, {Chu}, {Cinabro}, {Claire}, {Claver}, {Clowe}, {Connolly}, {Cook},
  {Cooke}, {Cooray}, {Covey}, {Culliton}, {de Jong}, {de Vries}, {Debattista},
  {Delgado}, {Dell'Antonio}, {Dhital}, {Di Stefano}, {Dickinson}, {Dilday},
  {Djorgovski}, {Dobler}, {Donalek}, {Dubois-Felsmann}, {Durech},
  {Eliasdottir}, {Eracleous}, {Eyer}, {Falco}, {Fan}, {Fassnacht}, {Ferguson},
  {Fernandez}, {Fields}, {Finkbeiner}, {Figueroa}, {Fox}, {Francke}, {Frank},
  {Frieman}, {Fromenteau}, {Furqan}, {Galaz}, {Gal-Yam}, {Garnavich},
  {Gawiser}, {Geary}, {Gee}, {Gibson}, {Gilmore}, {Grace}, {Green}, {Gressler},
  {Grillmair}, {Habib}, {Haggerty}, {Hamuy}, {Harris}, {Hawley}, {Heavens},
  {Hebb}, {Henry}, {Hileman}, {Hilton}, {Hoadley}, {Holberg}, {Holman},
  {Howell}, {Infante}, {Ivezic}, {Jacoby}, {Jain}, {R}, {Jedicke}, {Jee},
  {Garrett Jernigan}, {Jha}, {Johnston}, {Jones}, {Juric}, {Kaasalainen},
  {Styliani}, {Kafka}, {Kahn}, {Kaib}, {Kalirai}, {Kantor}, {Kasliwal},
  {Keeton}, {Kessler}, {Knezevic}, {Kowalski}, {Krabbendam}, {Krughoff},
  {Kulkarni}, {Kuhlman}, {Lacy}, {Lepine}, {Liang}, {Lien}, {Lira}, {Long},
  {Lorenz}, {Lotz}, {Lupton}, {Lutz}, {Macri}, {Mahabal}, {Mandelbaum},
  {Marshall}, {May}, {McGehee}, {Meadows}, {Meert}, {Milani}, {Miller},
  {Miller}, {Mills}, {Minniti}, {Monet}, {Mukadam}, {Nakar}, {Neill}, {Newman},
  {Nikolaev}, {Nordby}, {O'Connor}, {Oguri}, {Oliver}, {Olivier}, {Olsen},
  {Olsen}, {Olszewski}, {Oluseyi}, {Padilla}, {Parker}, {Pepper}, {Peterson},
  {Petry}, {Pinto}, {Pizagno}, {Popescu}, {Prsa}, {Radcka}, {Raddick},
  {Rasmussen}, {Rau}, {Rho}, {Rhoads}, {Richards}, {Ridgway}, {Robertson},
  {Roskar}, {Saha}, {Sarajedini}, {Scannapieco}, {Schalk}, {Schindler},
  {Schmidt}, {Schmidt}, {Schneider}, {Schumacher}, {Scranton}, {Sebag},
  {Seppala}, {Shemmer}, {Simon}, {Sivertz}, {Smith}, {Allyn Smith}, {Smith},
  {Spitz}, {Stanford}, {Stassun}, {Strader}, {Strauss}, {Stubbs}, {Sweeney},
  {Szalay}, {Szkody}, {Takada}, {Thorman}, {Trilling}, {Trimble}, {Tyson}, {Van
  Berg}, {Vanden Berk}, {VanderPlas}, {Verde}, {Vrsnak}, {Walkowicz},
  {Wandelt}, {Wang}, {Wang}, {Warner}, {Wechsler}, {West}, {Wiecha},
  {Williams}, {Willman}, {Wittman}, {Wolff}, {Wood-Vasey}, {Wozniak}, {Young},
  {Zentner}, \& {Zhan}}]{LSSTScienceBook}
{LSST Science Collaboration}, {Abell}, P.~A., {Allison}, J., {et~al.} 2009,
  arXiv e-prints, arXiv:0912.0201

\bibitem[{Marin {et~al.}(2013)Marin, Sucar, Gonzalez, \& Diaz}]{Marin2013}
Marin, M., Sucar, L., Gonzalez, J., \& Diaz, R. 2013, 438

\bibitem[{{McAlpine} {et~al.}(2016){McAlpine}, {Helly}, {Schaller}, {Trayford},
  {Qu}, {Furlong}, {Bower}, {Crain}, {Schaye}, {Theuns}, {Dalla Vecchia},
  {Frenk}, {McCarthy}, {Jenkins}, {Rosas-Guevara}, {White}, {Baes}, {Camps}, \&
  {Lemson}}]{McAlpine2016}
{McAlpine}, S., {Helly}, J.~C., {Schaller}, M., {et~al.} 2016, Astronomy and
  Computing, 15, 72

\bibitem[{{McInnes} {et~al.}(2018){McInnes}, {Healy}, \&
  {Melville}}]{McInnesUMAP2018}
{McInnes}, L., {Healy}, J., \& {Melville}, J. 2018, arXiv e-prints,
  arXiv:1802.03426

\bibitem[{Mittal {et~al.}(2019)Mittal, Soorya, Nagrath, \&
  Hemanth}]{Mittal2019}
Mittal, A., Soorya, A., Nagrath, P., \& Hemanth, D.~J. 2019, Earth Science
  Informatics, 13, 601

\bibitem[{{Naab} {et~al.}(2014){Naab}, {Oser}, {Emsellem}, {Cappellari},
  {Krajnovi{\'c}}, {McDermid}, {Alatalo}, {Bayet}, {Blitz}, {Bois}, {Bournaud},
  {Bureau}, {Crocker}, {Davies}, {Davis}, {de Zeeuw}, {Duc}, {Hirschmann},
  {Johansson}, {Khochfar}, {Kuntschner}, {Morganti}, {Oosterloo}, {Sarzi},
  {Scott}, {Serra}, {van de Ven}, {Weijmans}, \& {Young}}]{Naab2014}
{Naab}, T., {Oser}, L., {Emsellem}, E., {et~al.} 2014, \mnras, 444, 3357

\bibitem[{{Papovich} {et~al.}(2003){Papovich}, {Giavalisco}, {Dickinson},
  {Conselice}, \& {Ferguson}}]{Papovich2003}
{Papovich}, C., {Giavalisco}, M., {Dickinson}, M., {Conselice}, C.~J., \&
  {Ferguson}, H.~C. 2003, \apj, 598, 827

\bibitem[{{Pedrosa} \& {Tissera}(2015)}]{Pedrosa2015}
{Pedrosa}, S.~E. \& {Tissera}, P.~B. 2015, \aap, 584, A43

\bibitem[{{Peebles}(1969)}]{Peebles1969}
{Peebles}, P.~J.~E. 1969, \apj, 155, 393

\bibitem[{{Penoyre} {et~al.}(2017){Penoyre}, {Moster}, {Sijacki}, \&
  {Genel}}]{Penoyre2017}
{Penoyre}, Z., {Moster}, B.~P., {Sijacki}, D., \& {Genel}, S. 2017, \mnras,
  468, 3883

\bibitem[{{Planck Collaboration} {et~al.}(2014{\natexlab{a}}){Planck
  Collaboration}, {Ade}, {Aghanim}, {Alves}, {Armitage-Caplan}, {Arnaud},
  {Ashdown}, {Atrio-Barandela}, {Aumont}, {Aussel}, \& et~al.}]{Planck2014a}
{Planck Collaboration}, {Ade}, P.~A.~R., {Aghanim}, N., {et~al.}
  2014{\natexlab{a}}, \aap, 571, A1

\bibitem[{{Planck Collaboration} {et~al.}(2014{\natexlab{b}}){Planck
  Collaboration}, {Ade}, {Aghanim}, {Armitage-Caplan}, {Arnaud}, {Ashdown},
  {Atrio-Barandela}, {Aumont}, {Baccigalupi}, {Banday}, \&
  et~al.}]{Planck2014b}
{Planck Collaboration}, {Ade}, P.~A.~R., {Aghanim}, N., {et~al.}
  2014{\natexlab{b}}, \aap, 571, A16

\bibitem[{{Portillo} {et~al.}(2020){Portillo}, {Parejko}, {Vergara}, \&
  {Connolly}}]{Potrillo2020}
{Portillo}, S. K.~N., {Parejko}, J.~K., {Vergara}, J.~R., \& {Connolly}, A.~J.
  2020, \aj, 160, 45

\bibitem[{{Raddick} {et~al.}(2007){Raddick}, {Lintott}, {Schawinski}, {Thomas},
  {Nichol}, {Andreescu}, {Bamford}, {Land}, {Murray}, {Slosar}, {Szalay},
  {Vandenberg}, \& {Galaxy Zoo Team}}]{Raddick2007}
{Raddick}, J., {Lintott}, C.~J., {Schawinski}, K., {et~al.} 2007, in American
  Astronomical Society Meeting Abstracts, Vol. 211, American Astronomical
  Society Meeting Abstracts, 94.03

\bibitem[{{Reis} {et~al.}(2021){Reis}, {Rotman}, {Poznanski}, {Prochaska}, \&
  {Wolf}}]{Reis2021}
{Reis}, I., {Rotman}, M., {Poznanski}, D., {Prochaska}, J.~X., \& {Wolf}, L.
  2021, Astronomy and Computing, 34, 100437

\bibitem[{{Rosas-Guevara} {et~al.}(2015){Rosas-Guevara}, {Bower}, {Schaye},
  {Furlong}, {Frenk}, {Booth}, {Crain}, {Dalla Vecchia}, {Schaller}, \&
  {Theuns}}]{Rosas2015}
{Rosas-Guevara}, Y.~M., {Bower}, R.~G., {Schaye}, J., {et~al.} 2015, \mnras,
  454, 1038

\bibitem[{{Rosito} {et~al.}(2018){Rosito}, {Pedrosa}, {Tissera}, {Avila-Reese},
  {Lacerna}, {Bignone}, {Ibarra-Medel}, \& {Varela}}]{Rosito2018}
{Rosito}, M.~S., {Pedrosa}, S.~E., {Tissera}, P.~B., {et~al.} 2018, \aap, 614,
  A85

\bibitem[{{Rosito} {et~al.}(2021){Rosito}, {Pedrosa}, {Tissera}, {Chisari},
  {Dom{\'\i}nguez-Tenreiro}, {Dubois}, {Peirani}, {Devriendt}, {Pichon}, \&
  {Slyz}}]{Rosito2021}
{Rosito}, M.~S., {Pedrosa}, S.~E., {Tissera}, P.~B., {et~al.} 2021, \aap, 652,
  A44

\bibitem[{{Rosito} {et~al.}(2019{\natexlab{a}}){Rosito}, {Tissera}, {Pedrosa},
  \& {Lagos}}]{Rosito2019a}
{Rosito}, M.~S., {Tissera}, P.~B., {Pedrosa}, S.~E., \& {Lagos}, C.~D.~P.
  2019{\natexlab{a}}, \aap, 629, L3

\bibitem[{{Rosito} {et~al.}(2019{\natexlab{b}}){Rosito}, {Tissera}, {Pedrosa},
  \& {Rosas-Guevara}}]{Rosito2019b}
{Rosito}, M.~S., {Tissera}, P.~B., {Pedrosa}, S.~E., \& {Rosas-Guevara}, Y.
  2019{\natexlab{b}}, \aap, 629, A37

\bibitem[{{S{\'a}nchez} {et~al.}(2012){S{\'a}nchez}, {Kennicutt}, {Gil de Paz},
  {van de Ven}, {V{\'{\i}}lchez}, {Wisotzki}, {Walcher}, {Mast}, {Aguerri},
  {Albiol-P{\'e}rez}, \& {Alonso-Herrero}}]{CALIFA2012}
{S{\'a}nchez}, S.~F., {Kennicutt}, R.~C., {Gil de Paz}, A., {et~al.} 2012,
  \aap, 538, A8

\bibitem[{{S{\'a}nchez Almeida} {et~al.}(2010){S{\'a}nchez Almeida}, {Aguerri},
  {Mu{\~n}oz-Tu{\~n}{\'o}n}, \& {de Vicente}}]{Almeida2010}
{S{\'a}nchez Almeida}, J., {Aguerri}, J.~A.~L., {Mu{\~n}oz-Tu{\~n}{\'o}n}, C.,
  \& {de Vicente}, A. 2010, \apj, 714, 487

\bibitem[{{S{\'a}nchez Almeida} \& {Allende Prieto}(2013)}]{Almeida2013}
{S{\'a}nchez Almeida}, J. \& {Allende Prieto}, C. 2013, \apj, 763, 50

\bibitem[{{Sarmiento} {et~al.}(2021){Sarmiento}, {Huertas-Company}, {Knapen},
  {S{\'a}nchez}, {Dom{\'\i}nguez S{\'a}nchez}, {Drory}, \&
  {Falc{\'o}n-Barroso}}]{Sarmiento2021}
{Sarmiento}, R., {Huertas-Company}, M., {Knapen}, J.~H., {et~al.} 2021, \apj,
  921, 177

\bibitem[{{Scannapieco} {et~al.}(2008){Scannapieco}, {Tissera}, {White}, \&
  {Springel}}]{scan2008}
{Scannapieco}, C., {Tissera}, P.~B., {White}, S.~D.~M., \& {Springel}, V. 2008,
  \mnras, 389, 1137

\bibitem[{{Schaye} {et~al.}(2015){Schaye}, {Crain}, {Bower}, {Furlong},
  {Schaller}, {Theuns}, {Dalla Vecchia}, {Frenk}, {McCarthy}, {Helly},
  {Jenkins}, {Rosas-Guevara}, {White}, {Baes}, {Booth}, {Camps}, {Navarro},
  {Qu}, {Rahmati}, {Sawala}, {Thomas}, \& {Trayford}}]{Schaye2015}
{Schaye}, J., {Crain}, R.~A., {Bower}, R.~G., {et~al.} 2015, \mnras, 446, 521

\bibitem[{{Schaye} \& {Dalla Vecchia}(2008)}]{Schaye2008}
{Schaye}, J. \& {Dalla Vecchia}, C. 2008, \mnras, 383, 1210

\bibitem[{{Schulze} {et~al.}(2018){Schulze}, {Remus}, {Dolag}, {Burkert},
  {Emsellem}, \& {van de Ven}}]{Schulze2018}
{Schulze}, F., {Remus}, R.-S., {Dolag}, K., {et~al.} 2018, \mnras, 480, 4636

\bibitem[{{Scott} {et~al.}(2018){Scott}, {van de Sande}, {Croom}, {Groves},
  {Owers}, {Poetrodjojo}, {D'Eugenio}, {Medling}, {Barat}, {Barone},
  {Bland-Hawthorn}, {Brough}, {Bryant}, {Cortese}, {Foster}, {Green}, {Oh},
  {Colless}, {Drinkwater}, {Driver}, {Goodwin}, {Gunawardhana}, {Federrath},
  {Harischandra}, {Jin}, {Lawrence}, {Lorente}, {Mannering}, {O'Toole},
  {Richards}, {Sanchez}, {Schaefer}, {Sealey}, {Sharp}, {Sweet}, {Taranu}, \&
  {Varidel}}]{Scott2018}
{Scott}, N., {van de Sande}, J., {Croom}, S.~M., {et~al.} 2018, \mnras, 481,
  2299

\bibitem[{Selim {et~al.}(2016)Selim, Arabi, \& M.El}]{Selim2016}
Selim, I., Arabi, E., \& M.El, B. 2016, International Journal of Computer
  Applications, 137, 4

\bibitem[{{S{\'e}rsic}(1968)}]{Sersic1968}
{S{\'e}rsic}, J.~L. 1968, {Atlas de galaxias australes}

\bibitem[{{Springel} {et~al.}(2005){Springel}, {Di Matteo}, \&
  {Hernquist}}]{Springel+2005}
{Springel}, V., {Di Matteo}, T., \& {Hernquist}, L. 2005, \mnras, 361, 776

\bibitem[{{Springel} {et~al.}(2001){Springel}, {Yoshida}, \&
  {White}}]{springel2001}
{Springel}, V., {Yoshida}, N., \& {White}, S.~D.~M. 2001, \na, 6, 79

\bibitem[{{Storrie-Lombardi} {et~al.}(1992){Storrie-Lombardi}, {Lahav},
  {Sodre}, \& {Storrie-Lombardi}}]{Storrie1992}
{Storrie-Lombardi}, M.~C., {Lahav}, O., {Sodre}, L., J., \& {Storrie-Lombardi},
  L. 1992, in American Astronomical Society Meeting Abstracts, Vol. 181,
  American Astronomical Society Meeting Abstracts, 65.08

\bibitem[{{Tissera}(2000)}]{tissera2000}
{Tissera}, P.~B. 2000, \apj, 534, 636

\bibitem[{{Tissera} \& {Dominguez-Tenreiro}(1998)}]{Tissera1998}
{Tissera}, P.~B. \& {Dominguez-Tenreiro}, R. 1998, \mnras, 297, 177

\bibitem[{{Tissera} {et~al.}(2016{\natexlab{a}}){Tissera}, {Machado},
  {Sanchez-Blazquez}, {Pedrosa}, {S{\'a}nchez}, {Snaith}, \&
  {Vilchez}}]{Tissera2016}
{Tissera}, P.~B., {Machado}, R.~E.~G., {Sanchez-Blazquez}, P., {et~al.}
  2016{\natexlab{a}}, \aap, 592, A93

\bibitem[{{Tissera} {et~al.}(2016{\natexlab{b}}){Tissera}, {Pedrosa},
  {Sillero}, \& {Vilchez}}]{Tissera2016b}
{Tissera}, P.~B., {Pedrosa}, S.~E., {Sillero}, E., \& {Vilchez}, J.~M.
  2016{\natexlab{b}}, \mnras, 456, 2982

\bibitem[{{Tissera} {et~al.}(2019){Tissera}, {Rosas-Guevara}, {Bower}, {Crain},
  {del P Lagos}, {Schaller}, {Schaye}, \& {Theuns}}]{Tissera2019}
{Tissera}, P.~B., {Rosas-Guevara}, Y., {Bower}, R.~G., {et~al.} 2019, \mnras,
  482, 2208

\bibitem[{{Tissera} {et~al.}(2012){Tissera}, {White}, \&
  {Scannapieco}}]{tissera2012}
{Tissera}, P.~B., {White}, S.~D.~M., \& {Scannapieco}, C. 2012, \mnras, 420,
  255

\bibitem[{{Tohill} {et~al.}(2021){Tohill}, {Ferreira}, {Conselice}, {Bamford},
  \& {Ferrari}}]{Tohill2021}
{Tohill}, C., {Ferreira}, L., {Conselice}, C.~J., {Bamford}, S.~P., \&
  {Ferrari}, F. 2021, \apj, 916, 4

\bibitem[{{Tonini} {et~al.}(2016){Tonini}, {Mutch}, {Croton}, \&
  {Wyithe}}]{Tonini2016}
{Tonini}, C., {Mutch}, S.~J., {Croton}, D.~J., \& {Wyithe}, J.~S.~B. 2016,
  \mnras, 459, 4109

\bibitem[{{Uzeirbegovic} {et~al.}(2020){Uzeirbegovic}, {Geach}, \&
  {Kaviraj}}]{Uzeirbegovic2020}
{Uzeirbegovic}, E., {Geach}, J.~E., \& {Kaviraj}, S. 2020, \mnras, 498, 4021

\bibitem[{{van de Sande} {et~al.}(2021){van de Sande}, {Vaughan}, {Cortese},
  {Scott}, {Bland-Hawthorn}, {Croom}, {Lagos}, {Brough}, {Bryant}, {Devriendt},
  {Dubois}, {D'Eugenio}, {Foster}, {Fraser-McKelvie}, {Harborne}, {Lawrence},
  {Oh}, {Owers}, {Poci}, {Remus}, {Richards}, {Schulze}, {Sweet}, {Varidel}, \&
  {Welker}}]{vdSande2021}
{van de Sande}, J., {Vaughan}, S.~P., {Cortese}, L., {et~al.} 2021, \mnras,
  505, 3078

\bibitem[{{van de Ven} \& {van der Wel}(2021)}]{vandeven2021}
{van de Ven}, G. \& {van der Wel}, A. 2021, \apj, 914, 45

\bibitem[{{Veale} {et~al.}(2017){Veale}, {Ma}, {Thomas}, {Greene}, {McConnell},
  {Walsh}, {Ito}, {Blakeslee}, \& {Janish}}]{Veale2017}
{Veale}, M., {Ma}, C.-P., {Thomas}, J., {et~al.} 2017, \mnras, 464, 356

\bibitem[{{Vika} {et~al.}(2015){Vika}, {Vulcani}, {Bamford}, {H{\"a}u{\ss}ler},
  \& {Rojas}}]{Vika2015}
{Vika}, M., {Vulcani}, B., {Bamford}, S.~P., {H{\"a}u{\ss}ler}, B., \& {Rojas},
  A.~L. 2015, \aap, 577, A97

\bibitem[{{Walo-Mart{\'\i}n} {et~al.}(2020){Walo-Mart{\'\i}n},
  {Falc{\'o}n-Barroso}, {Dalla Vecchia}, {P{\'e}rez}, \&
  {Negri}}]{WaloMartin2020}
{Walo-Mart{\'\i}n}, D., {Falc{\'o}n-Barroso}, J., {Dalla Vecchia}, C.,
  {P{\'e}rez}, I., \& {Negri}, A. 2020, \mnras, 494, 5652

\bibitem[{{Weijmans} {et~al.}(2014){Weijmans}, {de Zeeuw}, {Emsellem},
  {Krajnovi{\'c}}, {Lablanche}, {Alatalo}, {Blitz}, {Bois}, {Bournaud},
  {Bureau}, {Cappellari}, {Crocker}, {Davies}, {Davis}, {Duc}, {Khochfar},
  {Kuntschner}, {McDermid}, {Morganti}, {Naab}, {Oosterloo}, {Sarzi}, {Scott},
  {Serra}, {Verdoes Kleijn}, \& {Young}}]{Weijmans2014}
{Weijmans}, A.-M., {de Zeeuw}, P.~T., {Emsellem}, E., {et~al.} 2014, \mnras,
  444, 3340

\bibitem[{{Wiersma} {et~al.}(2009){Wiersma}, {Schaye}, \&
  {Smith}}]{wiersma2009}
{Wiersma}, R.~P.~C., {Schaye}, J., \& {Smith}, B.~D. 2009, \mnras, 393, 99

\bibitem[{{Yoon} \& {Im}(2020)}]{Yoon2020}
{Yoon}, Y. \& {Im}, M. 2020, \apj, 893, 117

\end{thebibliography}

\begin{appendix} 

\section{Correlation between $\lambda_B$ and $\lambda_R$}
\label{app:spin}

\FloatBarrier

In this work, we have analysed the distributions of two spin parameters: $\lambda_B$ \citep{Bullock2001} and $\lambda_R$ \citep{Emsellem2007} for galaxies observed at different inclinations.
We remark that the former is obtained from the particle 3D distributions and therefore it does not depend on the inclination. 
This parameter ($\lambda_B$) plays a role in the description of the variation of the specific angular momentum as a function of the halo mass.
\cite{Bullock2001} propose it as an alternative definition of the spin parameter generalising that used before in the literature \citep[e.g.][]{Peebles1969}.
On the other hand, the spin parameter defined by \cite{Emsellem2007}, $\lambda_R$, despite also quantifying angular momentum, is computed from bidimensional kinematic information and observational descriptions of rotation and shape based on it are more accurate than those obtained from anisotropy diagrams \citep{Illingworth1977, Binney1978, Davies1983}.
\cite{Harborne2019} find an approximate functional relation between these two parameters from a set of N-body realization of galaxies.

We study the monotonic (increasing) relationship between these parameters by means of the Spearman correlation coefficient. In Table \ref{table:lambdas} we show these coefficients computed at different inclinations.
Although in all cases the low p-values ($\sim 0$) indicate a significant relationship, it can be appreciated that the coefficients decrease with the angle. This is related with the fact that the rotation becomes less noticeable at low inclinations. 

\begin{figure}
    \centering
    \includegraphics[width=0.242\textwidth]{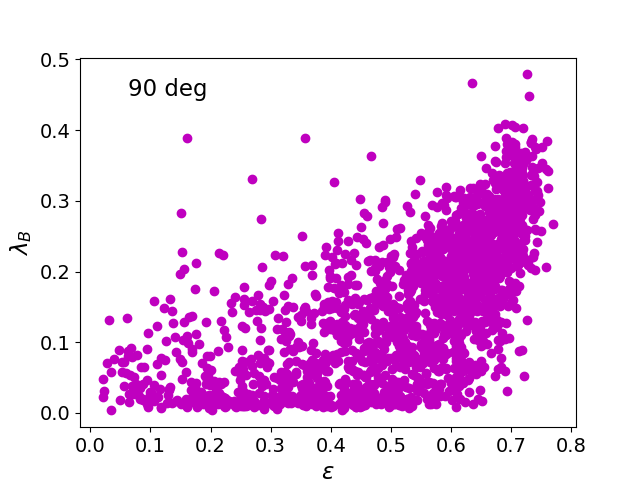}
    \includegraphics[width=0.242\textwidth]{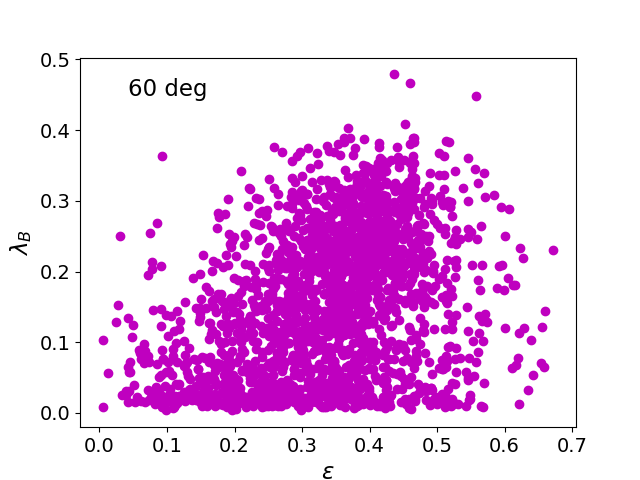} \\
    \includegraphics[width=0.242\textwidth]{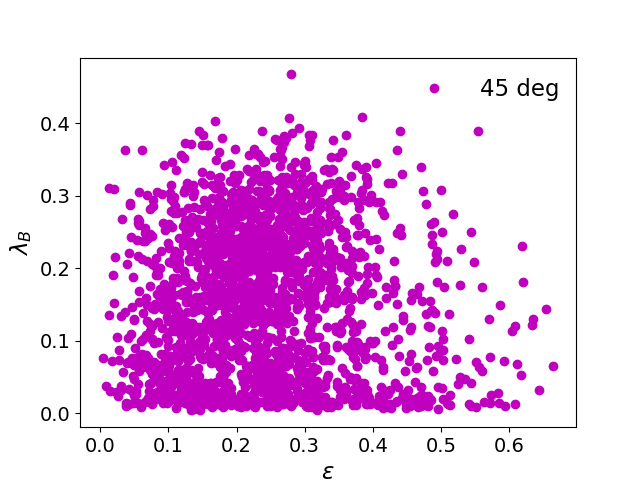} 
    \includegraphics[width=0.242\textwidth]{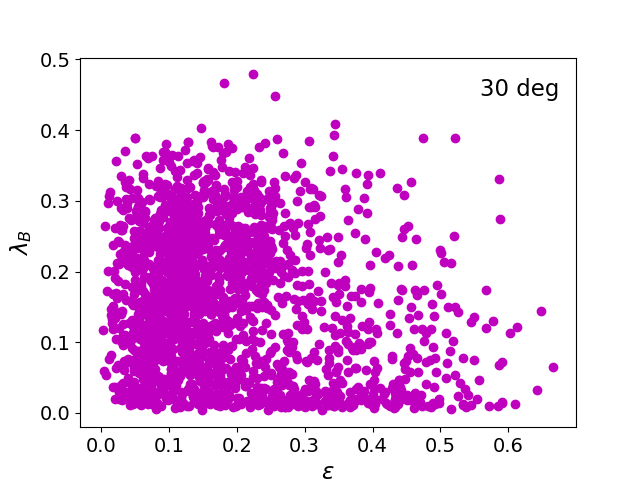} \\
    \includegraphics[width=0.242\textwidth]{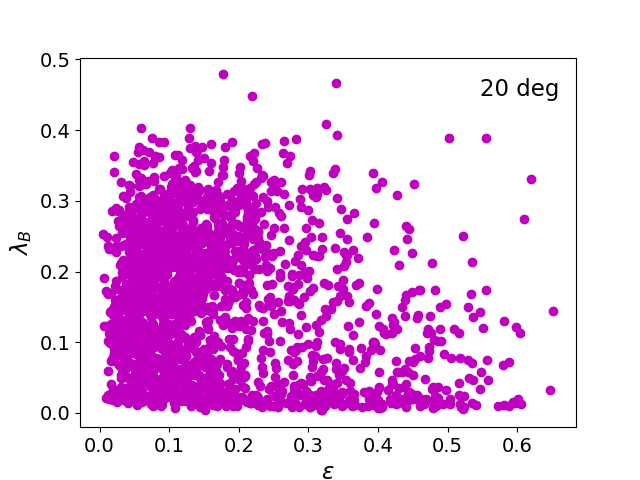}
    \includegraphics[width=0.242\textwidth]{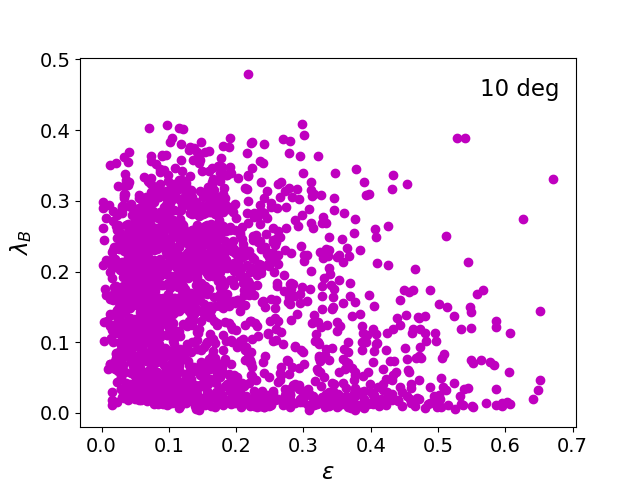} \\
    \includegraphics[width=0.242\textwidth]{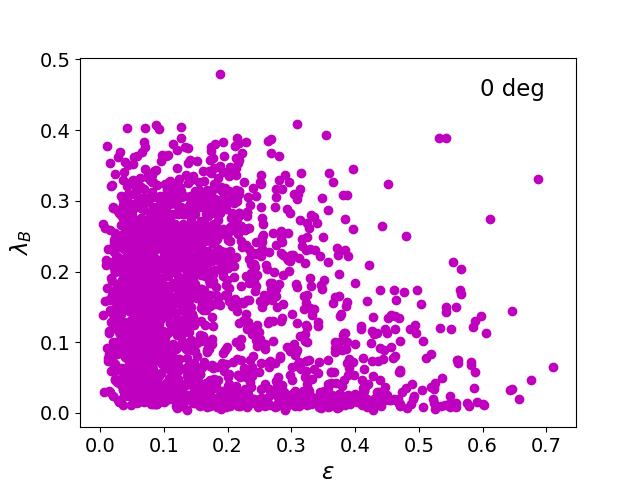}
    \caption{$\lambda_B$ as a function of $\varepsilon$ where ellipticities are computed at different inclinations. The decrease in the values of $\varepsilon$ with decreasing inclination indicates that galaxies are seen rounder when observed at low inclinations.}
    \label{fig:lambdaBeps}
\end{figure}

Figure \ref{fig:lambdaBeps} shows $\lambda_B$ as a function of $\varepsilon$ instead of the $\lambda_R-\varepsilon$ plane widely used to separate SRs and FRs. We measure $\varepsilon$ projected at different inclinations. 
It can be appreciated that at low inclinations galaxies are seen rounder than they would at high inclinations while $\lambda_B$ remains the same.
However, the correlation between both spin parameters may justify the use of $\lambda_B$ instead of $\lambda_R$ to analyse the relation between rotation and shape.
Because of the differences in the Spearman coefficients, the analysis would be more reliable at higher inclinations.

\begin{table}[h]
\caption{Spearman coefficients between $\lambda_B$ and $\lambda_R$ computed at different inclinations.} 
\label{table:lambdas}      
\centering
\footnotesize
\begin{tabular}{cc} 
\hline\hline   
Inclination & Spearman coefficient \\
\hline 
90 deg & 0.919 \\
60 deg & 0.907 \\
45 deg & 0.886 \\
30 deg & 0.833 \\
20 deg & 0.738 \\
10 deg & 0.490 \\
0 deg & 0.112 \\
  \hline
\end{tabular} \\
\end{table}

\FloatBarrier

\clearpage

\section{Dimensionality of the output space}
\label{app:dimensions}

In this Appendix, we explore the intrinsic dimensionality of the problem and discuss how the method would be impacted if more than two dimensions are considered in the clustering algorithm. 

\subsection{Analysis of the number of dimensions}

\FloatBarrier

Throughout this work, we focus on bidimensional projections on which we applied the HDBSCAN clustering algorithm. It is reasonable to be concerned about the loss of information when reducing very high dimensional data to only two components.
However, we show that \nombre is able to produce galaxy groups that are meaningful without using labels. 
Furthermore, classical works about galaxy kinematics \citep[e.g.][]{Emsellem2007, Emsellem2011, cappellarireview2016, vdSande2021} divide SRs and FRs using only two dimensions ($\varepsilon$ and $\lambda_R$) and those classifications are proven to be insightful for subsequent studies.

Because UMAP lacks the interpretability of the dimensions of the embeddings, as well as a quantification of the explained variance \citep{McInnesUMAP2018}, we turn to PCA to obtain an approximation of the information retained by the dimensionality reduction process. PCA has been previously used to study galaxy morphology. For instance, \cite{Uzeirbegovic2020} applied PCA to galaxy images and found that 85 per-cent of the variance could be explained using just two components. 

In Table~\ref{table:dimensions} we show the percentage of variance that a PCA algorithm would explain for the line-of-sight velocity maps (Sec.~\ref{sec:vlos}) and the set of all the types of kinematic maps (Sec.~\ref{sec:call}).
This linear algorithm leads to non-negligible information loss with the explained variance ranging from 47.6 per-cent to 68.6 per-cent. 
However, a non-linear method such as UMAP outperforms other algorithms in the preservation of global structure and may lead to better performance compared to PCA, for example when using classifiers such as k-nearest neighbours \citep{McInnesUMAP2018}.

\begin{table}[h]
\caption{Analysis of the explained variance (\% var) for two and ten principal components (PCs) and number of PCs needed to explain 75 per-cent of the variance for the PCA algorithm applied to the sample of (a)  the line-of-sight velocity maps (Experiment 1) and (b) the three types of kinematic maps (Experiment 2) for different inclinations.} 
\label{table:dimensions}
\resizebox{\columnwidth}{!}{
\begin{tabular}{lllllll}
\hline \hline
 & \multicolumn{3}{c}{Experiment 1} & \multicolumn{3} {c}{Experiment 2} \\ \hline
 & \multicolumn{1}{l}{90°}  & \multicolumn{1}{l}{60°} & 45°   & \multicolumn{1}{l}{90°} & \multicolumn{1}{l}{60°} & 45° \\
 \begin{tabular}[c]{@{}l@{}}\% var\\ 2 PCs\end{tabular} & \multicolumn{1}{l}{47.6\%} & \multicolumn{1}{l}{55.3\%} & 50.8\% & \multicolumn{1}{l}{60.1\%} & \multicolumn{1}{l}{60.5\%} & 59.0\% \\
 \begin{tabular}[c]{@{}l@{}}\% var\\ 10 PCs\end{tabular}        & \multicolumn{1}{l}{50.5\%} & \multicolumn{1}{l}{57.6\%} & 53.0\% & \multicolumn{1}{l}{67.1\%} & \multicolumn{1}{l}{68.6\%} & 67.9\% \\
 \begin{tabular}[c]{@{}l@{}}N$^{\circ}$ PCs\\ to explain\\ 75\% var\end{tabular} & \multicolumn{1}{l}{141} & \multicolumn{1}{l}{121} & 146 & \multicolumn{1}{l}{98} & \multicolumn{1}{l}{81} & 89 \\
 \hline
\end{tabular}
}
\end{table}

\FloatBarrier

\subsection{Clustering on a 10-dimensional UMAP embedding}

\FloatBarrier

\begin{figure}[h]
    \centering
    \includegraphics[width=0.237\textwidth]{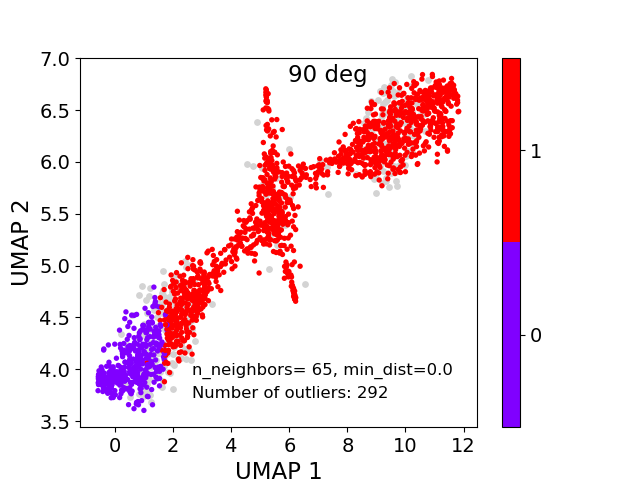}
    \includegraphics[width=0.237\textwidth]{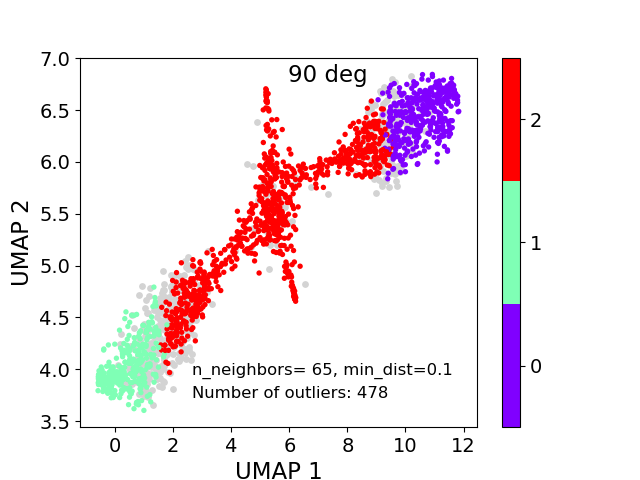} \\
    \includegraphics[width=0.237\textwidth]{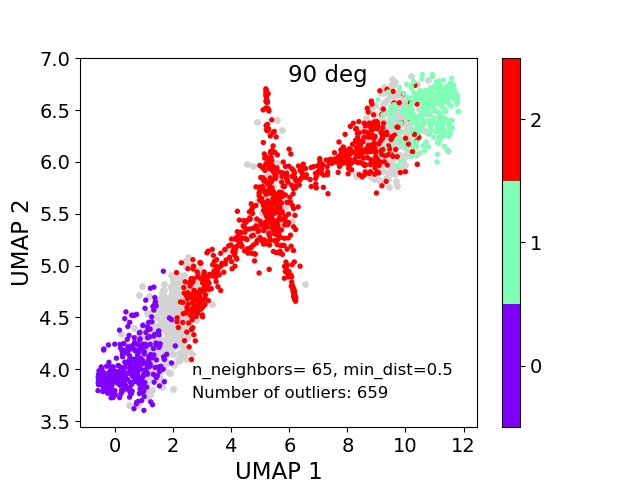} 
    \includegraphics[width=0.237\textwidth]{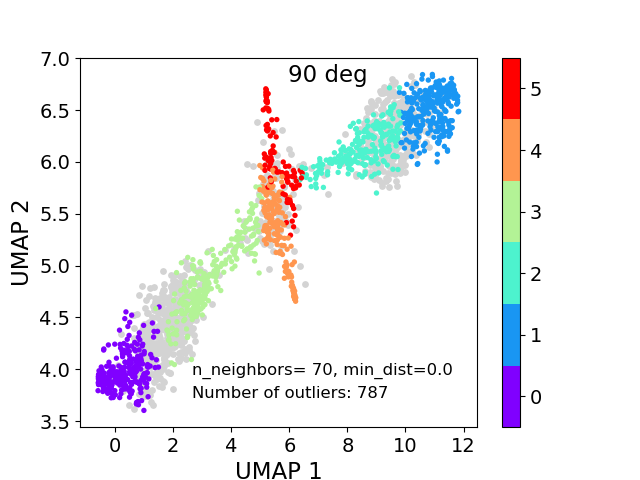} \\
    \includegraphics[width=0.237\textwidth]{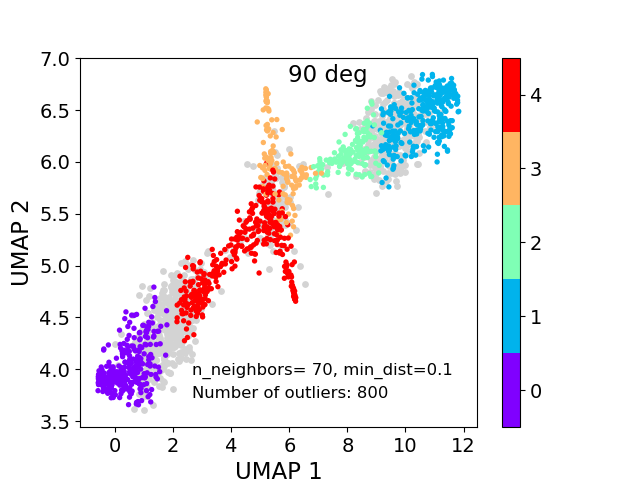}
    \includegraphics[width=0.237\textwidth]{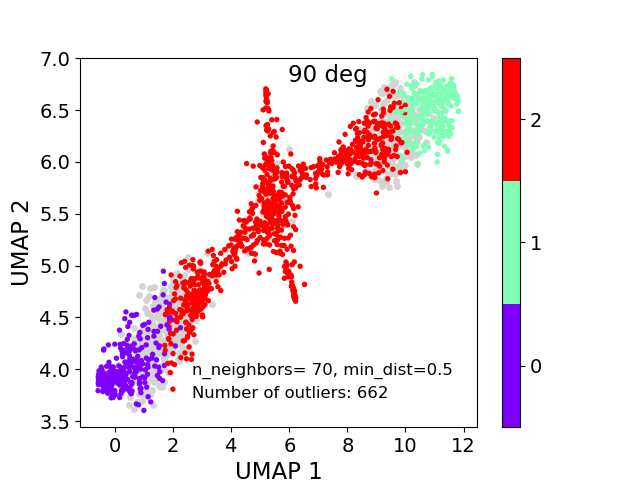} \\  \includegraphics[width=0.237\textwidth]{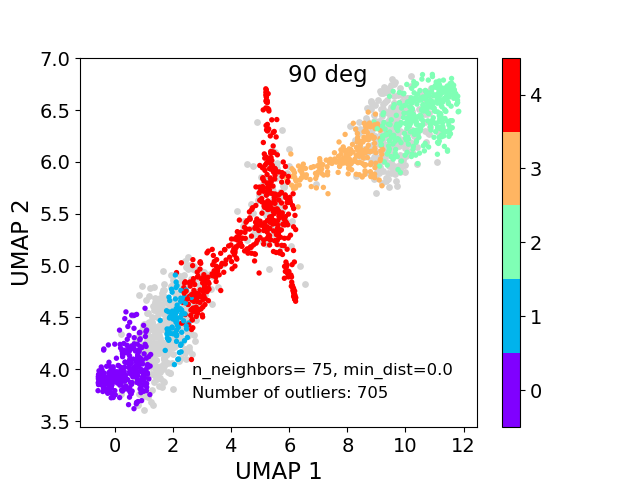}
    \includegraphics[width=0.237\textwidth]{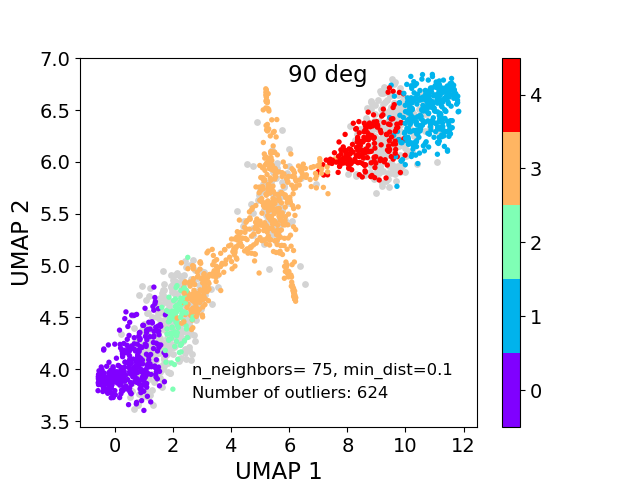}
    \includegraphics[width=0.237\textwidth]{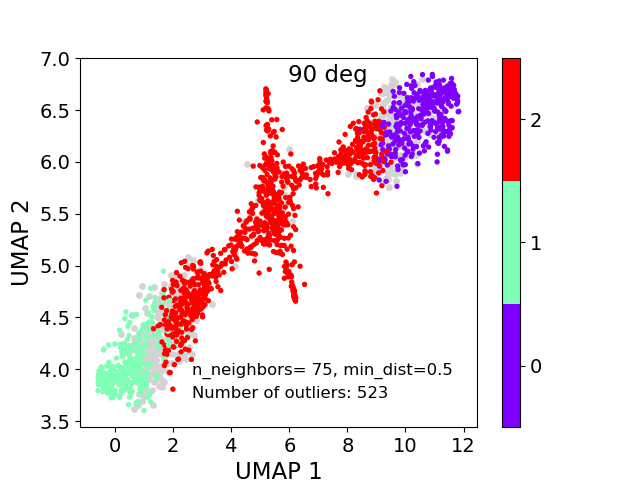}  
    \caption{Bidimensional UMAP projection coloured according to HDBSCAN clustering applied to a 10 dimensional UMAP embdedding for the set of line-of-sight velocity maps of edge-on galaxies (Sec.~\ref{sec:vlos1}) including the outliers as grey dots. We also include the values of n\_neighbors and min\_dist as well as the number of outliers.}
    \label{fig:clust10}
\end{figure}

To explore the effects of using a larger number of dimensions for the clustering part of this work, we conduct experiments using 10-dimensional embeddings.
In Fig.~\ref{fig:clust10}, we show the bidimensional projections computed in Sec.~\ref{sec:vlos1} coloured by the HDBSCAN clustering of the 10-dimensional UMAP embeddings.
We use a small (hyper)parameter grid that includes the ones utilized for our analysis (Table~\ref{table:params}, Experiment 1).
We modify n\_neighbors, min\_dist, and set n\_components = 10.
In all cases, we find a large number of outliers, especially those in which we observe more than three clusters.
In general, we find that using ten components instead of two does not improve the quality of the clustering significantly or make it more meaningful or robust.
As mentioned in Sec.~\ref{sec: method}, the selection of (hyper)parameters is not straightforward and a larger number of dimensions appears to make the clustering more sensitive to small changes of (hyper)parameter values.

Although it would be worthwhile to explore the potential improvements we would obtain by clustering higher dimensional embeddings, the method we present here leads to meaningful classifications by using bidimensional projections.
This exploration shall be analysed in a future work and may be helpful in data leveraging and generalization of \nombre. 

\FloatBarrier

\end{appendix}

\end{document}